\documentclass[twocolumn, amsmath,amssymb,
aps,dbfloatfix,floats]{revtex4}
\usepackage{dcolumn}
\usepackage{bm}
\usepackage[pdftex]{graphicx}
\usepackage{float}
\usepackage{url}
\usepackage{hyperref}
\usepackage{xstring}
\usepackage{tabularx}
\usepackage{dblfloatfix}

\def\KdS{Kerr--de~Sitter}

\def\be{\begin{equation}} 
\def\ee{\end{equation}}
\def\beq{\begin{equation}} 
\def\eeq{\end{equation}}
\def\bea{\begin{eqnarray}} 
\def\eea{\end{eqnarray}}
\def\din{\,\mathrm{d}} 
\def\dbe{\mathrm{d}}
\def\bet{\begin{tabular}} \def\ent{\end{tabular}}
\def\cale{{\cal E}} 
\def\sign{\,\mathrm{sign\,}}

\begin{document}

\title{Spherical photon orbits in the field of Kerr naked singularities}

\author{Daniel Charbul\'{a}k}
\email	{daniel.charbulak@fpf.slu.cz}
\author{Zden\v{e}k Stuchl\'{i}k}
\email	{zdenek.stuchlik@fpf.slu.cz}

\address{Institute of Physics and Research Centre of Theoretical Physics and Astrophysics, 
	Faculty of Philosophy and Science, 
	Silesian university in Opava, 
	Bezru\v{c}ovo n\'{a}m. 13, CZ-746 01 Opava, Czech Republic}

\date{\today}

\begin{abstract}
For the Kerr naked singularity (KNS) spacetimes, we study properties of spherical photon orbits (SPOs) confined to constant Boyer-Lindquist radius $r$. Some new features of the SPOs are found, having no counterparts in the Kerr black hole (KBH) spacetimes, especially stable orbits that could be pure prograde/retrograde, or with turning point in the azimuthal direction. At $r>1$ ($r<1$) the covariant photon energy $\cale > 0$ ($\cale < 0$), at $r=1$ there is $\cale = 0$. All unstable orbits must have $\cale > 0$. It is shown that the polar SPOs can exist only in the spacetimes with dimensionless spin $a < 1.7996$. Existence of closed SPOs with vanishing total change of the azimuth is demonstrated. Classification of the KNS and KBH spacetimes in dependence on their dimensionless spin $a$ is proposed, considering the properties of the SPOs. For selected types of the KNS spacetimes, typical SPOs are constructed, including the closed paths. It is shown that the stable SPOs intersect the equatorial plane in a region of stable circular orbits of test particles, depending on the spin $a$. Relevance of this intersection for the Keplerian accretion discs is outlined and observational effects are estimated. 
\end{abstract}

\maketitle

\section{Introduction}\label{intro}

Large effort has been devoted to studies of null geodesics in the gravitational field of compact objects, because the null geodesics govern motion of photons that carry information on the physical processes in strong gravity to distant observers, giving thus direct signatures of the extraordinary character of the spacetime around compact objects that influences both the physical processes and the photon motion. The most detailed studies were devoted to the Kerr geometry that is assumed to describe the spacetime of black holes governed by the Einstein gravity \cite{Cal-deF:1972:NuoCim:,Bar:1973:BlaHol:,Cun-Bar:1973:ApJ:,Bic-Stu:1976:BULAI:,Bao-Stu:1992:ApJ:,Vie:1993:ASTRA:,Fan-Cad-deF:}. Extension to the photon motion in the most general black hole asymptotically flat spacetimes, governed by the Kerr-Newman geometry, has been treated, e.g., in \cite{Stu:1981:BAIC:,Sche-Stu:2009:IJMPD:,Sche-Stu:2009:GRG:,Zak:2014:,Gre-Per-Lam:2015:IJMPD:,Abd-Ami-Ahm-Gho:2016:PHYSR4:}. The case of spherical photon orbits in the Reissner-Nordstr\"{o}m(-de Sitter) spacetimes has been intensively studied in \cite{Zak:2018:EPJC:}. However, the recent cosmological tests indicate presence of dark energy, probably reflected by a relict cosmological constant, that could play a significant role in astrophysical processes \cite{Stu:2005:MODPLA:, Stu-Sche:2011:JCAP:, Far:2016:PDU:,Stu-Hle-Nov:2016:PHYSR4:, Stu-etal:2017:JCAP:, Arra:2017:Universe:}. Therefore, the photon motion in the field of Kerr-de Sitter black holes, where the cosmological horizon exists along with the static radius giving the limit on the free circular motion \cite{Stu:1983:BULAI:,Stu-Hle:1999:PHYSR4:}, has been studied in \cite{Stu-Hle:2000:CLAQG:,Lak-Abd:2011:PHYSR4:,Char-Stu:2017:EPJC:,Stu-Char-Sche:2018:EPJC:,Per-Bis-Tsu:2018:PHYSR4:,Bis-Tsu:2018:arxiv:}; extension to spacetimes containing a charge parameter has been discussed in \cite{Bla-Stu:2016:PHYSR4:,Stu-Bla-Sche:2017:PHYSR4:,Eir-Sen:2018:EPJC:}. There is a large number of studies related to the photon motion in generalizations of the Einstein theory, e.g., for regular black hole spacetimes of the Einstein theory combined with non-linear electrodynamics \cite{Stu-Sche:2015:IJMPD:,Sche-Stu:2015:JCAP:}, and for black holes in alternative approaches to gravity \cite{Sche-Stu:2009:IJMPD:, Stu-Sche:2014:CLAQG:, Arraut:2013bqa:, Abd-Ahm-Dad-Ata:2017:PHYSR4:, Tos-Bam-Ahm-Abd-Stu:2017:EPJC:, Bam:2017:RMP:}. 

A crucial role in the photon motion is played by the spherical photon orbits, i.e., photons moving along orbits of constant (Boyer-Lindquist) radial coordinate -- their motion constants govern the local escape cones of photons related to any family of observers, and specially the shadow (silhouette) of black holes located in front of a radiating source \cite{Bar:1973:BlaHol:}, e.g. an orbiting accretion disk \cite{Lum:1979:ASTRA:,Marc:1996:CLAQG:,Doe-etal:2008:Nat:,Bro-Loeb:2009:ApJ:,Stu-Sche:2010:CLAQG:,Stu-Char-Sche:2018:EPJC:}. The properties of the spherical photon orbits outside the outer horizon of Kerr black holes were studied in \cite{Teo:2003:GenRelGrav:}. Spherical photon orbits in the field of Kerr-de Sitter black holes were discussed in \cite{Char-Stu:2017:EPJC:,Stu-Char-Sche:2018:EPJC:} - in this case, photons with negative covariant energy could be relevant, in contrast to the case of pure Kerr black holes where only spherical photons with positive covariant energy enter the play \cite{Teo:2003:GenRelGrav:}. 

Recently, growing interest in Kerr naked singularity spacetimes is demonstrated, mainly due to the possibility of existence of Kerr superspinars proposed in the framework of String theory by Ho\v{r}ava and his co-workers, with interior governed by the String theory and exterior described by the Kerr naked singularity geometry \cite{Gim-Hor:2009:PhysLetB:,Stu-Sche:2010:CLAQG:,Stu-Sche:2012:CLAQG:}. The presence of Kerr superspinars in active galactic nuclei or in microquasars could give clear signatures in the ultra-relativistic collisional processes \cite{Stu-Sche:2013:CLAQG:}, in the high-frequency quasiperiodic oscillations in Keplerian disks \cite{Tor-Stu:2005:ASTRA:,Kot-etal:2014:ASTRA:,Kot-etal:2017:ASTRA:}, or in the Lense-Thirring precession effects \cite{Chak-etal:2017:PHYSR4:}. The instability of test fields in the Kerr naked singularity backgrounds has been studied in \cite{Dot-Gle-Ran-Vuc:2008:CLAQG:,Car-Pan-Cad-Cav:2008:CLAQG:}, possible stabilizing effects were demonstrated for the Kerr superspinars in \cite{Nak-Jos-Guo-Koc-Tag-Har-Pat-Kro:2017:arxiv:1707.07242v1:}. The classical instability of Kerr naked singularity (superspinar) spacetimes, converting them to black holes due to standard Keplerian accretion, has been shown to be slow enough in order to enable observation of primordial Kerr superspinars -- at least at cosmological redshifts larger then $z=2$ \cite{Stu-Hle-Tru:2011:CLAQG:}. The Kerr naked singularity spacetimes could be applied also in description of the exterior of the superspinning quark stars with spin violating the black hole limit $a=1$ \cite{Stu-Sche-Sra-Tor:2017:ACTA:}. The optical phenomena related to the Kerr naked singularity (superspinar) spacetimes were treated in \cite{Stu:1980:BULAI:,Stu-Sche:2010:CLAQG:,Stu-Sche:2012:CLAQG:,Sche-Stu:2013:JCAP:}, and in more general case including the influence of the cosmological constant in \cite{Stu-Hle:2000:CLAQG:,Char-Stu:2017:EPJC:,Stu-Char-Sche:2018:EPJC:}. 

Here we focus our attention to the properties of the spherical photon orbits in the Kerr naked singularity spacetimes, generalizing thus the study of spherical photon orbits in the Kerr black hole spacetimes \cite{Teo:2003:GenRelGrav:}. For the spherical photon orbits we give the motion constants in dependence on their radius and dimensionless spin, and present detailed discussion of their latitudinal and azimuthal motion. We introduce detailed classification of the Kerr spacetimes according to the properties of the spherical photon orbits, including the stability of the spherical orbits and the role of the spherical photon orbits with negative energy relative to infinity, extending thus an introductory study in more general Kerr-de Sitter spacetimes \cite{Char-Stu:2017:EPJC:}. We also discuss astrophysically important interplay of the spherical photon orbits and the Keplerian accretion disks, with matter basically governed by the circular geodesic motion; we shortly discuss possible observational effects related to the irradiation of the Keplerian disks by the photons following the spherical orbits. 

\section{The Kerr spacetimes} 

The line element of the Kerr spacetime is in the standard Boyer-Lindquist spheroidal coordinates $t,r,\theta,\phi$, with geometric system of units ($c = G = 1$), described by the well known formula
\bea
\dbe s^2 =
&-&\frac{\Delta}{ \Sigma}\left(\dbe t - a \sin^2\theta \din \phi\right)^2\nonumber \\
&+& \frac{ \sin^2\theta}{\Sigma}\left[a \din t - \left(r^2 +   a^2\right)\din \phi \right]^2 \nonumber \\
&+&\frac{\Sigma}{\Delta}\din r^2 + \Sigma\din \theta^2, \label{line}
\eea
where
\bea
\Delta&=&r^2 + a^2 - 2Mr,\\
\Sigma&=&r^2 + a^2\cos^2\theta. \label{rho}
\eea
Here $M$ is the gravitational mass parameter of the Kerr spacetime, and $a$ is its angular momentum per unit gravitational mass. Without any loss of generality we can assume the parameter $a$ to be positive.\\ 
If $a<M$, the Kerr spacetime describes black holes that posses two pseudosingularities (horizons) determined by the condition
\be
\Delta =0,
\ee
located at radii
\be
r_{\pm}= M\pm \sqrt{M^2-a^2}.\label{horzns}
\ee
If $a=M$, the horizons coincide at $r=M$ and the spacetime describes an extreme Kerr black hole \cite{Bar:1973:BlaHol:}. In the present paper, we focus ourselves to the case $a>M$ corresponding to the Kerr naked singularity spacetimes where no horizons exist. In all the Kerr spacetimes, the physical singularity is located at $r=0, \theta=\pi/2$. The stationary limits of the Kerr spacetimes, determined by the condition $g_{tt}=0$, give the boundary of the so called ergosphere where extraction of rotational energy is possible due to the Penrose process \cite{Mis-Tho-Whe:1973:Gravitation:}. At $r<0$, allowed in the Kerr spacetimes, the so called causality-violation region exists \cite{Car:1973:BlaHol:} -- the Kerr superspinars are constructed in such a way that both the causality-violation region and the physical singularity are removed, and it is assumed that they are substituted by an interior regular solution governed by the String theory \cite{Gim-Hor:2009:PhysLetB:,Stu-Sche:2010:CLAQG:}. 

In order to demonstrate clearly behaviour of the Kerr spacetimes at region close to the physical singularity, it is convenient to use the ("flat") Kerr-Schild coordinates that are connected to the (spheroidal) Boyer-Lindquist coordinate $r$ by the relations 
\bea
       x^2 + y^2 = (r^2 + a^2)\sin^{2}\theta  ,  z^2 = r^2 \cos^{2}\theta 
\eea
and enable proper visualization of the Kerr spacetime in the innermost regions. 

\section{Carter equations of geodesic motion}

The symmetries of the Kerr spacetimes imply existence of the time Killing vector $\xi_{(t)}=\partial/\partial t$ and the axial Killing vector $\xi_{(\phi)}= \partial/\partial \phi$. Then the projections $\cale=-\xi_{(t)}\cdot p$ (energy) and $\Phi=\xi_{(\phi)}\cdot p$ (angular momentum about the $\phi$-axis) of a particle four-momentum $p=\frac{\din x^{\mu}}{\dbe \lambda}$ are constants of the particle motion. Here, $x^{\mu}$ are the coordinate components of the four-momentum and $\lambda$ is an affine parameter. Two another motion constants are the rest energy $m$ of the particle ($m=0$ for photons) and $Q$, Carter's constant connected with total angular momentum of the particle. The motion of photons can be described by the Carter equations 
\bea
(\Sigma \frac{\din r}{\din \lambda})^2&=&R \label{Cart_R1}\\
(\Sigma \frac{\din \theta}{\din \lambda})^2&=& W \label{Cart_Th1}\\
\Sigma \frac{\din \phi}{\din \lambda}&=&\frac{a P}{\Delta}-a \cale+\frac{\Phi}{\sin^2\theta} \label{Cart_Phi1}\\ 
\Sigma \frac{\din t}{\din \lambda}&=& \frac{r^2+a^2}{\Delta}P-a(a \cale \sin^2\theta-\Phi)\label{Cart_t1},
\eea
where the functions $P,R,W$ are defined by the relations 
\bea
P&=&\cale (r^2+a^2)-a\Phi  \\  \label{P}
R&=&P^2-\Delta [Q+(\Phi-a\cale)^2]  \\   \label{R}
W&=&Q-\cos^2\theta[\frac{\Phi^2}{\sin^2\theta}-a^2\cale^2]. \label{W}
\eea
It is convenient to use following rescaling $\lambda \to \lambda'=\lambda \cale$ and put $M=1$, i.e., express the radius and time (line element) in units of gravitational mass $M.$ Further, it is usual to introduce the substitution 
\bea
m&=&\cos^2\theta,\\ \label{m_cos2}
\din m&=&2\mbox{sign}(\theta-\pi/2)\sqrt{m(1-m)}\mathrm{d}\theta \nonumber,
\eea
which enables one to replace dealing with the trigonometric functions by the algebraic ones. The Carter equations then take the form 
\bea
(\Sigma \dot{r})^2&=& R(r)\label{Cart_R}\\
& \equiv& (r^2-a\ell+a^2)^2-\Delta [(a-\ell)^2+q]\nonumber \\
(1/2\Sigma \dot{m})^2&=&M(m)\label{Cart_M}\\
&\equiv& m[q+(a^2-\ell^2-q)m-a^2m^2]. \nonumber \\
\Delta \Sigma \dot {\phi}&=&2ar+\frac{\ell}{1-m}(\Sigma-2r) \label{Cart_Phi}\\ 
\Delta \Sigma \dot{t}&=& \Sigma (r^2+a^2)-2ar[\ell-a(1-m)]\label{Cart_t}.
\eea
Here the dot indicates differentiation with respect to the rescaled affine parameter $\lambda'$. We have introduced new parameters (assuming $\cale \neq 0$)
\bea
      q=\frac{Q}{\cale^2} \quad \mathrm{and}\quad \ell=\frac{\Phi}{\cale}, 
\eea      
where $\ell$ is the impact parameter \footnote{In case of more complex spacetimes with non-zero cosmological constant and possibly non-zero electric charge of the gravitating source it is more convenient to introduce the modified impact parameter $X=\ell-a,$ which simplifies the radial equation of motion \ref{Cart_R} and its discussion, as shown in \cite{Char-Stu:2017:EPJC:}, where the more general case of the Kerr-de~Sitter spacetimes was considered. However, its introduction is unnecessary in this paper.}. Introduction of the Carter constant $Q$ enables simple classification of the latitudinal motion in all the Kerr spacetimes, as $Q=0$ governs the equatorial motion with $\theta = \pi/2$, $Q>0$ governs the so called orbital motion crossing the equatorial plane \footnote{The orbital motion is allowed also for $q=0$ and $l^2<a^2$ when the motion in the equatorial plane is unstable \cite{Bic-Stu:1976:BULAI:}}, while $Q<0$ governs so called vortical motion when crossing of the equatorial plane is forbidden \cite{Cal-deF:1972:NuoCim:,Bic-Stu:1976:BULAI:,Stu:1983:BULAI:}. It is crucial for the purposes of the present paper that in both Kerr black hole \cite{Teo:2003:GenRelGrav:} and Kerr (-de Sitter) naked singularity \cite{Char-Stu:2017:EPJC:} spacetimes the spherical orbits with $r=const$ must be of the orbital type, having thus necessarily $Q>0$.

\section{Spherical photon orbits}

The SPOs have a crucial role in characterization of the Kerr spacetimes as they govern the shadows of Kerr black holes or Kerr superspinars, and could have influence on the behaviour of the Keplerian disks or more complex accretion structures due to effect of self-illumination \cite{Stu:1981:BAIC:,Bao-Stu:1992:ApJ:,Stu-Sche:2010:CLAQG:}. 

\subsection{Covariant energy of photons following spherical orbits}

The photons (test particles) whose motion is governed by the Carter equations of motion can be in the classically allowed positive-root states where they have positive energy as measured by local observers, and, equivalently, they evolve to future ($\din t/\din \lambda > 0$), or in the classically forbidden negative-root states with negative energy measured by local observers, evolving into past ($\din t/\din \lambda < 0$) \cite{Mis-Tho-Whe:1973:Gravitation:}. Above the outer horizon of a Kerr black hole, the situation is simple and all photons on the SPOs have positive covariant energy $\cale>0$ and are in the positive-root states. However, the situation is more complex in the case of Kerr naked singularities. 

For distinguishing of the positive-root and negative-root states in the case of Kerr naked singularities, and under the inner horizon of Kerr black holes, we can use the time equation of motion (\ref{Cart_t}); alternatively, the projection of the photon $4$-momentum on the time-like tetrad vector of physical observers can be used for this purpose \cite{Bic-Stu-Bal:1989:BAC:,Char-Stu:2017:EPJC:,Stu-Char-Sche:2018:EPJC:}. The time equation can be written in the form 
\beq
       \Delta \Sigma \frac{\din t}{\din \lambda}= \cale \left\lbrace \Sigma(r^2+a^2)-2ar[\ell-a(1-m)]\right\rbrace . \label{PRS}
\eeq
Now it is clear that in the regions of spherical orbits where the bracket on the r.h.s. is positive, we have positive-root states for $\cale>0$, while at regions where the bracket takes negative values, the positive-root states must have the energy relative to infinity $\cale<0$. Notice that in the case of $\cale<0$ the photons with $\Phi<0$ have positively-valued impact parameter $\ell=\frac{\Phi}{\cale}$. 

We present the results of the determination of the covariant energy of the SPOs later, and then we use them in the classification of the Kerr spacetimes where this property is considered as one of the criteria of the classification. 

\subsection{Motion constants of spherical photon orbits}

The simultaneous solution of the equations
\be
R(r)=0,\quad dR/dr=0,\label{cond_s_o}
\ee
where $R(r)$ denotes the r.h.s. of the equation (\ref{Cart_R}), yields the motion constants of the spherical orbits \footnote{There exists a second family of solution of these equations, but it is not physically relevant -- see \cite{Teo:2003:GenRelGrav:,Char-Stu:2017:EPJC:}}
\be
q=q_{sph}(r;a^2)\equiv-\frac{r^3}{a^2}\frac{r(r-3)^2-4a^2}{(r-1)^2},\label{qsph}
\ee
\be
\ell=\ell_{sph}(r;a^2)\equiv \frac{r^3-3r^2+a^2r+a^2}{a(1-r)}.\label{lsph}
\ee
The functions (\ref{qsph}),(\ref{lsph}) determine the constants of motion of the SPOs in dependence on their radius, hence, they represent the basic characteristics of the spherical orbits and require careful analysis. 

\subsection{Existence of spherical orbits}

First, we shall devote attention to the function in Eq.(\ref{qsph}) giving limits on the parameter q. Discussion of this function was already performed in \cite{Teo:2003:GenRelGrav:} for the case of the Kerr black holes, here we extend the discussion to the case of the Kerr naked singularities, including also the properties of the SPOs under the inner horizon of Kerr black holes that were not discussed in \cite{Teo:2003:GenRelGrav:}. 
The function $q_{sph}(r;a)$ is well defined for any radius $r\neq1,$ while it diverges at $r=1$ for $a\neq1$, with $$\lim_{r\rightarrow 1} q_{sph}=\mp \infty\quad \mathrm{as}\quad a\lessgtr1$$.  

In the case of extreme Kerr black holes, one can find by substituting $a=1$ into the conditions (\ref{cond_s_o}) that the function in (\ref{qsph}) should be replaced by
\be
q_{sph}(r;a=1)=(4-r)r^3,
 \label{qsph_ex_case}
\ee
which is now continuous even for $r=1$, c. f. \cite{Teo:2003:GenRelGrav:}. As we shall see bellow, only non-negative values of the parameter $q$ permit the motion of constant radius \footnote{The same applies to a more general case of the \KdS\ spacetimes, see \cite{Char-Stu:2017:EPJC:}.}. 

The zeros of (\ref{qsph}) determining the photon equatorial circular orbits are given by the condition 
\be
    a = a_{co}(r)\equiv\frac{1}{2}\sqrt{r(r-3)^2}. \label{a2eq}
\ee
For $a=0$ this function gives the SPO around the Schwarzschild black hole, located at $r=3$; if $0<a<1$, it determines one circular orbit under the inner black hole horizon with radius
\be
r_{ph0}=2\{1-\cos[\frac{\pi}{3}-\frac{1}{3}\arccos(2a^2-1)]\}, \label{rphin}
\ee
and two circular orbits located above the outer black hole horizon, which are the main subject of astrophysical interest. The inner one being co-rotating, the outer being counter-rotating, both being unstable with respect to radial perturbations. Their radii $r_{ph+}, r_{ph-},$ where $r_{ph+}<r_{h-},$ can be expressed by the relation \cite{Bar:1973:BlaHol:}
\be
r_{ph\pm}=2\{1+\cos[\frac{2}{3}\arccos(\mp a)]\}. \label{bh_circ_orbs}
\ee

Above the outer horizon of the KBH spacetimes, the SPOs are thus located at radii 
\be
      r_{ph+}<r<r_{ph-} . 
\ee       
In the extreme KBH case $a=1$, the counter-rotating orbit is located at $r=4$, while the co-rotating orbit shares the same (Boyer-Lindquist) radius $r=1$ with the black hole horizon, although they are in fact separated by a non-zero proper radial distance \cite{Bar:1973:BlaHol:}. 

In the Kerr naked singularity spacetimes, only the counter-rotating equatorial photon orbit exists. For  $a>1$, we express the photon orbit radii $r_{ph-}$ by the expression \cite{Stu:1980:BULAI:}
\be
r_{ph-}=2\{1+\cosh[\frac{1}{3}\arg \cosh(2a^2-1)]\}. \label{ns_circ_orb}
\ee
The co-rotating circular photon orbit should be (formally) located at $r=0, \theta=\pi/2$, representing thus the limit of unstable circular equatorial orbits of test particles; however, such an orbit has no physical meaning as it coincides with the physical singularity of the Kerr spacetimes. 
Therefore, in the KNS spacetimes, the SPOs can be located at all radii 
\be
      0<r_{sph}\leq r_{ph-} . 
\ee

\subsection{Stability of spherical photon orbits} 

Stability of the spherical null geodesics against radial perturbations is governed by the condition $d^2R/dr^2<0$ considered in the loci of the spherical orbits, i.e., by the extrema of the function $q_{sph}(r;a)$. The function $q_{sph}(r;a)$ has one local extreme $q_{ex}=27$ located at $r=3$, independently of the rotational parameter $a$. This extreme is a local maximum for $0<a<3,$ while for $a>3$ it becomes a minimum. The significance of this extreme, as follows from (\ref{Cart_Phi}), is that the photon orbit at $r=3$ intersects the equatorial plane perpendicularly, i.e., $\dot{\phi}(r=3,\theta=\pi/2)=0$ (c.f. \cite{Teo:2003:GenRelGrav:}). 

For the stability criterion, another extreme of the function $q_{sph}(r;a)$ is relevant that is determined by the condition 
\be
     a = a_{stab}(r)\equiv \sqrt{(r-1)^3+1},\label{stability}
\ee
which is implied also directly by the condition $d^2R/dr^2=0$ determining the marginal stability of the spherical orbits. The stability condition $d^2R/dr^2<0$ can be thus written in the form $a>a_{stab}(r)$. The function $a_{stab}(r)$ governing the marginally stable SPOs is increasing with increasing $r$, having an inflexion point at the special radius $r=1$; its behaviour is demonstrated in Figure 1. It is immediately clear that the marginally stable spherical orbits are located at $r>1$ in KNS spacetimes, while in the KBH spacetimes it must be located under the inner horizon $r<r_{-}<1$. The radius of the marginally stable SPO, $r_{ms\pm}$, can be directly expressed by the simple formulas
\be
r_{ms-} = 1 - (1-a^2)^{1/3}\quad \mathrm{for} \quad a^2\leq1  \label{rms-}
\ee
or 
\be
r_{ms+} = 1 + (a^2 - 1)^{1/3}\quad \mathrm{for} \quad a^2\geq1. \label{rms+}
\ee
The stable (unstable) spherical orbits are located for given parameter $a$ at $r<r_{ms\pm}$ ($r>r_{ms\pm}$). 

\subsection{Polar spherical photon orbits}
 
Now we consider behaviour of the function $\ell_{sph}(r;a)$. In the case of non-extreme Kerr BH spacetimes, this function is monotonically decreasing in the stationary region above the outer horizon and its point of discontinuity at $r=1$ is hidden between the black hole horizons. In the extreme KBH case $a=1$, the equation  (\ref{cond_s_o}) reduces to the form having no discontinuity 
\be
\ell_{sph}=r(2-r)+1. \label{X_sph_ex}
\ee
For the KNS spacetimes, $a>1$, there is $$\lim_{r\to 1^{\mp}}\ell_{sph}=\pm \infty$$ and the discontinuity occurs at $r=1$. 

In order to find the special case of spherical orbits covering whole the range of the latitudinal coordinate, reaching thus the symmetry axis at $\theta = 0$, we have to find when the function $\ell_{sph}(r;a)$ takes the significant value of $\ell_{sph}(r;a)=0$ corresponding to photons with zero angular momentum, since only such photons can reach the symmetry axis. We adhere notation introduced in \cite{Char-Stu:2017:EPJC:} and denote the radii of the polar SPOs crossing the symmetry axis by $r_{pol}$. These radii can be found by solving the equation 
\be
     a = a_{pol}(r) \equiv \sqrt{\frac{(3-r)r^2}{r+1}}. \label{a2pol}
\ee
The function $a_{pol(r)}$ has zeros at $r=0$ and at $r=3$, and a local maximum at $r_{pol}=\sqrt{3}$ for 
\be
     a = a_{pol(max)} \equiv \sqrt{6\sqrt{3}-9} = 1.17996 . 
\ee
The solution of Eq.(\ref{a2pol}) can be written in whole the relevant region of the rotation spacetime parameter $a \in (0,a_{pol(max)})$ in the form
\be
r_{pol\pm}=1+2\sqrt{1-\frac{a^2}{3}}\cos[\frac{\pi}{3}\pm\frac{1}{3}\arccos \frac{a^2-1}{(1-\frac{a^2}{3})^\frac{2}{3}}], \label{rpol12}
\ee
where $r_{pol+}<r_{pol-}.$

In the case of KBH spacetimes, $0<a<1$, only the formula for $r_{pol-}$ is relevant and it gives the only polar SPO in the stationary region, where $r_{pol-}<3$ \cite{Teo:2003:GenRelGrav:}. In the KNS spacetimes with $1<a<a_{pol(max)}=1.17996$, two polar SPOs exist at the radii $r_{pol\pm}$ given by Eq.(\ref{rpol12}). For $a=a_{pol(max)}$ these radii coalesce at $r_{pol}=\sqrt{3}$. For $a > a_{pol(max)}$ no polar SPOs exist. Combining Eq. (\ref{qsph}) and Eq. (\ref{a2pol}) one can find the values of parameter $q$ of the polar orbits to be given by
\be
q=q_{pol}(r)\equiv \frac{r^2(r+3)}{r-1}. \label{fqpol}
\ee
The graph of the function (\ref{fqpol}) is depicted in Figure \ref{Fig_qpol}. Its minimum value is $q_{pol(min)}=19.3923$ for $r=r_{pol}$ and in the limit points of its definition range it is $$q_{pol}(r)\to\infty\quad \mathrm{as}\quad r\to 1^{+}$$ and $$q_{pol}(r=3)=27.$$ We shall still mention the polar SPOs in section devoted to latitudinal motion.

\begin{figure}[H]
	\centering
	\includegraphics[scale=1]{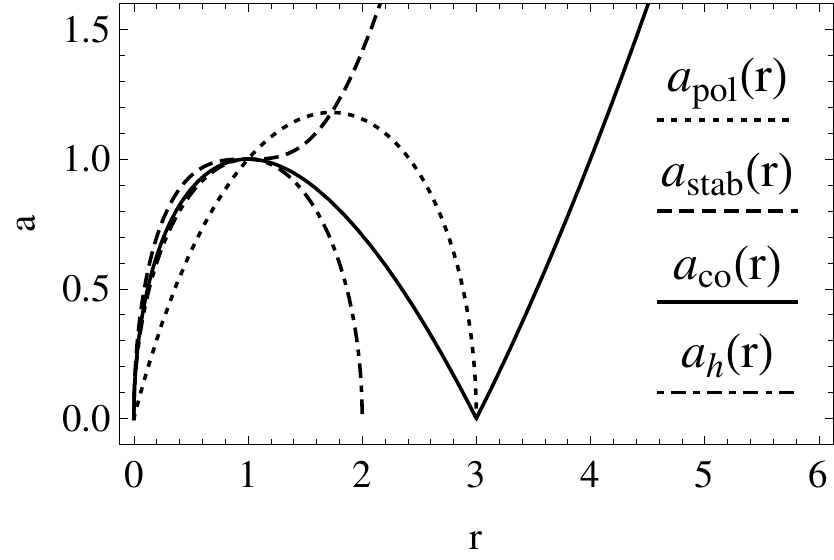}
	\caption{Characteristic functions $a(r)$ determining the behaviour of the function $q_{sph}$. They govern existence and stability of the SPOs and existence of polar spherical orbits. The event horizons for the black hole spacetimes are given by the function $a_{h}(r),$ the function $a_{co}(r)$ determines the equatorial circular photon orbits, the function $a_{pol}(r)$ determines loci of the polar SPOs with $\ell=0$ crossing the rotary axis. The function $a_{stab}(r)$ governs the stability of the SPOs -- the orbits above/bellow the curve $a_{stab}(r)$ are stable/unstable. 
}\label{Fig_ar}
\end{figure}

\begin{figure*}[h]
	\centering
	\includegraphics[width=\textwidth]{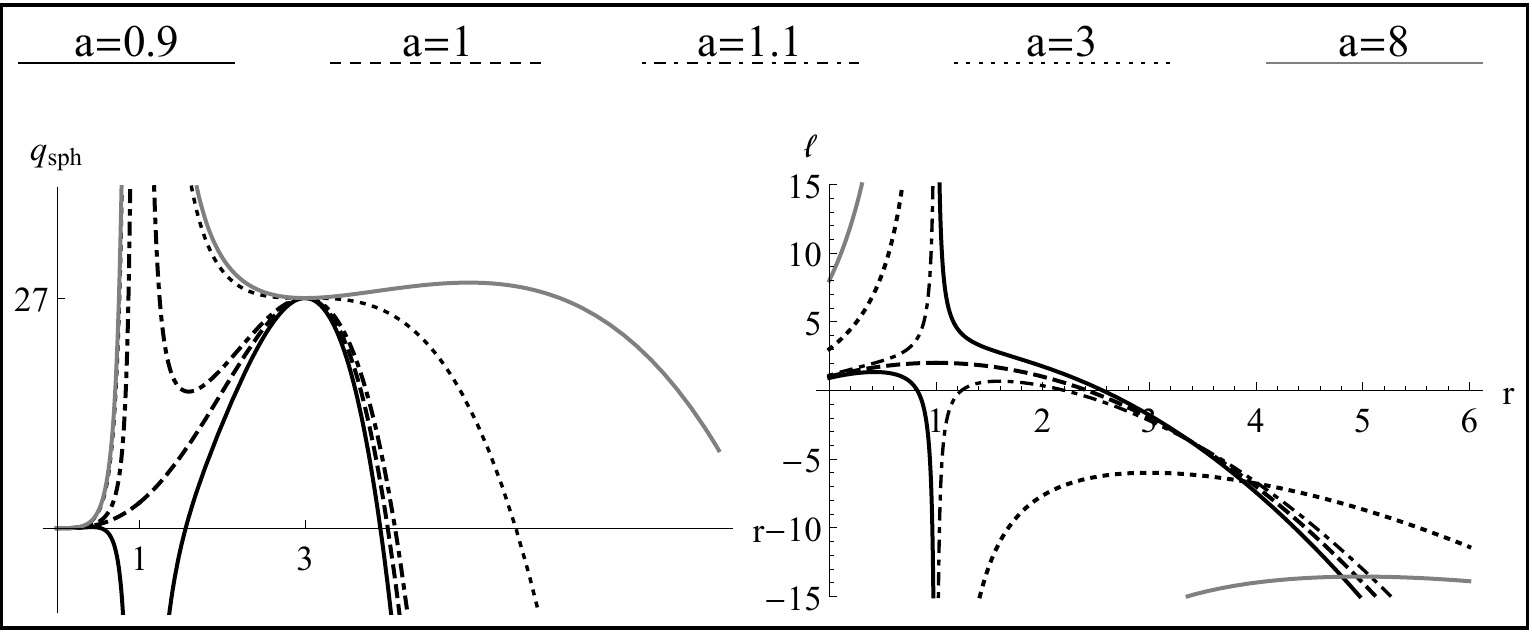}
	\caption{Depiction of the functions $q_{sph}(r;a)$ (left) and $\ell_{sph}(r;a)$ (right) for appropriately chosen values of the dimensionless spin $a$ giving characteristic types of their behaviour. }\label{Fig_qsph_lsph}
\end{figure*}

\begin{figure*}[h]
	\centering
	\begin{tabular}{cc}
		\includegraphics[width=0.5\textwidth]{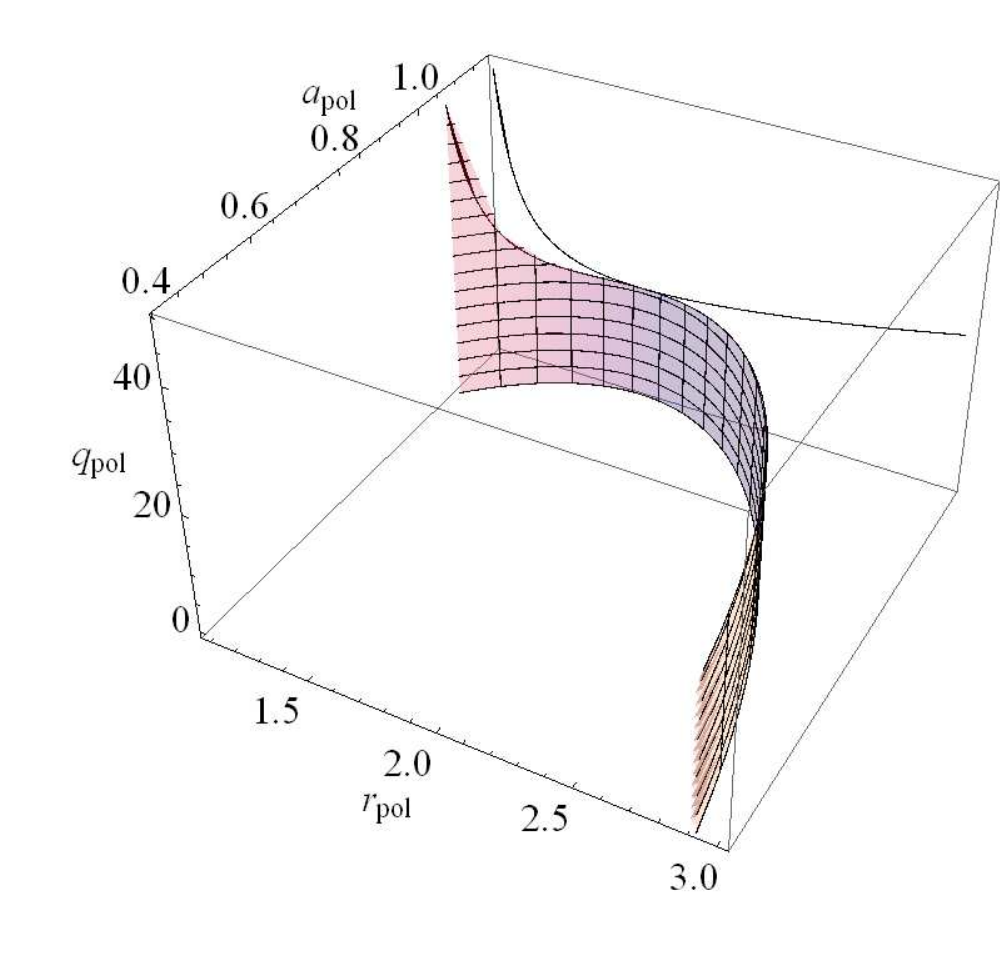}&\includegraphics[width=0.48\textwidth]{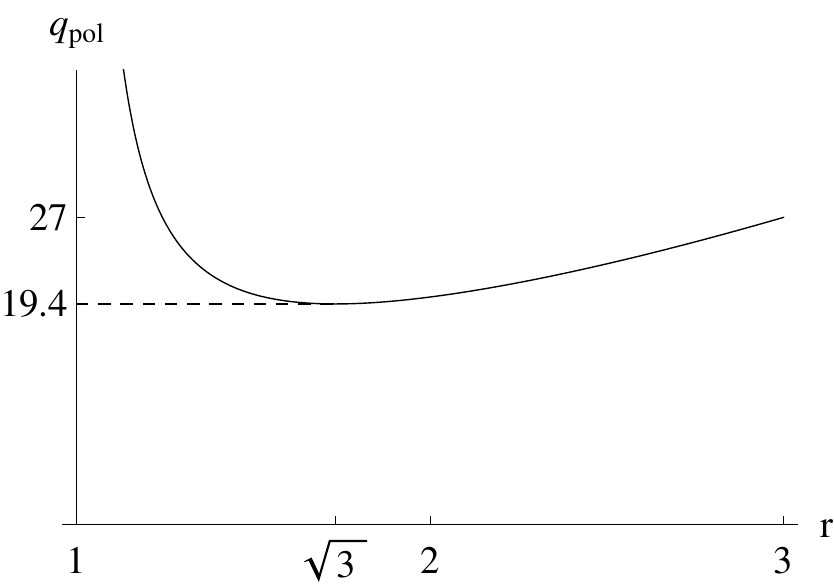}
	\end{tabular}	
	\caption{Dependence of the parameter $q_{pol}$ of the polar spherical orbits on its radii $r_{pol}$ depicted with corresponding dimensionless spin $a_{pol}$ (left) and its projection $q_{pol}(r_{pol})$ onto the $(r-q)$ plane (right).  
	}\label{Fig_qpol}
\end{figure*}

The local extrema of the function $\ell_{sph}(r;a)$ are determined by the relation (\ref{stability}), i.e., they are located at the same radii as the local extrema of the function $q_{sph}(r;a)$. 
The functions (\ref{a2eq}),(\ref{stability}), (\ref{a2pol}) are, together with the function
\be
 a_{h}(r)\equiv \sqrt{2r-r^2}, \label{horizons}
\ee
determining the loci of the event horizons, illustrated in Fig. \ref{Fig_ar}. Behaviour of the functions $q_{sph}(r;a)$ and $\ell_{sph}(r;a)$ is demonstrated for typical values of the dimensionless spin parameter $a$ in Fig. \ref{Fig_qsph_lsph}.

\subsection{Spherical photon orbits with negative energy}

For further discussion it is now necessary to determine the loci of the SPOs with negative energy $\cale<0$. Substituting $\ell=\ell_{sph}$ into the time equation (\ref{PRS}), we arrive to equation
\be
\Delta \Sigma \frac{\din t}{\din \lambda}= \cale \cdot T(r;m,a), \label{PRS1}
\ee
where
\be
T(r;m,a)\equiv \frac{r^2(r+3)+a^2m(r-1)}{r-1}.\label{T}
\ee
\clearpage
Clearly, the equation (\ref{PRS1}) describes the motion of photons with positive energy $\cale>0$ directed to the future ($\din t/\din \lambda>0$) on spherical orbits at radii $r>1$ \footnote{Here we restrict our discussion on the stationary region $\Delta>0.$}. On the contrary, for $0<r<1$ the condition $\din t/\din \lambda>0$ demands $\cale<0$ and $m<m_{z\cale}(r;a),$ or $\cale>0$ and $m>m_{z\cale}(r;a)$, where
\be 
m_{z\cale}(r;a)\equiv \frac{2r(3-r^2)}{a^2(1-r)}.\label{m_zen}
\ee
In the region of interest, $0<r<1$, the function $m_{z\cale}(r;a)$ is increasing, and $m_{z\cale}(0;a)=0$ and $m_{z\cale}(r;a)\to \infty$ as $r\to 1$. As we shall see bellow, at the interval $0<r<1$, the inequality $m>m_{z\cale}(r;a)$ is inconsistent with the reality condition for the latitudinal motion, hence the latter alternative is irrelevant and the range $0<r<1$ corresponds to region of the SPOs with $\cale < 0$. Of course, this region  must be hidden under the inner black hole horizon in case of the Kerr black holes. From discussion of the stability of the SPOs it follows that all the SPOs with $\cale < 0$ must be stable against radial perturbations. Of course, photons from the stable SPOs cannot escape to infinity due to a perturbation. 

\subsection{Spherical photon orbits with $\cale = 0$}

As we have shown above, the radii $0<r<1$ correspond to range of SPOs with negative energy, which suggests that there are SPOs with zero energy at radii $r=1$. It is supported by the fact that the functions (\ref{lsph}, \ref{qsph}, \ref{T}) diverge at this radii. For this reason, let us consider the Carter equations of motion for the photons (\ref{Cart_R1}-\ref{Cart_Phi1}) with explicitly expressed energy $\cale$ and revise the discussion assuming $\cale=0$. The conditions (\ref{cond_s_o}) then imply $r=1$ as expected, and the ratio of the motion constants of the SPO with $\cale = 0$ in the KNS spacetimes reads
\be
\Phi^2/Q=a^2-1 . \label{ratioPhi_Q}
\ee
Notice that the impact parameters $\ell$ and $q$ are not defined in this special case. In the extreme KBH spacetimes, $\Phi=0$ is required for arbitrary $Q>0$ for the spherical orbits at $r=1$. 

\section{Trajectories of photons on the spherical null geodesics}

In order to construct trajectories of the photons following the spherical null geodesics, we have to discuss in detail the latitudinal and azimuthal motion at the $r=const$ surfaces. In the context of the spherical motion of photons the natural question arises, what is the range of the latitudinal coordinate in dependence on the allowed motion constants and the dimensionless spin of the Kerr spacetime. Simultaneously, the important question is on the possible existence and number of the turning points of the azimuthal motion. 

\subsection{Latitudinal motion}

The latitudinal motion can be of the so called orbital type, where the photons oscillate between two latitudes $\theta_0,$ $\pi-\theta_0$, crossing repeatedly the equatorial plane or even being confined to the equatorial plane \footnote{This is the case of the equatorial circular orbits characterized by the value $q=0$ that can be regarded as a special case of the spherical orbits.}, or of the so called vortical type, where the photons oscillate 'above' or 'bellow' the equatorial plane between two pairs of cones coaxial with the symmetry axis of the spacetime, with latitudes $\theta_1, \theta_2$, $\theta_1< \theta_2$ and $\pi-\theta_1,\pi- \theta_2$. The special case is the vortical motion along the symmetry axis $\theta=0$, or the motion at any constant latitude -- such photons are called PNC photons and have a generic role in the Kerr spacetimes \cite{Bic-Stu:1976:BULAI:}. Now one can ask, which of these types is possible for the spherical photon motion. 

First, we demonstrate that for the spherical orbits there is $q\geq0$ necessarily, i.e., the motion is of the orbital type. This can be shown easily using the Carter equation of the latitudinal motion \ref{Cart_M}), from which it can be immediately seen that for $q<0$ the condition $M(m)\geq0$ ensuring the existence of the latitudinal motion is fulfilled only if 
\be
a^2 - q - l^2 > 0. \label{a2_q_l2}
\ee
On the other hand, we can show that $a^2 - q_{sph}(r;a) + l^{2}_{sph}(r;a) < 0$ so that the possibility $q<0$ must be rejected \cite{Teo:2003:GenRelGrav:, Char-Stu:2017:EPJC:}. 

In fact, there is even more restrictive condition for the motion constants than that given by (\ref{a2_q_l2}). To show this, let us express the reality condition $M(m)\geq0$ using linearity in parameter $q$ by
\be
q \geq q_m(m;a,\ell) \equiv m(\frac{\ell^2}{1-m}-a^2). \label{q_min}
\ee
The equality gives the turning points in the latitudinal coordinate. Notice that using the coordinate $m = cos^2\theta$, we have to restrict the range of the solutions to $m \in \left\langle 0;1\right\rangle$. Of course, there is obvious zero of (\ref{Cart_M}) given by $m=0$, emerging due to the used substitution, which indicates just a transit through the equatorial plane. Behaviour of the function $q_m(m;a,\ell)$ in the limit points of the interval $\left\langle 0;1\right\rangle$ is as follows: $$q_m(0;a,\ell)=0,$$ $$\lim_{ m \to 1}q_{m}(m;a,\ell)=\infty.$$ Another zero of $q_{m}(m;a,\ell)$ is at $$m=\frac{a^2-\ell^2}{a^2}$$ for $\left| \ell \right| \leq a.$ The local extrema can be expressed from the condition $\din q/\din m=0$ in an implicit form 
\be
\ell^2=a^2(1-m)^2,
\ee
or, equivalently
\be
\ell=\ell_{min\pm}\equiv \pm a(1-m). \label{l_min} 
\ee
Clearly, the extrema exist for $|\ell|\leq a.$ The subscript 'min' in (\ref{l_min}) indicates that these extrema must be minima, as follows from the inequality $\din^2q_m/\din m^2=2a^2/(1-m)>0$. The values of these minima read $q_{min}=-(a-l)^2$ for $0\leq \ell \leq a,$ or  $q_{min}=-(a+l)^2$ for $-a \leq \ell \leq 0.$ The above results can be written in a compact form

\begin{equation}
q\geq q_{m(min)} \equiv \left\{
\begin{array}{l}
-(|l| - a)^2,\quad \mbox{for}\quad |l| < a;\\
\\
0, \quad\mbox{for}\quad |l| \geq  a. \label{qm(min)}
\end{array}\right.
\end{equation}
This is the relation giving stronger limitation on the impact parameter $\ell$ in case $q<0$ than that given by (\ref{a2_q_l2}).

\begin{figure}[hb]
	\centering
	\includegraphics[scale=1]{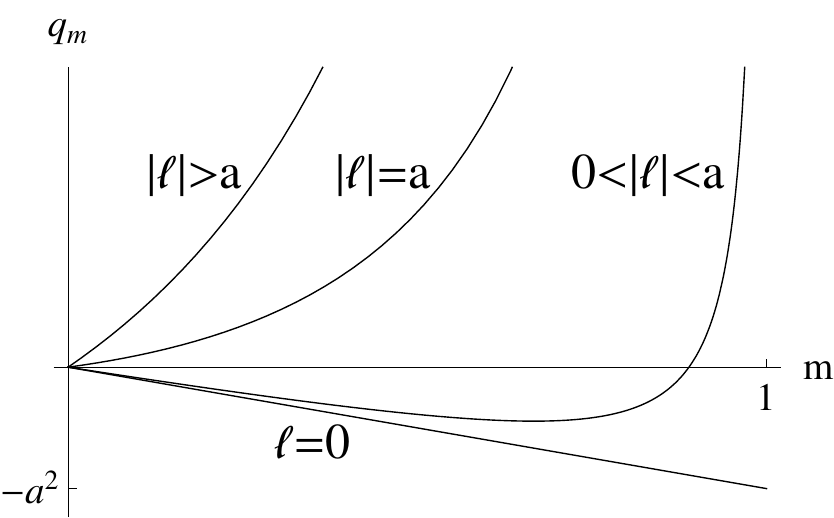}
	\caption{Typical behaviour of the function $q_{m}(m;a,\ell)$ for appropriately chosen values of the parameter $\ell$.}\label{Figure qmin}
\end{figure}
From the behaviour of the function $q_{m}(m;a,\ell)$, which is shown in Fig. \ref{Figure qmin}, it can be seen that the vortical motion exists for negative values of the parameter $q$. According to the relation (\ref{qm(min)}), the negative values of the parameter $q$ allowing the latitudinal motion exist for $|\ell|\leq a$, where the lowest value is $q=-a^2$ and occurs for $\ell=0$. However, as can be verified by calculation, at radii where $|l_{sph}|\leq a$, there is $q_{sph}>0$, which confirms that the vortical motion of constant radius is impossible for the spherical orbits.\\

The latitudinal turning points of the SPOs are given by zeros of the r. h. s. of Eq. (\ref{Cart_M}) with substitution of $\ell=\ell_{sph}$. The maximum latitude reached by a photon following a spherical null geodesic at a particular radius can be inferred from the relation $$m=m_{\theta}(r;a),$$ where the latitudinal turning function (relevant in the range $m \in (0,1)$) is defined by 
\be
m_{\theta}(r;a) \equiv \frac{r^2}{a^2}\frac{4a^2-9r+6r^2-r^3}{r^3-3r+ 2a^2+2\sqrt {\Delta(2r^3-3r^2+a^2) }}. \label{latp}
\ee
The function $m_{\theta}(r;a)$ is real everywhere in the stationary region. Its zeros are determined by the function $a_{co}(r)$. For KBHs ($0<a<1$), the function $m_{\theta}(r;a)$ has two local maxima (Fig. \ref{Fig_lat_az}a); one under the inner black hole horizon, located at radius determined by the function $a_{stab}(r)$, i.e., at $r=r_{ms-}$. 
The second maximum of $m_{\theta(max)}=1$ is determined by the function $a_{pol}(r)$. In extreme KBHs case where $a=1$ (Fig. \ref{Fig_lat_az}b), the function $m_{\theta}(r;a)$ has one local maximum $m_{\theta(max)}=1$ given by $a_{pol}(r)$. For fixed rotation parameter $a$, the radii $r_{pol\pm}$ are given implicitly by relation (\ref{a2pol}) and explicitly by Eq.(\ref{rpol12}). The corresponding impact parameter $q_{pol}=q_{sph}(r_{pol}(a),a)$ is represented in Fig. \ref{Fig_qpol}; the other motion constant, $l_{pol}=0$ by definition.

For the KNS spacetimes with $a<a_{pol(max)}$ (Fig. \ref{Fig_lat_az}c), two maxima exist at $m_{\theta(max)}=1$ that are given by the function $a_{pol}(r)$, and one local minimum located at $r=r_{ms+}$. If $a=a_{pol(max)}$, the three extrema coalesce into maximum $m_{\theta(max)}=1$ (Fig. \ref{Fig_lat_az}d). In these KNS spacetimes thus polar SPOs can exist for properly chosen motion constants at the properly chosen radii. 

For the KNS spacetimes with $a>a_{pol(max)}$ (Fig. \ref{Fig_lat_az}e,f), the function $m_{\theta}(r;a)$ has a local maximum at $m_{\theta(max)}<1$ at $r=r_{ms+}$. In such KNS spacetimes the polar SPOs cannot exist. \\

\subsection{Azimuthal motion}

Finally, let us consider the relations of the function $m_{\theta}(r;a)$ with a function $m_{\phi}(r;a)$ determining the latitude at which a turning point of the azimuthal motion occurs. The equation (\ref{Cart_Phi}), after performing substitution $\ell=\ell_{sph}$, can be written in the form
\be
\Sigma \frac{\din \phi}{\din \lambda}=\cale \cdot \varPhi(r;a,m), \label{dot_phi_sph}
\ee
where
\be
\varPhi(r;a,m) \equiv \frac{r^2(r-3)+a^2m(r+1)}{a(1-m)(1-r)} . \label{varPhi}
\ee
The condition $\din \phi/\din \lambda \geq0$ implies $$m\leq m_{\phi}(r;a),$$ where the azimuthal turning function is defined as 
\be
m_{\phi}(r;a)\equiv \frac{(3-r)r^2}{a^2(r+1)}. \label{aztp}
\ee
The change of sign in denominator of (\ref{varPhi}), while crossing the divergence point $r=1$, now plays no role, since, as follows from the preceding discussion, in order to have $\din t/\din \lambda>0$, the relation $\cale/(1-r)<0$ holds at any radius $r$ since $\cale<0 (\cale>0)$ at $r<1 (r>1)$. Therefore, the motion in the $\phi$-direction is fully governed by the function (\ref{aztp}) and the inequalities presented above. If we compare the expressions in (\ref{a2pol}) and (\ref{aztp}), we see that 
\be
a_{pol}(r)=a \sqrt{m_{\phi}(r;a)}.\label{a2pol_mphi}
\ee
The turning point of the azimuthal motion thus exist for all KBHs and KNSs. The functions $m_{\theta}(r;a)$and $m_{\phi}(r;a)$ have common points at $r=1$, where they have value $m_{\theta}(1;a)= m_{\phi}(1;a)=1/a^2$, and at the local extrema determined by the curve $a_{pol}(r)$.  As is evident from Eq.(\ref{a2pol_mphi}), the function $m_{\phi}(r;a)$ has a local maximum at $r=\sqrt{3}=r_{pol}$ with $m_{\phi(max)}=(6\sqrt{3}-9)/a^2=a^2_{pol(max)}/a^2$.\newpage The position of the SPOs with turning point of the azimuthal motion is represented in Fig. \ref{Figure 1ref}. The latitudinal angle where the azimuthal turning point occurs is given by the relation 
\be
\theta_{turn(\phi)}(r;a) = \pm \arccos(\sqrt\frac{(3-r)r^2}{a^2(r+1)}). \label{thaztp}
\ee
The existence of the azimuthal turning points enables existence of "oscillatory" orbits with change of the azimuthal angle for half period in the latitudinal motion $\Delta\phi(a) = 0$.
\begin{widetext}
\onecolumngrid 
 \begin{figure}[h]
	\centering
	\begin{tabular*}{0.9\textwidth}{c@{\extracolsep{\fill}}c}
		\includegraphics[width=6.5cm]{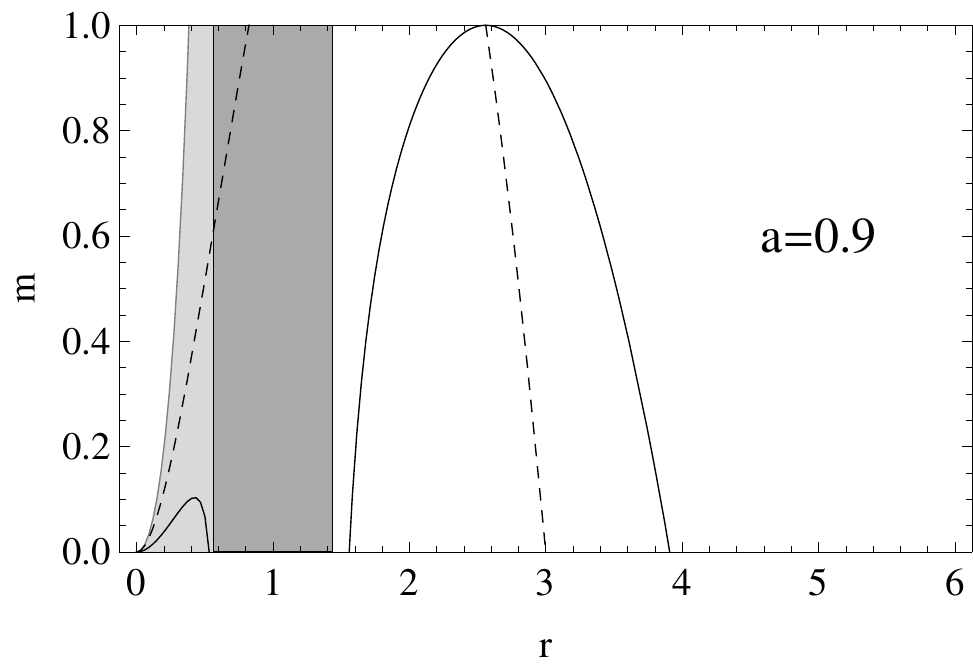}&\includegraphics[width=6.5cm]{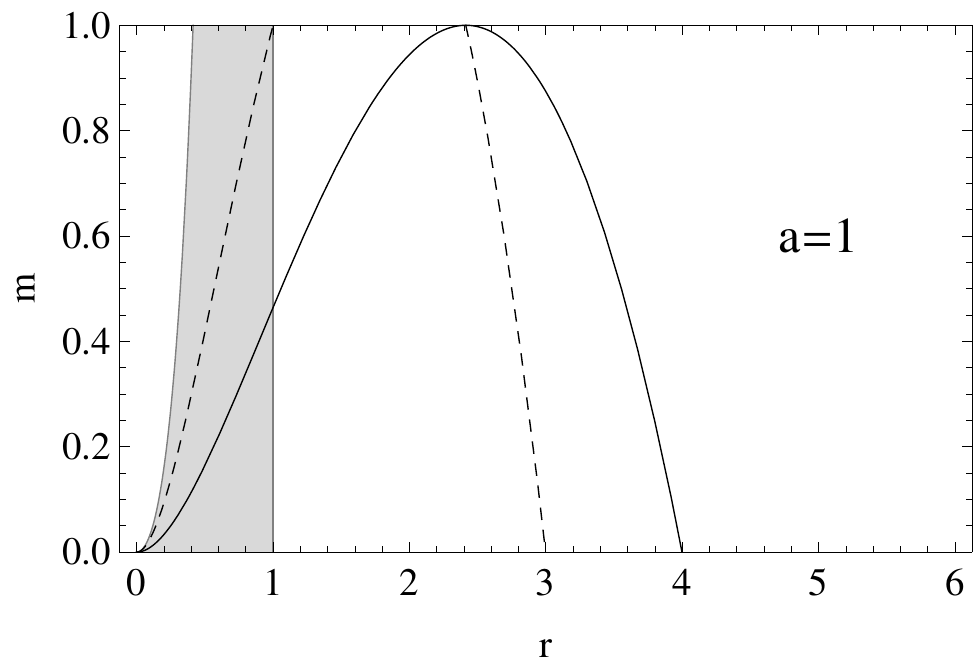}\\
		(a)&(b)\\	
		\includegraphics[width=6.5cm]{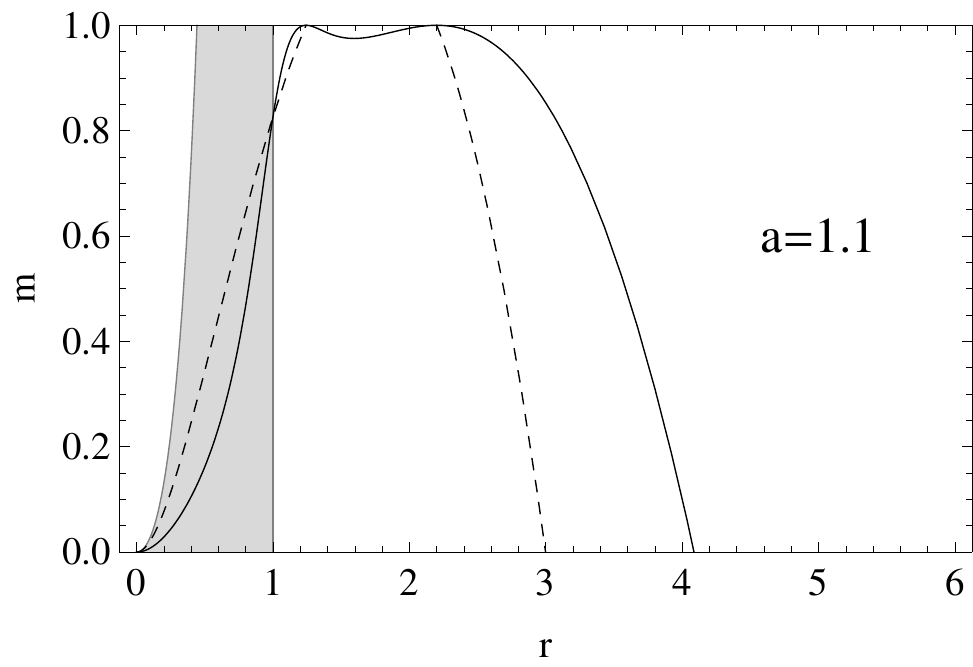}&
		\includegraphics[width=6.5cm]{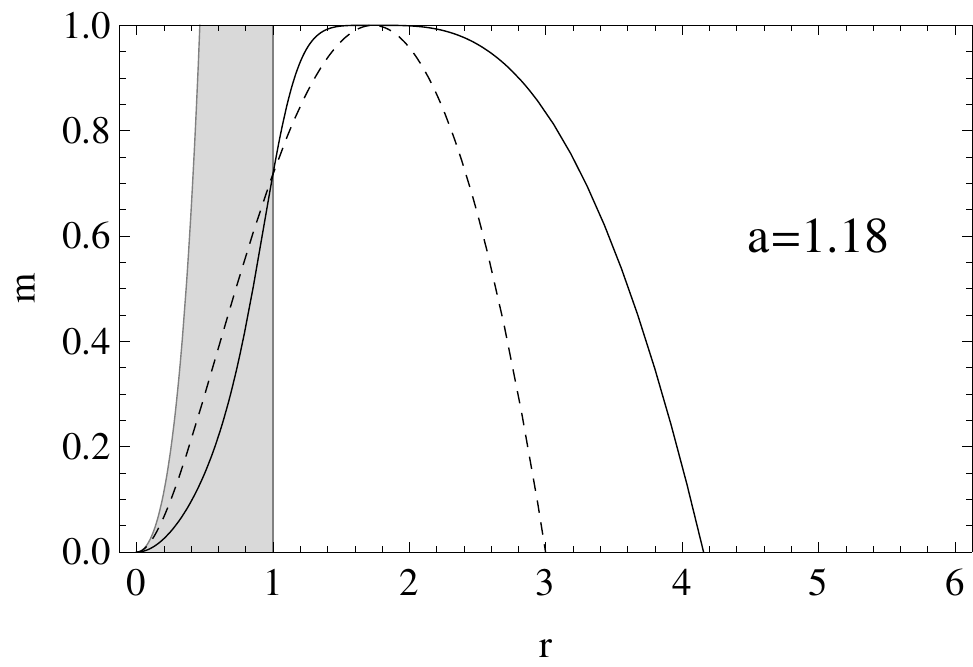}\\
		(c)&(d)\\	
		\includegraphics[width=6.5cm]{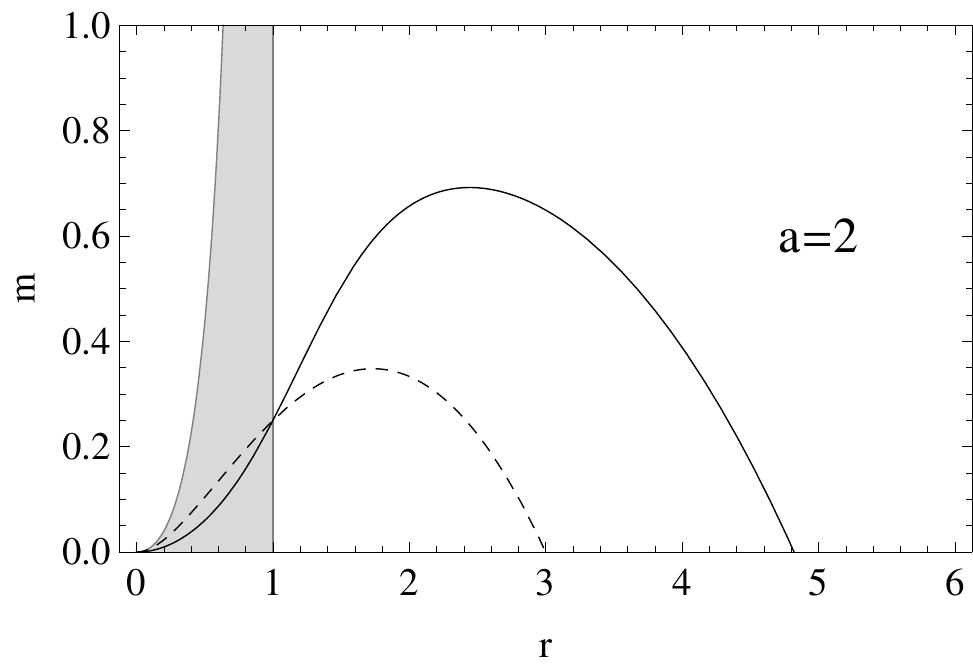}&\includegraphics[width=6.5cm]{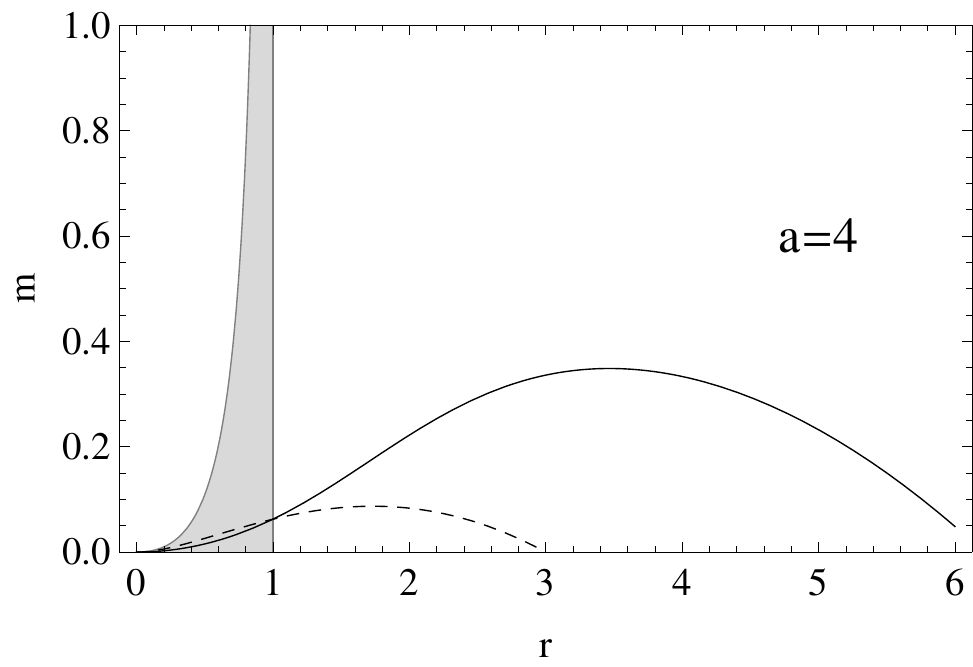}\\
		(e)&(f)
	\end{tabular*}
	\caption{The graphs of the characteristic functions $m_{\theta}(r;a)$ (full black curve), $m_{\phi}(r;a)$ (dashed black curve) and $m_{z\cale}(r;a)$ (grey curve) are given for appropriately chosen values of the spin parameter $a$, demonstrating all qualitatively different cases of their behaviour. Grey vertical line is an asymptote of $m_{z\cale}(r;a)$. Dark shading demarcates the dynamic region $\Delta<0$, light shading corresponds to negative energy orbits. Region of the prograde orbits with $d\phi/d \lambda >0$ is located under the curve $m_{\phi}(r;a)$. Therefore, all the SPOs with negative energy are prograde.}
	\label{Fig_lat_az} 
\end{figure}
\twocolumngrid
\clearpage
\end{widetext}
\subsection{Shift of nodes}
Now we examine the dragging of the nodes, i.e., we thus determine the azimuthal angle between the two points, where the ascending, i.e., the 'northwards' directed parts of the photon track intersects the equatorial plane. For this purpose we need to evaluate the change in the azimuthal coordinate $\Delta \phi$ by integrating $d\phi/dm$ expressed from the Carter`s equations \ref{Cart_M}, \ref{Cart_Phi} with $\ell=\ell_{sph}(r;a)$, $q=q_{sph}(r;a)$ inserted. \footnote{The more general case of the function $M(m)$ with non-zero cosmological parameter was studied in detail in \cite{Char-Stu:2017:EPJC:}. } The exact expression for the nodal shift $\Delta \phi$ was found in \cite{Teo:2003:GenRelGrav:} -- with our notification it reads
\bea
\Delta \phi&=&\frac{4}{\sqrt{m_{+}-m_{-}}} [\frac{2r-a\ell_{sph}}{\Delta} K(\frac{m_{+}}{m_{+}-m_{-}})+ \nonumber \\
&&\frac{\ell}{a(1+m_{+})} \Pi(\frac{m_{+}}{m_{+}-1},\frac{m_{+}}{m_{+}-m_{-}}) ].\label{delphi}
\eea
Here $m_{\pm}$ are the positive and negative roots of the function $M(m)$, while 
\be
K(s)=\int \limits_0^{\pi/2}(1-s\sin^2\theta)^{-1/2} d\theta \label{K(m)}
\ee
and
\be 
\Pi(v,s)=\int \limits_0^{\pi/2}(1-v\sin^2\theta)^{-1}(1-s\sin^2\theta)^{-1/2} d\theta \label{Pi(n,m)}
\ee
are the complete elliptic integrals of the first and third kind, respectively. The dependence of the nodal shift $\Delta \phi$ on the radial coordinate qualitatively differs in dependence on the presence of the polar orbits. If they are present, discontinuities occur at the radii $r_{pol\pm}$ -- see Figs.\ref{shift_nodes}a-d. The jump drop is always $4\pi$, and the single points, which are inherent values in the discontinuity points $r_{pol\pm}$, are halfway between the endpoints of the interrupted curve. This is a simple consequence of the singular behaviour of the Boyer-Lindquist coordinates at the poles, as explained in \cite{Teo:2003:GenRelGrav:}. For $a_{pol(max)}$ these discontinuities coalesce at $r=r_{pol}$, for which the nodal shift function $\Delta \phi< 0$ (Fig. \ref{shift_nodes}e).
\vspace{1cm}
\begin{widetext}
	\onecolumngrid
	
 \begin{figure*}[htbp]
 	\centering
 	\begin{tabular}{ccc}
 	\includegraphics[scale=0.66]{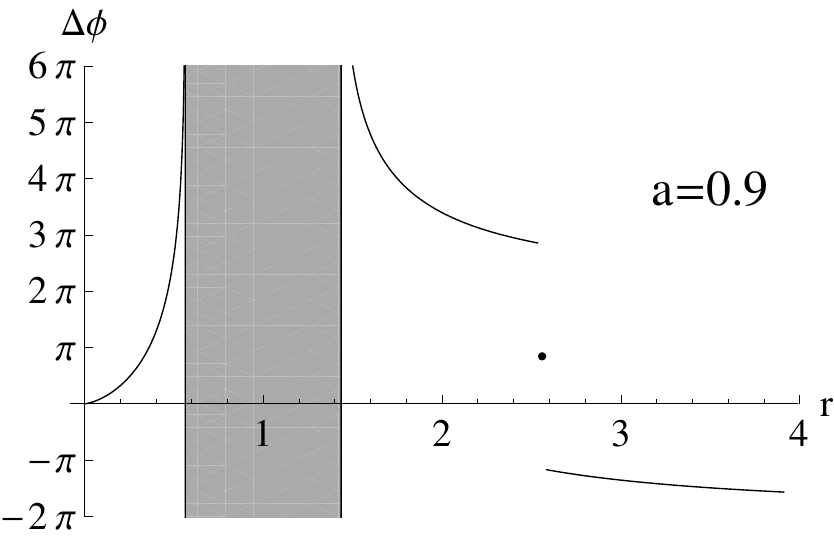}&\includegraphics[scale=0.66]{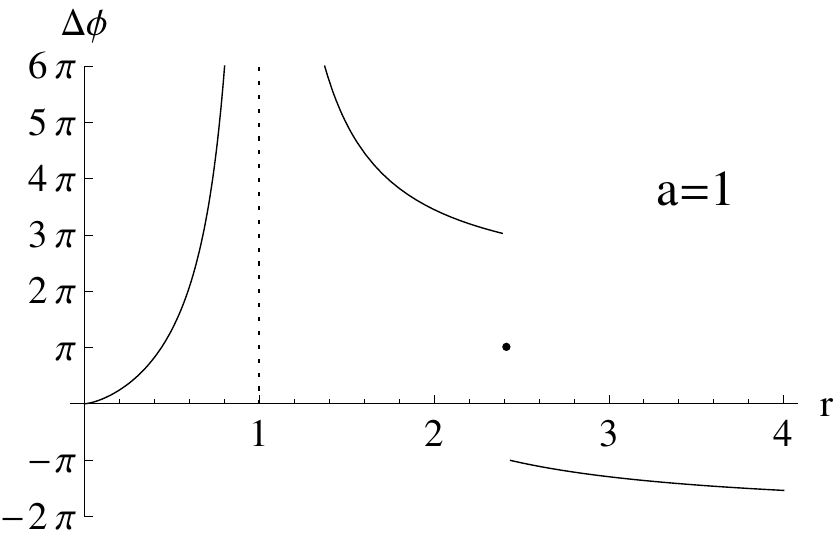}&\includegraphics[scale=0.66]{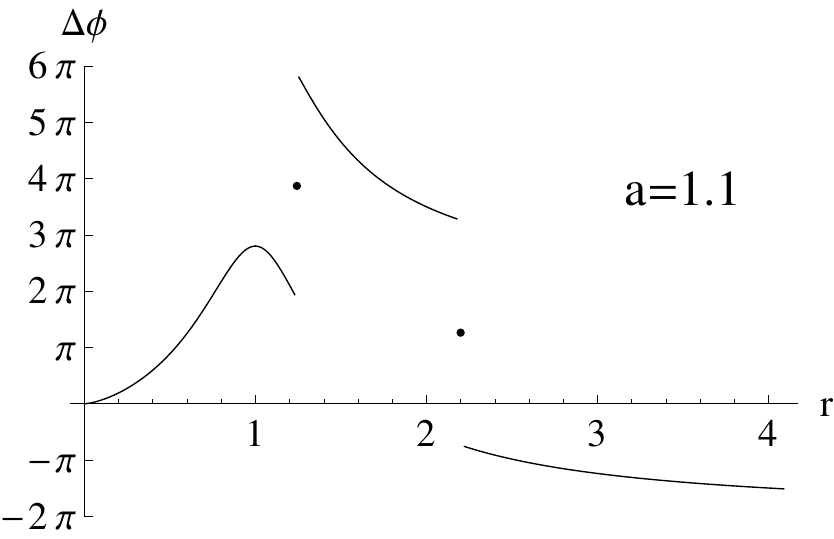}\\
 	(a)&(b)&(c)\\
 	\includegraphics[scale=0.66]{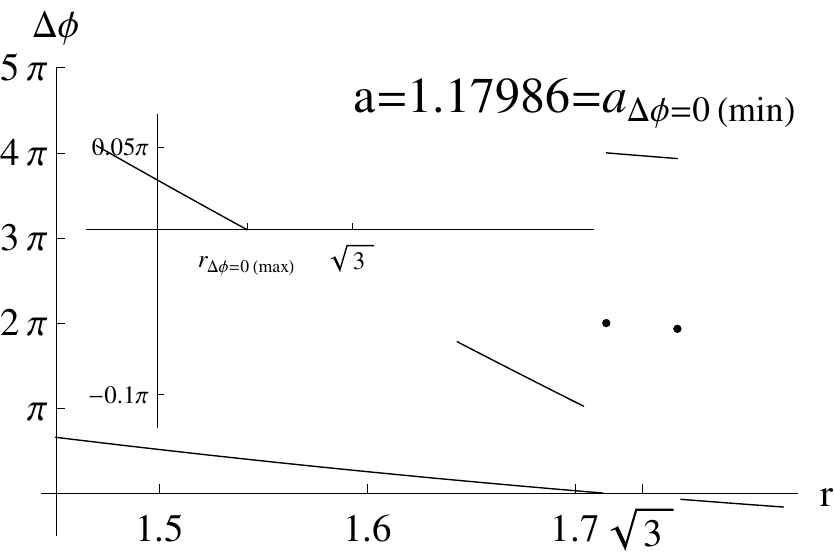}&\includegraphics[scale=0.66]{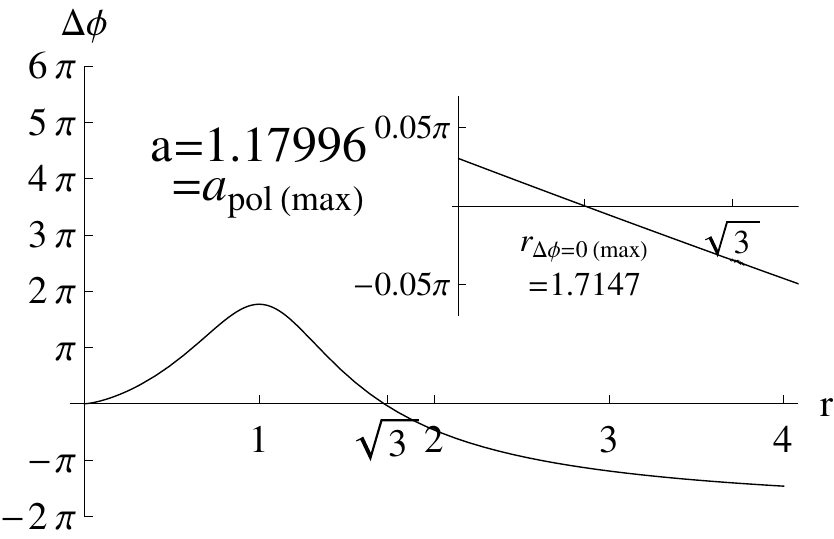}&\includegraphics[scale=0.66]{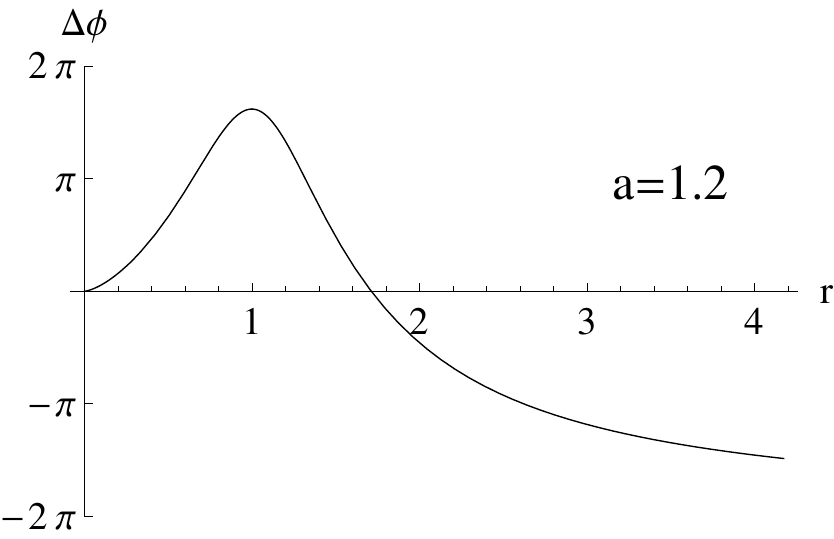}\\
 	(d)&(e)&(f)	
 	\end{tabular}
 	\caption{Dependence of the node shift $\Delta\phi$ on the radius of the spherical orbit. A discontinuity occurs at $r=r_{pol\pm}$ the jump change is $2\pi$ between the single point and each of the end points of the curve, c. f. \cite{Teo:2003:GenRelGrav:}. For $a=a_{pol(max)}=1.18$ the points of the discontinuity coalesce at $r=r_{pol}=\sqrt{3}.$ For $a>a_{pol(max)}$ the function is continuous. The zero of $\Delta\phi$ first occurs for parameter $a=a_{\Delta\phi=0(min)}=1.179857$ slightly lower than $a_{pol(max)};$ it indicates an existence of an oscillating orbit, which forms a boundary between prograde ($\Delta\phi>0$) and retrograde orbits ($\Delta\phi<0.$) For $a<a_{pol(max)}$ the boundary is at $r=r_{pol-}.$}\label{shift_nodes}
 \end{figure*} 
\twocolumngrid
\end{widetext}

\subsection{Periodic orbits}

\begin{figure}[hb]
	\centering
	\includegraphics[width=0.5\textwidth]{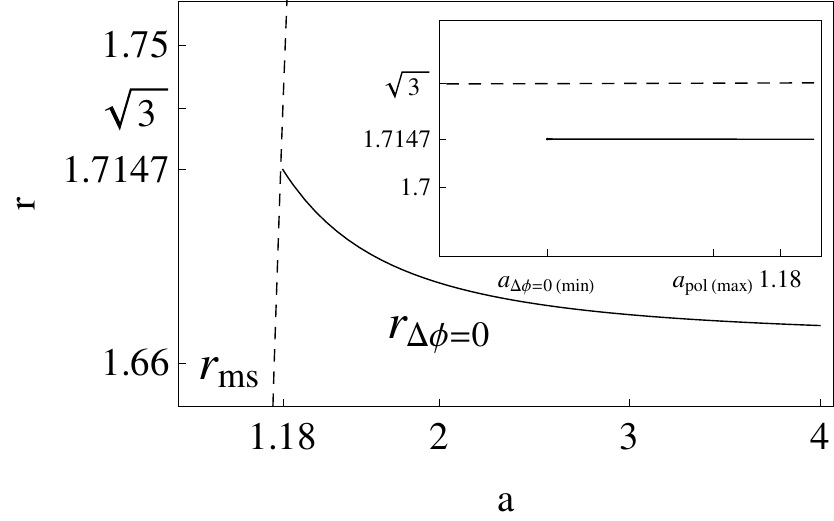}
	\caption{Radii $r_{\Delta\phi=0}$ of orbits with zero nodal shift $\Delta \phi$ in dependence on the spin parameter $a$ and comparison with the loci of marginally stable orbits $r_{ms}.$ For given dimensionless spin $a$ the radii of the stable orbits are $r<r_{ms}(a).$ Hence all such orbits are stable. Note that the detail is approx. $10^4$-times horizontally stretched.   
	}\label{Fig_rzs_a}
\end{figure}

The nodal shift function becomes to be continuous for the KNS spacetimes with $a>a_{pol(max)}$. Therefore, in KNS spacetimes with $a>a_{pol(max)} \sim 1.18$, there exist "oscillatory" trajectories with $\Delta \phi=0$, i. e., the photons in such trajectories are following a closed path with finite extent in azimuth, ending at the starting point. They are of octal-like shape  (see, e. g., Fig. \ref{clos_orb} case $k=0$). However, a detailed calculation using numerical procedure reveals that the radius $r_{\Delta\phi=0}$ of the zero nodal shift $\Delta \phi =0$ first occurs for $a=a_{\Delta\phi=0(min)}\equiv 1.179857$ at $r=r_{\Delta\phi=0(max)}=1.71473$ (see Fig. \ref{shift_nodes}d). In Fig. \ref{Fig_rzs_a} we show that the function $r_{\Delta\phi=0}(a)$ is descending, hence the subscript '$\Delta\phi=0(max)$'. Notice that the value of $a_{\Delta\phi=0(min)}$ is only very slightly lower than $a_{pol(max)}$. Since the point $r_{\Delta\phi=0(max)}$ and corresponding point of the $4\pi$-discontinuity $r_{pol+}(a_{\Delta\phi=0(min)})$ are infinitesimally separated, we can claim that the SPO at $r_{pol+}(a_{\Delta\phi=0(min)})$ is a special case of polar oscillatory orbit. We illustrate its trajectory explicitly in Fig. \ref{phot_trace_rzs_azs} and for comparison we give illustration of the orbit at $r=r_{pol}$ for $a=a_{pol(max)}$.  

\begin{figure}[hb]
	\centering
	\includegraphics[width=0.5\textwidth]{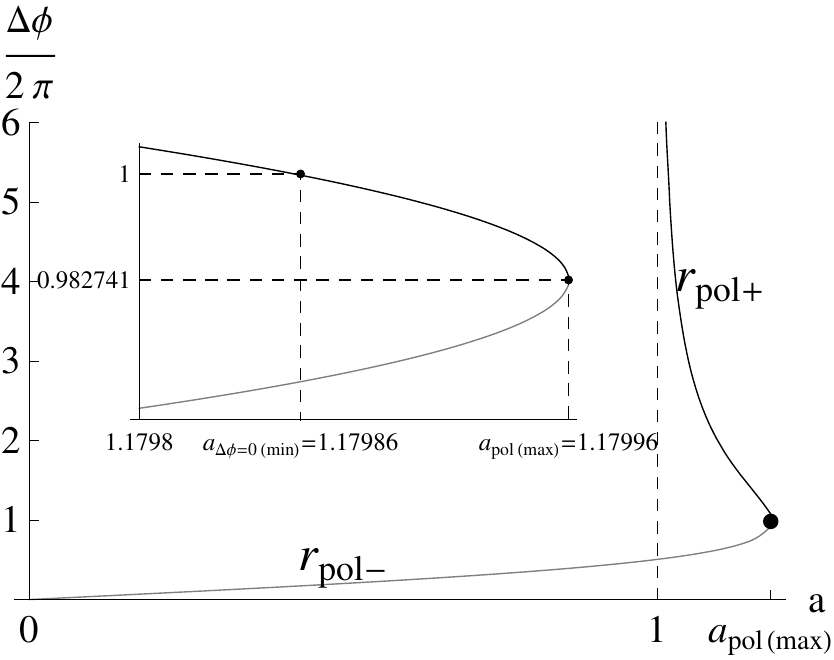}
	\caption{Shift of nodes for the polar spherical orbits $r_{pol\pm}$ represented by the isolated points in Fig. \ref{shift_nodes} in independence on the dimensionless spin $a.$ 
	}\label{Fig_snrp12}
\end{figure}

\begin{figure*}[h]
	\begin{tabular}[t]{|c|c|c|}
		\hline
		\includegraphics[height=4cm]{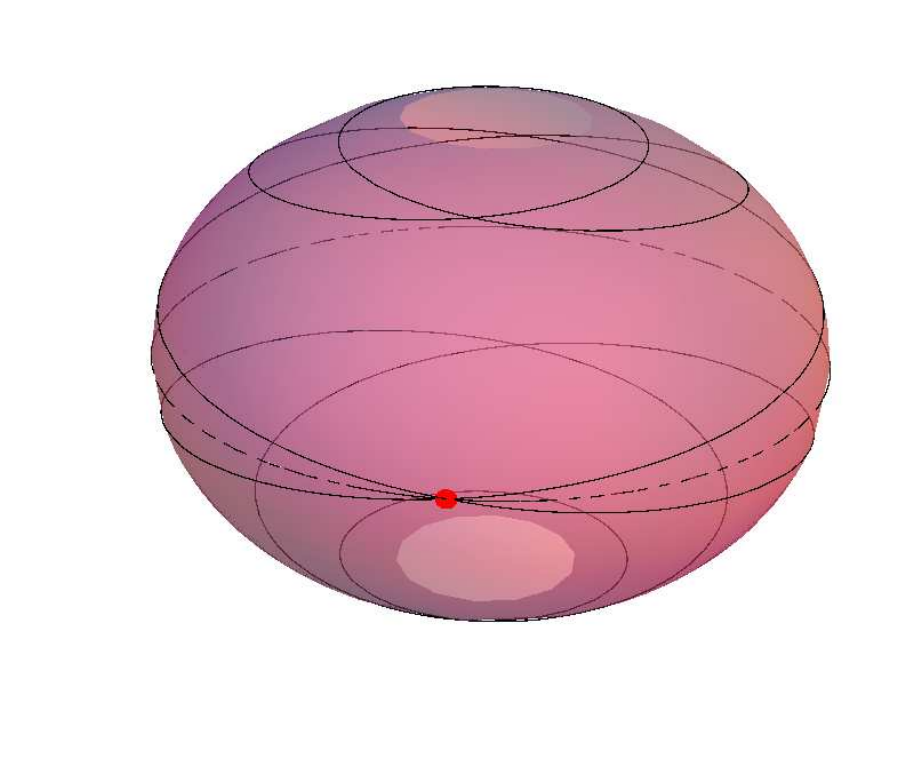}&\includegraphics[height=4cm]{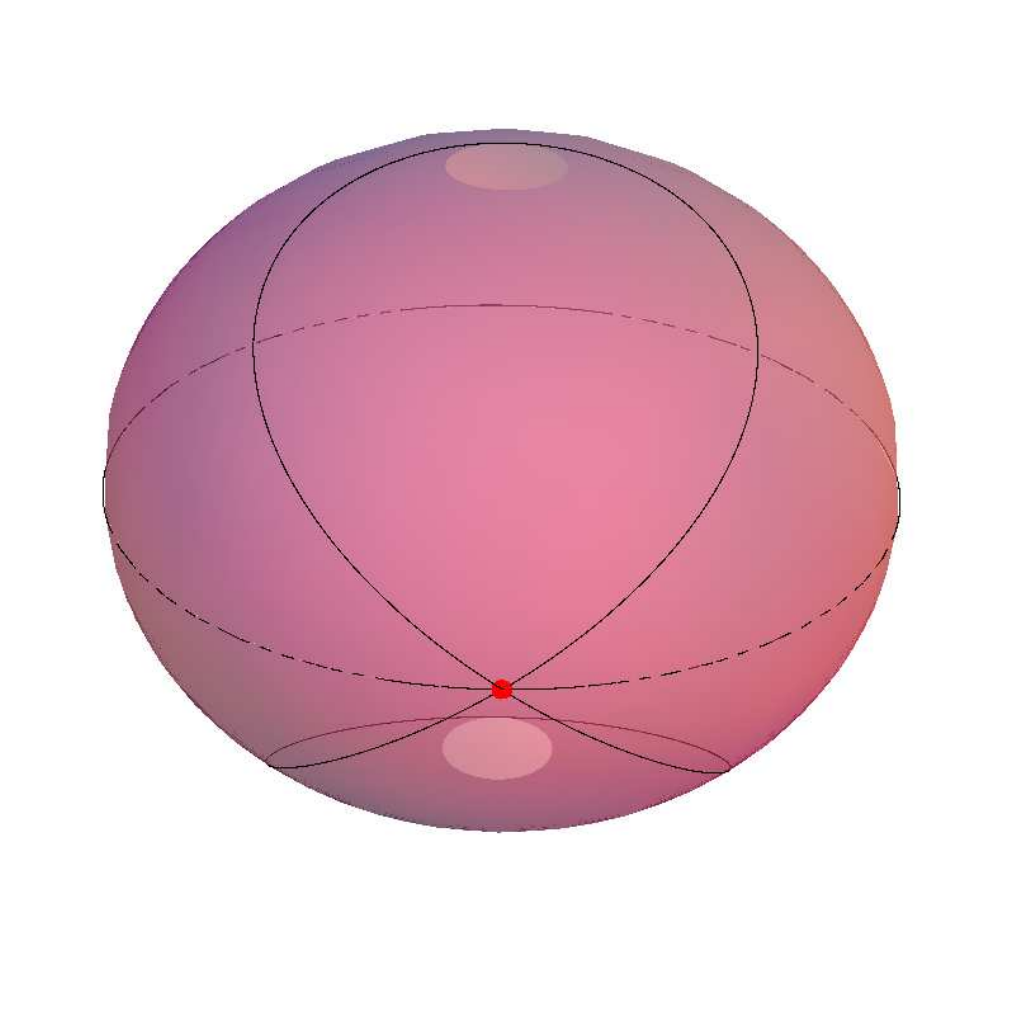}&\includegraphics[height=4cm]{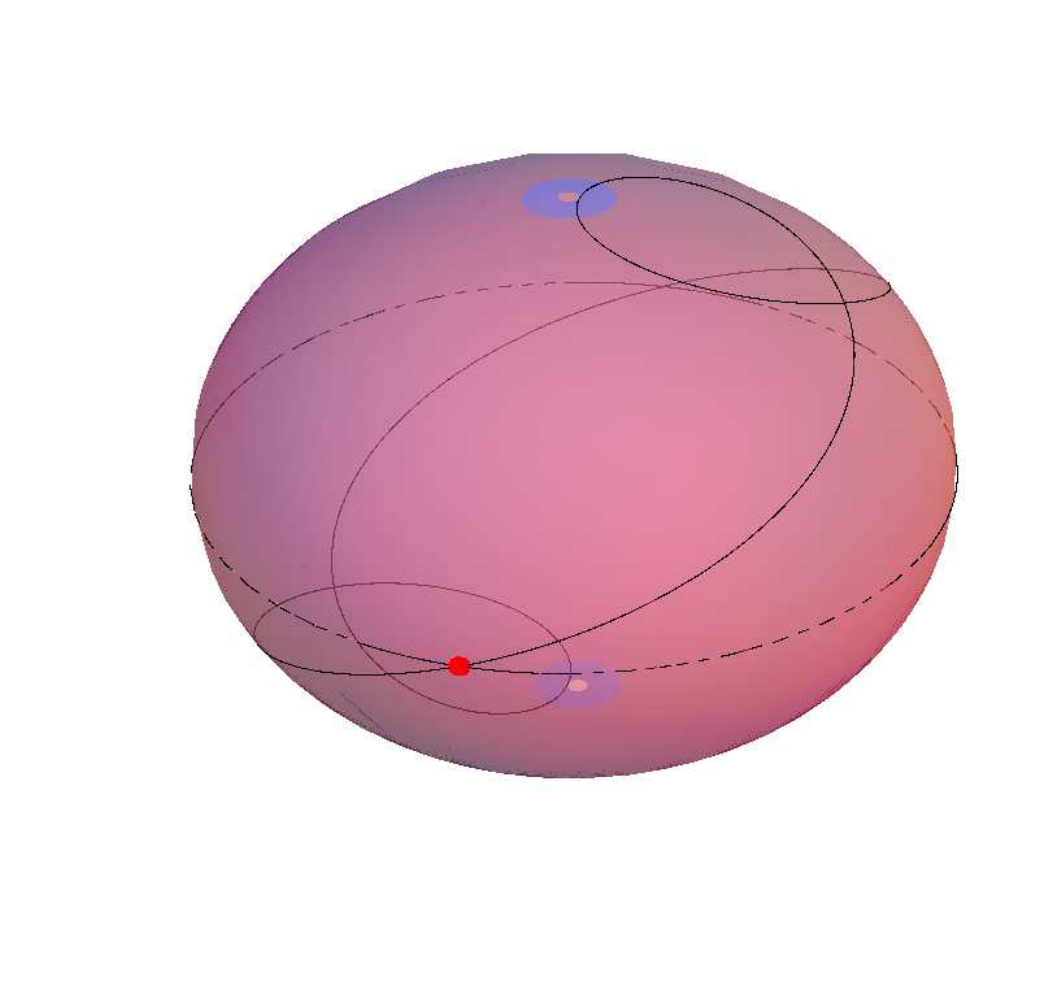}\\
		$k=6$&$k=2$&$k=1, \cale >0$\\
		\hline
		\includegraphics[height=4cm]{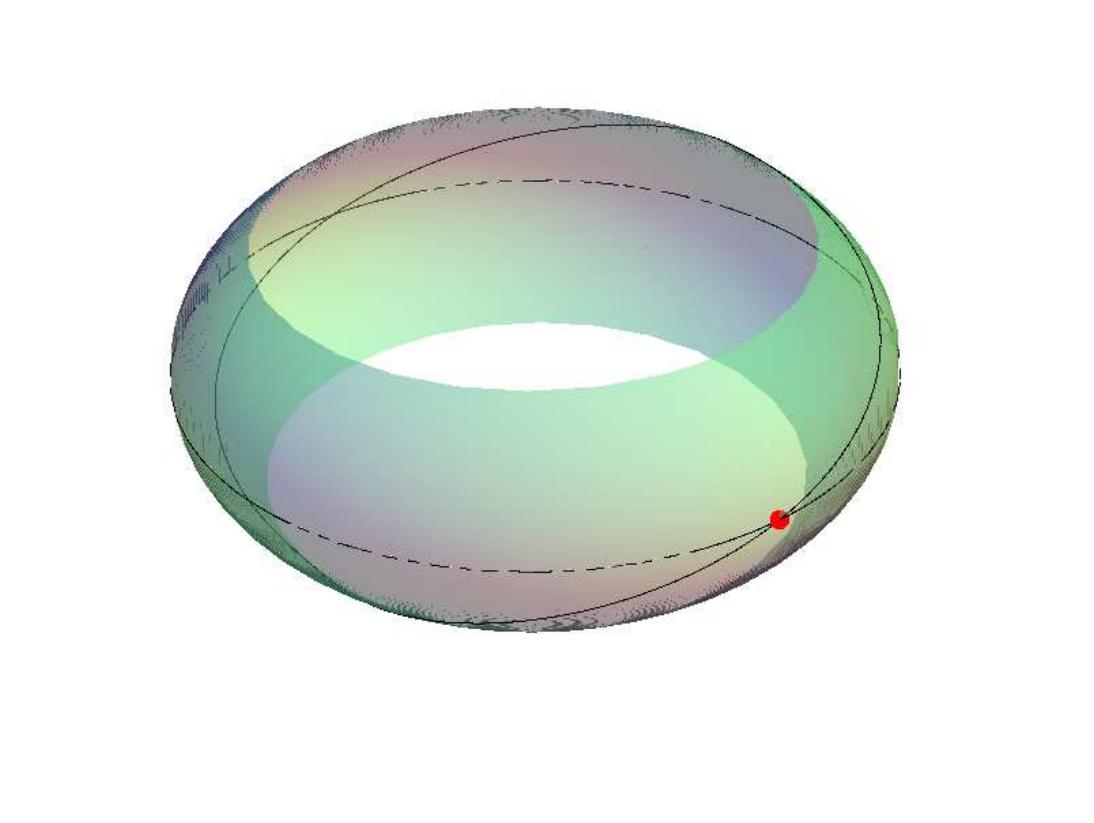}&\includegraphics[height=4cm]{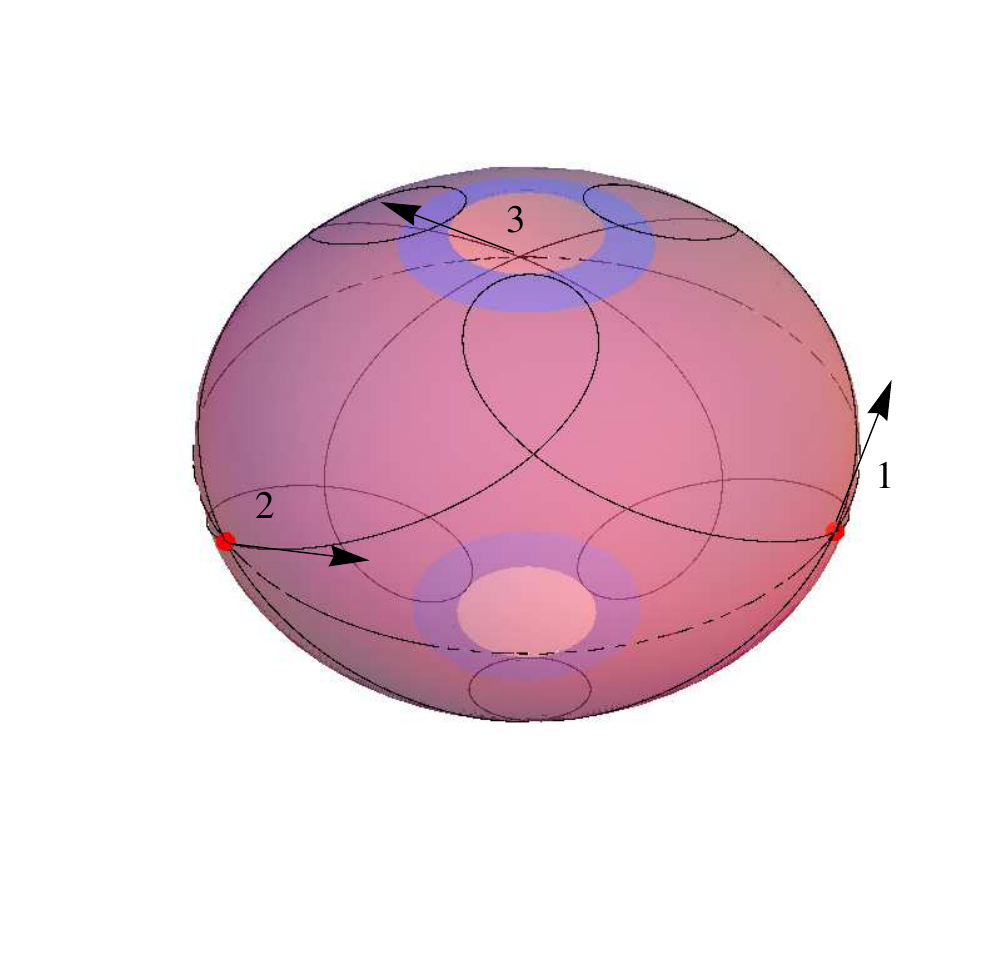}&\includegraphics[height=4cm]{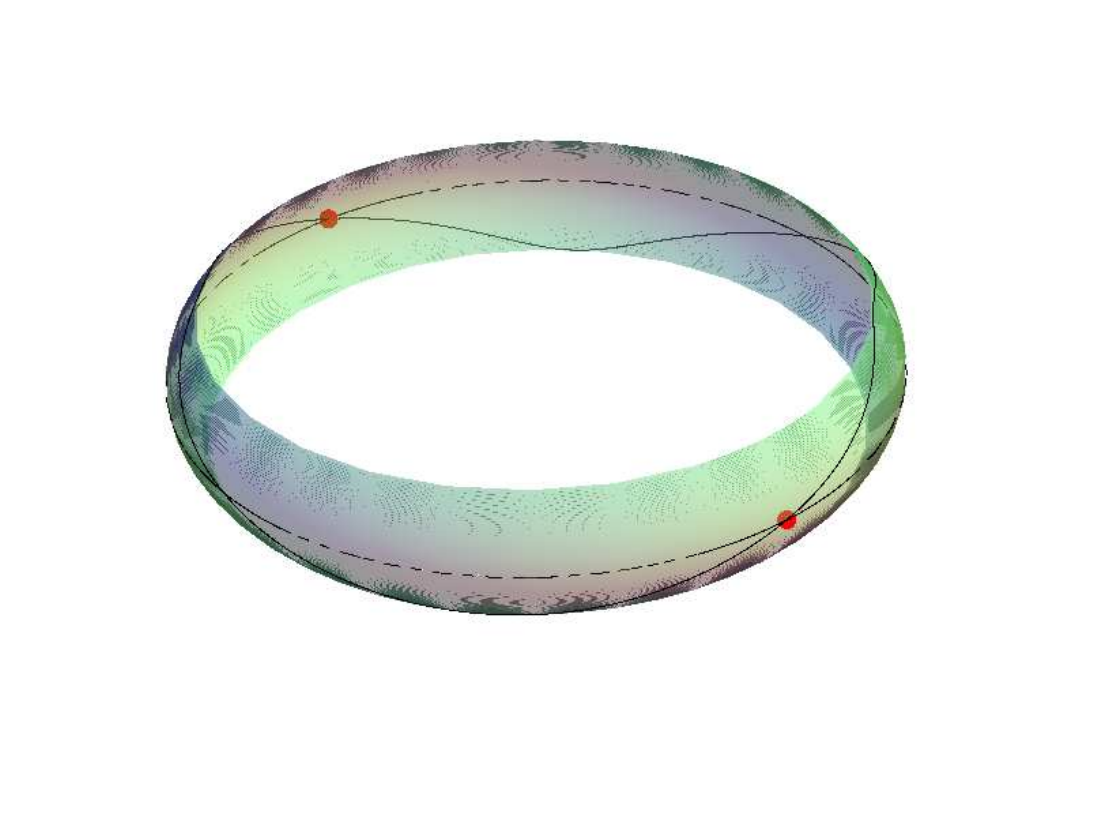}\\
		$k=1, \cale <0$&$k=2/3$&$k=1/2, \cale <0$\\
		\hline
		\includegraphics[height=4cm]{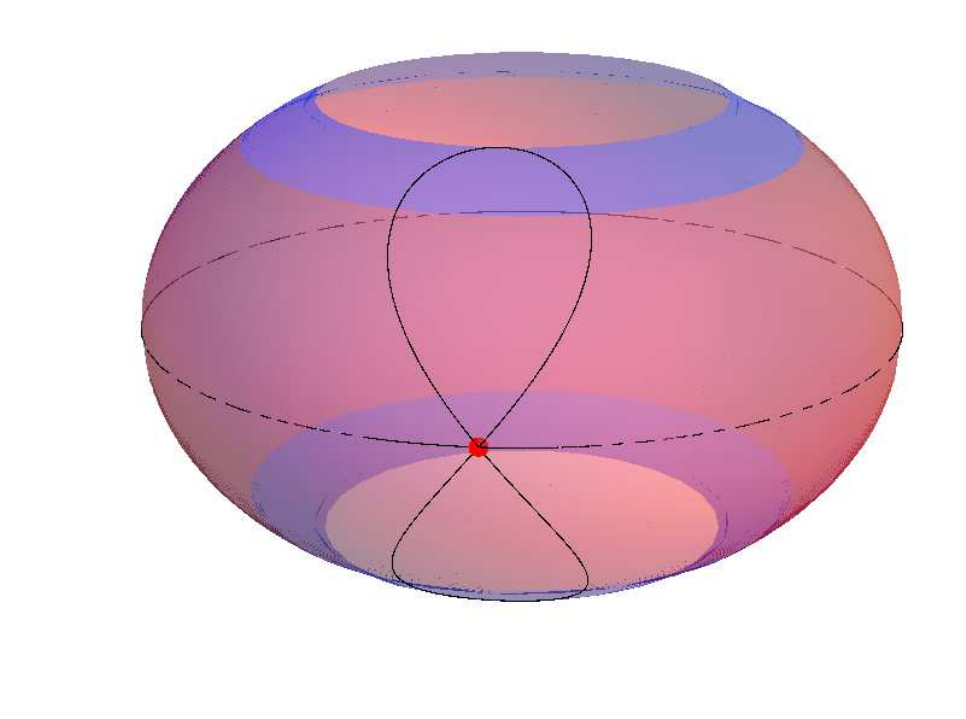}&\includegraphics[height=4cm]{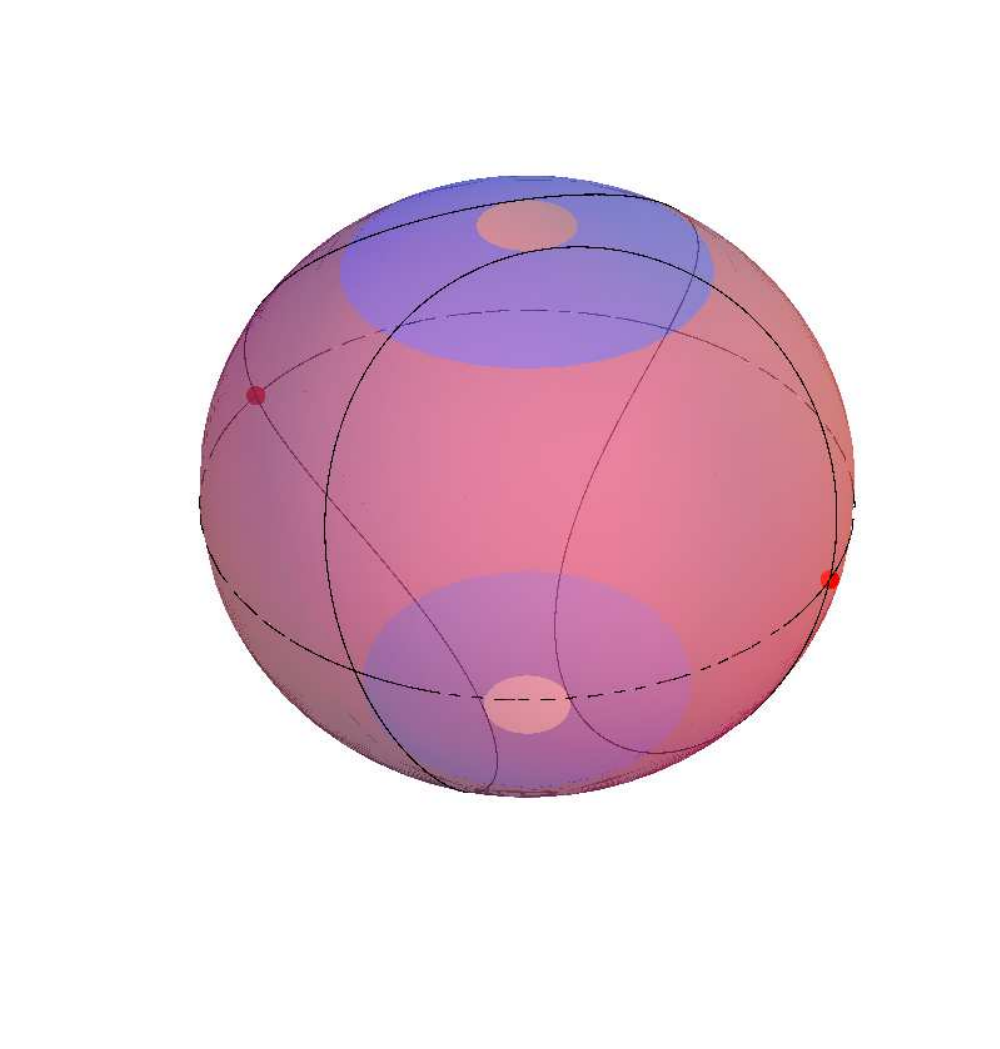}&\includegraphics[height=4cm]{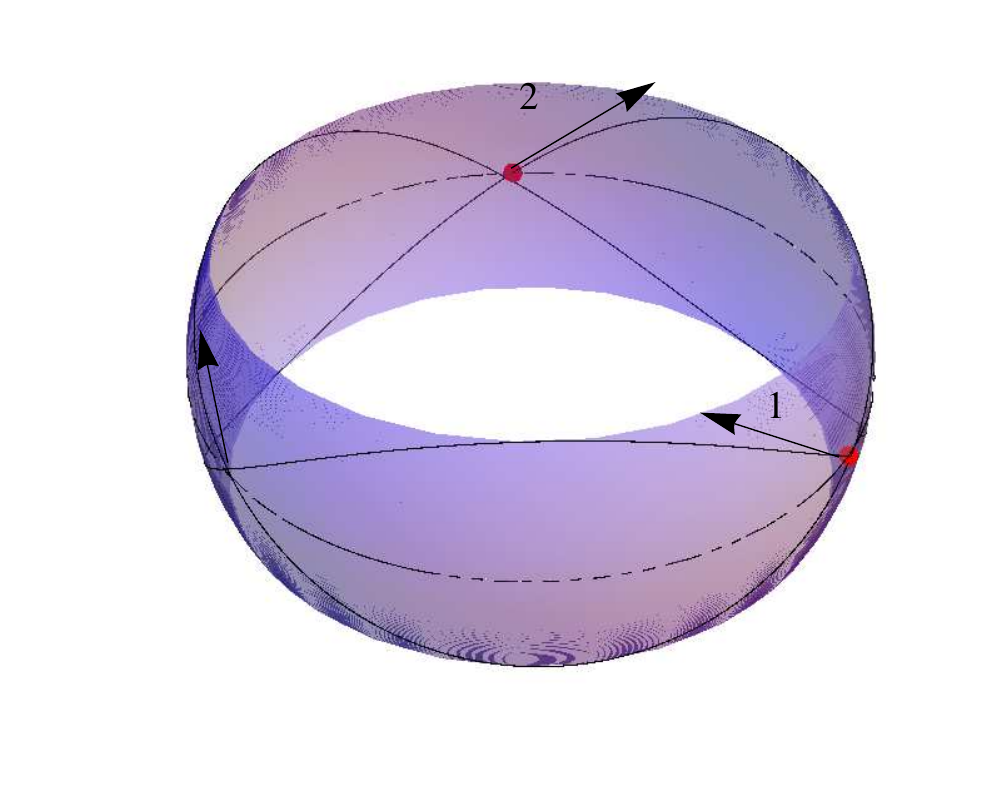}\\
		$k=0$&$k=-1/2$&$k=-2/3$\\
		\hline
			\end{tabular}

	\caption{Various types of the periodic SPOs characterized by $k$ - the number of revolutions about the $\phi$ - axis per one latitudinal oscillation. Here and hereafter we shall depict in green the sphere of the photon orbit with a negative energy, and in red/blue that parts of the spheres, where the photon moves in the positive/negative azimuthal direction, having a positive energy. Figures include cases of the prograde orbits ($k>0$), both with and without the turning point of the azimuthal motion, the oscillating orbit ($k=0$), and the retrograde orbits ($k<0$), with and without turning point in the azimuthal motion. Cases $|k|>>1$ correspond to orbits with multiple revolution about the spin axis per one latitudinal oscillation, tending to have a helix-like shape. However, the case $k\leq-1$ can not occur, hence such orbits are solely prograde. Details, including the minimum attained latitude $\theta_{min}$, and latitude $\theta_{\phi}$ at which the turning point in the $\phi$-direction occurs, sign of the photon's energy $\cale$ and the time period of the orbit (to be discussed bellow),  characterizing the orbits are presented in the attached Table \ref{tab_clos_orbs}.}\label{clos_orb}
\end{figure*}

\begin{figure*}[h]
	\centering
	\begin{tabular}{|c|c|}
		\hline
		\includegraphics[width=0.5\textwidth]{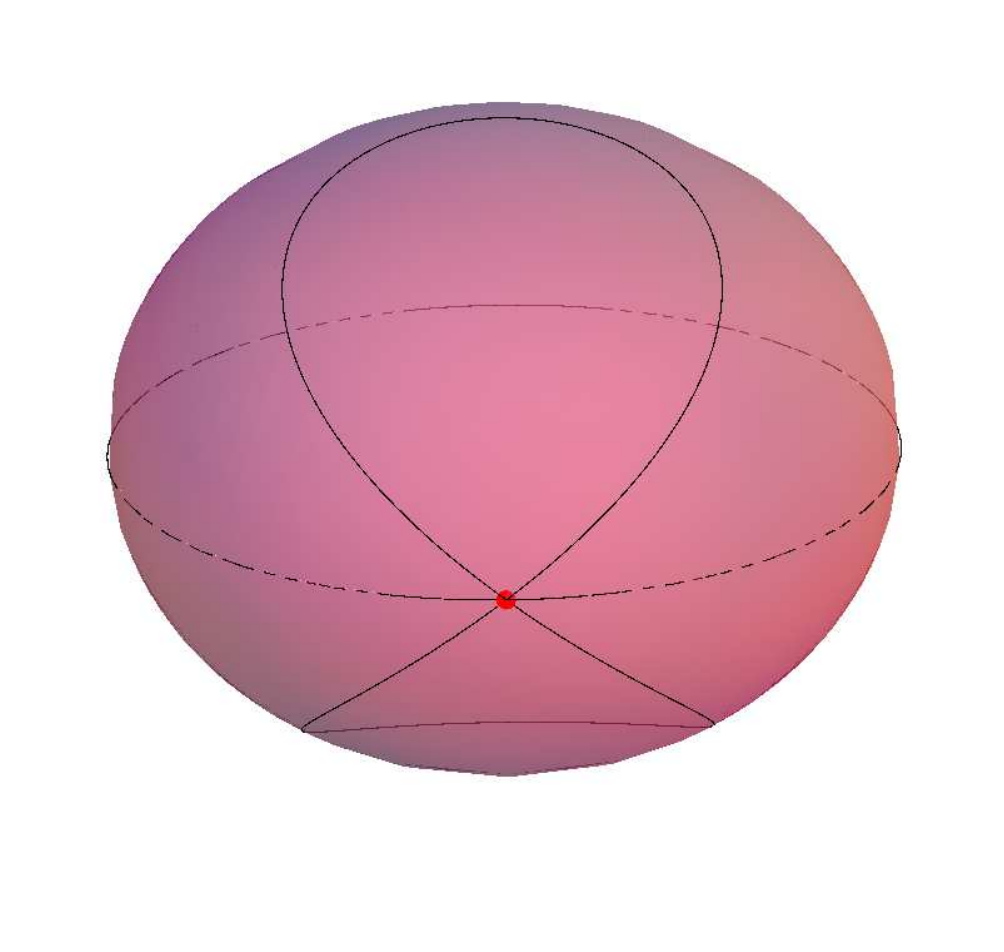}&\includegraphics[width=0.5\textwidth]{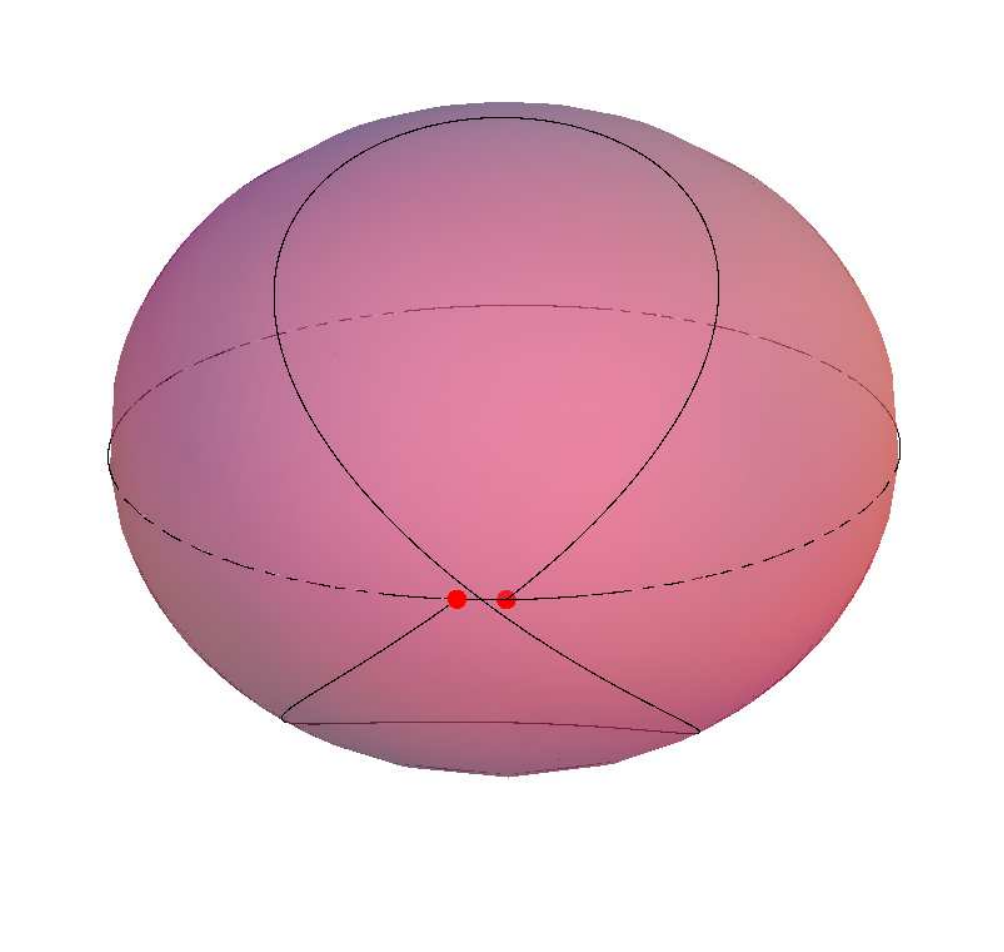}\\
		(a)&(b)\\
		\hline
	\end{tabular}
\caption{Illustration of the special cases of the polar SPOs: oscillatory orbit with $a=1.17986=a_{\Delta\phi=0(min)}$ at $r=1.7147=r_{\Delta\phi=0(max)}$ (Fig. a) and the SPO with $a=1.17996=a_{pol(max)}$ at the coalescing radius $r=r_{pol+}=r_{pol-}=\sqrt{3}=r_{pol}$ (Fig. b). Note that the latter orbit is globally retrograde (c. f. detail in Fig.\ref{shift_nodes}e).  }\label{phot_trace_rzs_azs}
\end{figure*}

\begin{figure*}[h]
	\centering
	\begin{tabular}{|c|c|c|}
		\hline
		\includegraphics[width=0.33\textwidth]{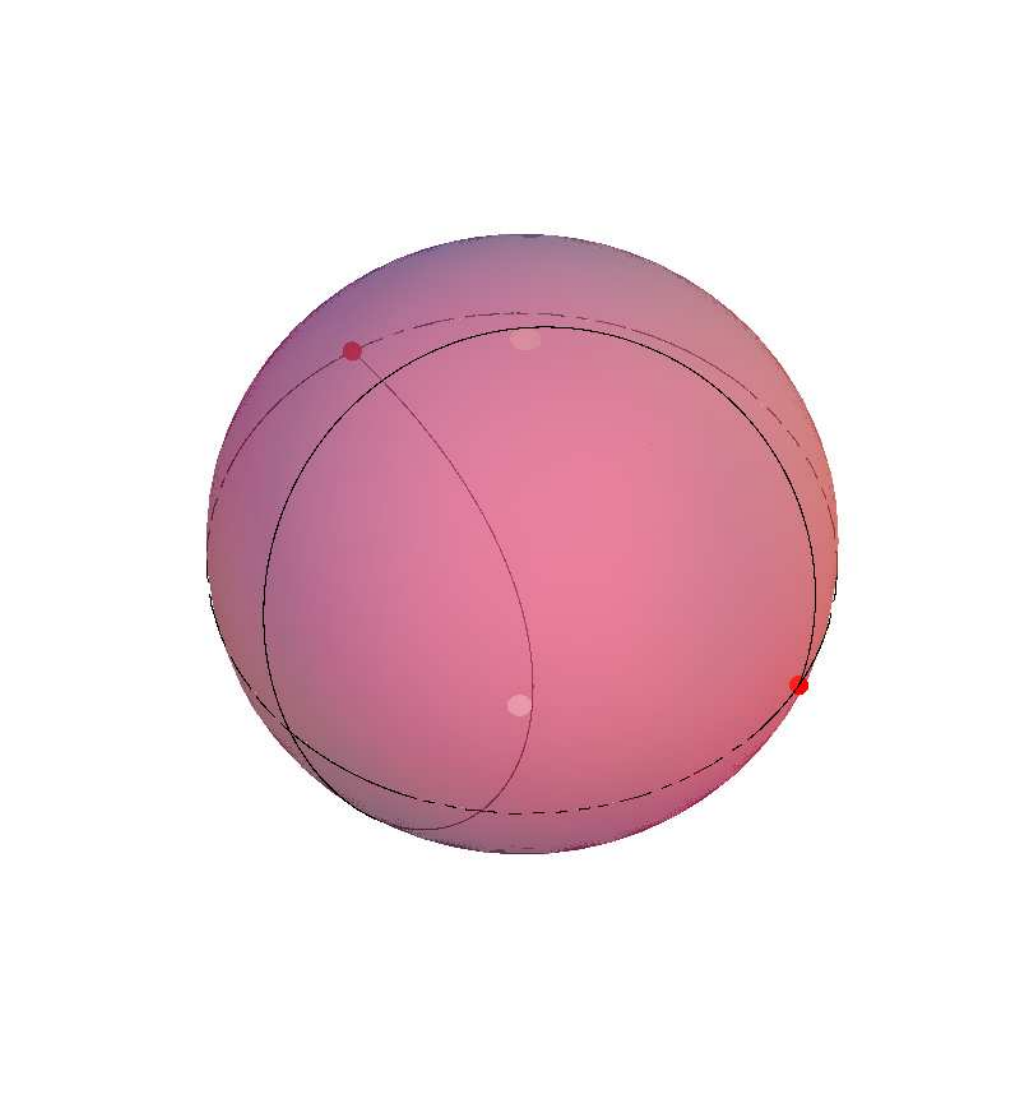}&\includegraphics[width=0.33\textwidth]{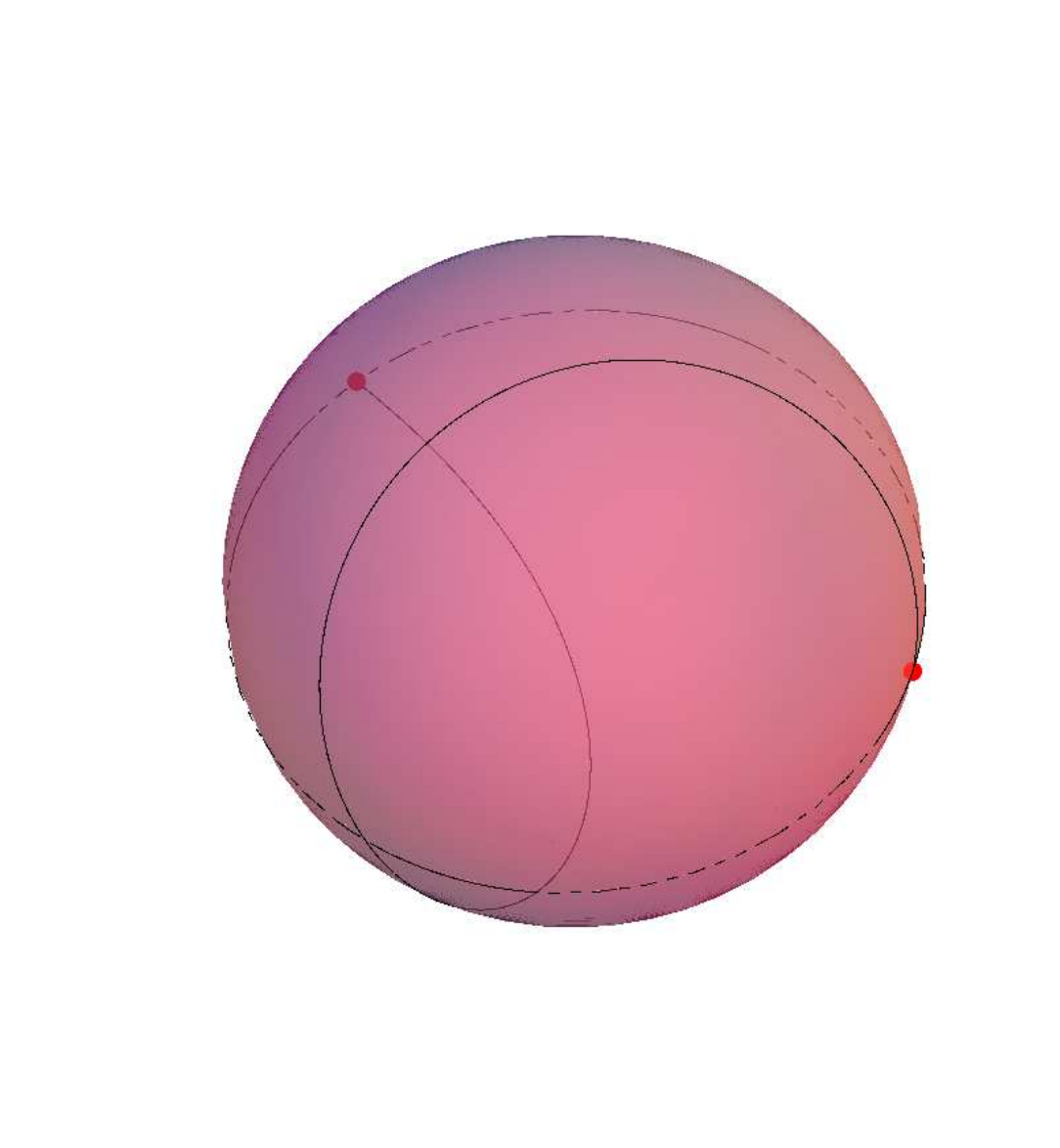}&\includegraphics[width=0.33\textwidth]{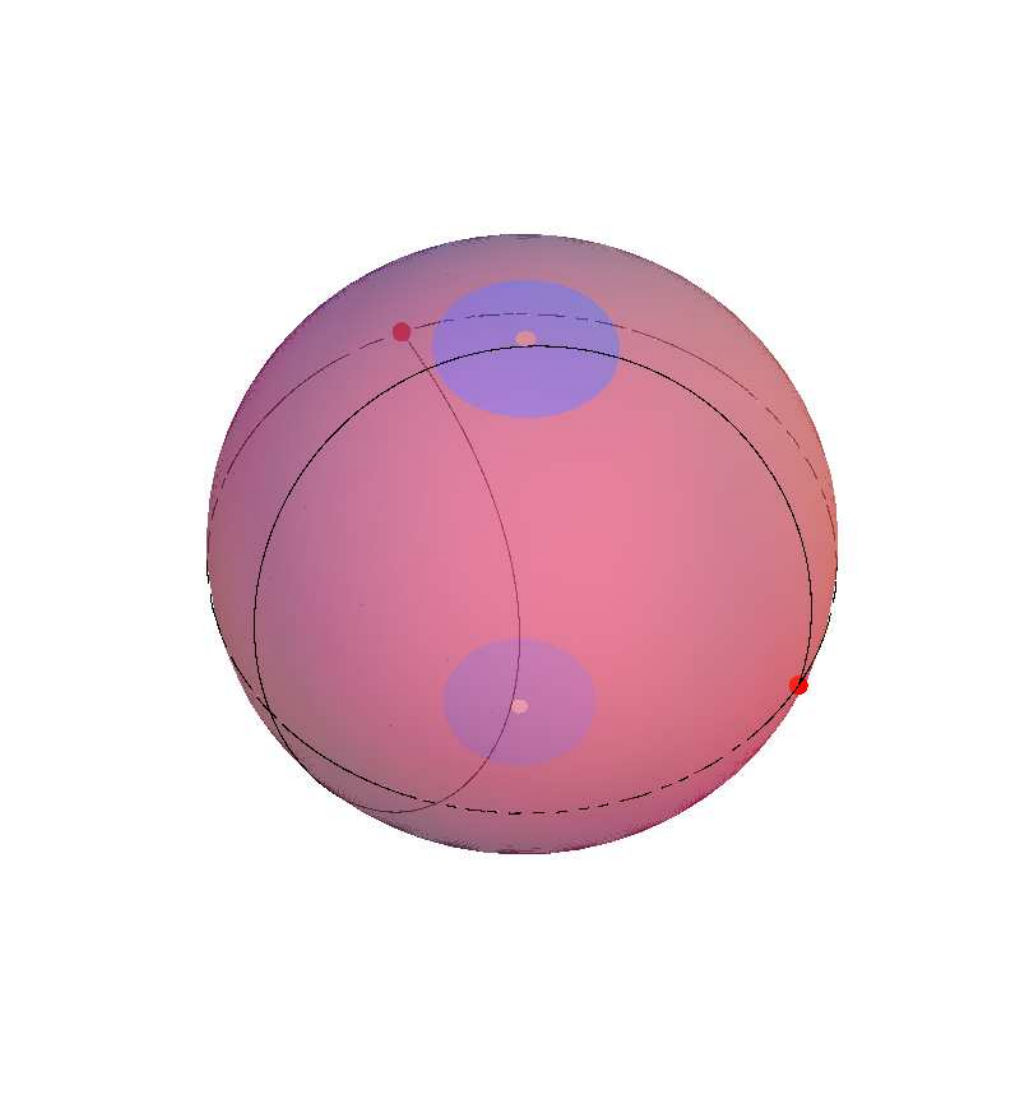}\\
		$r=2.5$, $\Delta \phi=518.4^\circ$&$r=2.56=r_{pol-}$, $\Delta \phi=151.7^\circ$&$r=2.6$, $\Delta \phi=-212.5^\circ$\\
		\hline
	\end{tabular}
	\caption{Demonstration of the $4\pi$ jump drop in value of the $\Delta \phi$ function  when overpassing the discontinuity point at $r=r_{pol-}$ in case of the Kerr BH with the spin parameter $a=0.9$ (see Fig. \ref{shift_nodes}(a)). In general, the SPOs are globally prograde at $r<r_{pol-}$ and globally retrograde at $r>r_{pol-}$, however, there is not zero nodal shift at $r=r_{pol-}$. The jump of the shift is a simple consequence of a singular behaviour of the $\phi$- coordinate at the poles $\theta=0, \pi$.  }\label{spoBH_rpolm}
\end{figure*}

In the Fig. \ref{Fig_snrp12}, the nodal shift is shown for the polar spherical orbits $r_{pol\pm}$, i. e., for the isolated points in Fig. \ref{shift_nodes}, in dependence on the spin parameter $a$. As follows from the above, varying $a$, the first occurrence of the zero nodal shift appears when there is $\Delta \phi[r_{pol+}(a)]=2\pi$. According to Figs. \ref{shift_nodes}-\ref{Fig_snrp12}, we can summarize that for the very tiny interval $a_{\Delta\phi=0(min)}\leq a \leq a_{pol(max)}$ there exist orbits with $\Delta \phi <0,$ i. e., globally retrograde, with radii $r_{\Delta\phi=0(max)}<r<r_{pol+}(a)$, followed by orbits with $\Delta \phi >0$, i. e., globally prograde, at $r_{pol+}(a)<r<r_{pol-}(a)$, and again orbits globally retrograde with $r>r_{pol-}$. Otherwise, the SPOs of this class of the KNS spacetimes possesses no new essential features in comparison with the other cases, therefore, its character is explained sufficiently.\\
The radius of the oscillatory orbits $r_{\Delta\phi=0}$ is the only point dividing the trajectories which are globally prograde ($\Delta \phi>0$), and those that are globally retrograde ($\Delta \phi<0$) in case $a>a_{pol(max)}$. For the KNS spacetimes with $a_{\Delta\phi=0(min)}<a<a_{pol(max)}$ the role of such dividing points play the radii $r_{\Delta\phi=0}$ and the discontinuity points $r_{pol\pm}$. For $a<a_{\Delta\phi=0(min)}$ only the radius $r_{pol-}$ is the divider, however, the oscillatory orbits are not possible (c.f. orbits in the BH background with $a=0.9$ at $r\lesssim r_{pol-}$, $r=r_{pol-}$ and $r\gtrsim r_{pol-}$ in Fig. \ref{spoBH_rpolm}). The function $r_{\Delta\phi=0}(a)$ is compared in Fig. \ref{radfunc} with another relevant functions. 

Of course, another closed paths corresponding to the general periodic orbits occur whenever 
\be
     n\Delta \phi =2m\pi . 
\ee
In Fig.\ref{clos_orb}, we present basic types of the closed SPOs for several ratios $k=m/n$, giving the cases of prograde ($k>0$), oscillate ($k=0$), and retrograde ($k<0$) orbits. Of course, of special astrophysical relevance could be the oscillate orbits with $\Delta \phi=0$, as they immediately return to the starting point at the azimuthal coordinate related to distant static observers. 

Further, we consider a possibility of orbits having a plurality of revolutions about the spacetime symmetry axis per one latitudinal oscillation ($|k|>1$), which are of a helix-like shape, and that of greater number of latitudinal oscillations per one revolution about the axis ($|k|<1$), both with, or without a turning point in the $\phi$-direction. Note that the cases $k\leq -1$ cannot be realized, since from Eq. (\ref{delphi}) it follows that the nodal shift $\Delta \phi>-2\pi$. This is demonstrated in Fig. \ref{Fig_delphi} on the left, which depicts the shift of nodes at $r\to r_{ph-}$.

\begin{table*}[h]
	\caption{Characteristics of the periodic spherical photon orbits in Fig.\ref{clos_orb} .}\label{tab_clos_orbs}
	\begin{tabularx}{\textwidth}{XXXXXXXXX}
		\hline
		$k$&$a$&$r$&$\ell$&$q$&$\theta_{min}$&$\theta_{\phi}$&$\sign \cale$&$\Delta t$\\
		\hline
		\hline
		$6$&$1.01$&$1.11$&$1.58$&$13.15$&$22.9^\circ$&-&$+1$&$74.76$\\
		$2$&$1.1$&$1.72$&$0.62$&$16.42$&$8.5^\circ$&-&$+1$&$30.13$\\
		$1$&$1.1$&$1.22$&$-0.16$&$30.32$&$1.6^\circ$&$6.7^\circ$&$+1$&$32.98$\\
		$1$&$1.1$&$0.77$&$3.2$&$6.85$&$49.8^\circ$&-&$-1$&$14.08$\\
		$2/3$&$1.18$&$1.24$&$-1.47$&$41.44$&$12.5^\circ$&$21.2^\circ$&$+1$&$26.36$\\
		$1/2$&$1.1$&$0.53$&$2.25$&$0.92$&$64.6^\circ$&-&$-1$&$5.69$\\
		$0$&$\sqrt{3}$&$1.69$&$-3.63$&$30.80$&$32.3^\circ$&$47.1^\circ$&$+1$&$19.36$\\
		$-1/2$&$1.1$&$2.52$&$-0.72$&$24.38$&$8.2^\circ$&$32.2^\circ$&$+1$&$29.45$\\
		$-2/3$&$\sqrt{3}$&$4.34$&$-7.31$&$8.70$&$67.5^\circ$&-&$+1$&$34.30$\\		
		\hline
		
	\end{tabularx}	
\end{table*}

\clearpage

\subsection{Spherical photon orbits with zero energy and their nodal shift}

For the SPOs with zero energy, located at radii $r=1$, the constants of motion have to fulfil the relation $\Phi^2/Q=a^2-1$. The latitudinal equation (\ref{Cart_Th1}) with the substitution $m=\cos^2\theta$ then reads
\be
(1/2\Sigma \frac{\din m}{\din \lambda})^2=M'(m;a,Q)\equiv Qm(1-a^2m),\label{Cart_M1}
\ee
and the azimuthal equation (\ref{Cart_Phi1}) reads 
\be
\Sigma \frac{\din \phi}{\din \lambda}=\varPhi'(m;a,\Phi)\equiv\Phi\frac{1-a^2m}{(a^2-1)(m-1)}.\label{Cart_Phi1m}
\ee
Therefore, the turning points of the latitudinal and the azimuthal motion coalesce at $m=1/a^2$, in concordance with the result for intersection of the functions $m_{\theta}(r;a)$, $m_{\phi}(r;a)$, as we have found earlier. In this case, the point with $d\phi/d\lambda = 0$ is not a real azimuthal turning point, it is only the point of vanishing of the azimuthal velocity, as shown in Fig.\ref{phot_paths_zen}.

\begin{figure}[hb]
	\centering
	\includegraphics[width=8cm,height=5cm]{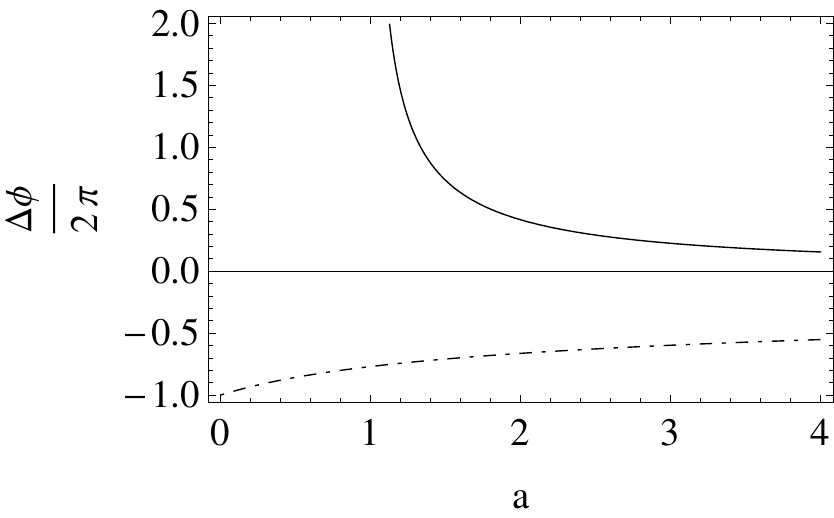}
	\caption{Behaviour of the function (\ref{Del_Phi1}) in dependence on the rotational parameter $a,$ depicted as the full curve, which represents the shift of nodes $\Delta\phi$ of orbits with zero energy. The dot-dashed curve is a depiction of the lowest negative shift of nodes of a retrogressive spherical orbits approaching the position of the equatorial circular counter-rotating orbit at $r=r_{ph-}.$ It can be seen that the increase of the rotational parameter $a$ weakens the negative shift due to strong dragging of the spacetime. The shift is always $\Delta\phi>-2\pi.$
	}\label{Fig_delphi}
\end{figure}

The complete change of the azimuthal coordinate per one latitudinal oscillation, denoted $\Delta \phi$, is now given by the relation 
\be
\Delta \phi=2\pi(\frac{a}{\sqrt{a^2-1}}-1),\label{Del_Phi1}
\ee
which corresponds to values of the local maxima of the function (\ref{delphi}). Behaviour of this function is shown in Fig. \ref{Fig_delphi}. 

\begin{widetext}
	\onecolumngrid

\begin{figure*}[h]
	
	\centering
	\begin{tabular}[t]{|c|c|}
		\hline
		\includegraphics[width=0.45\textwidth]{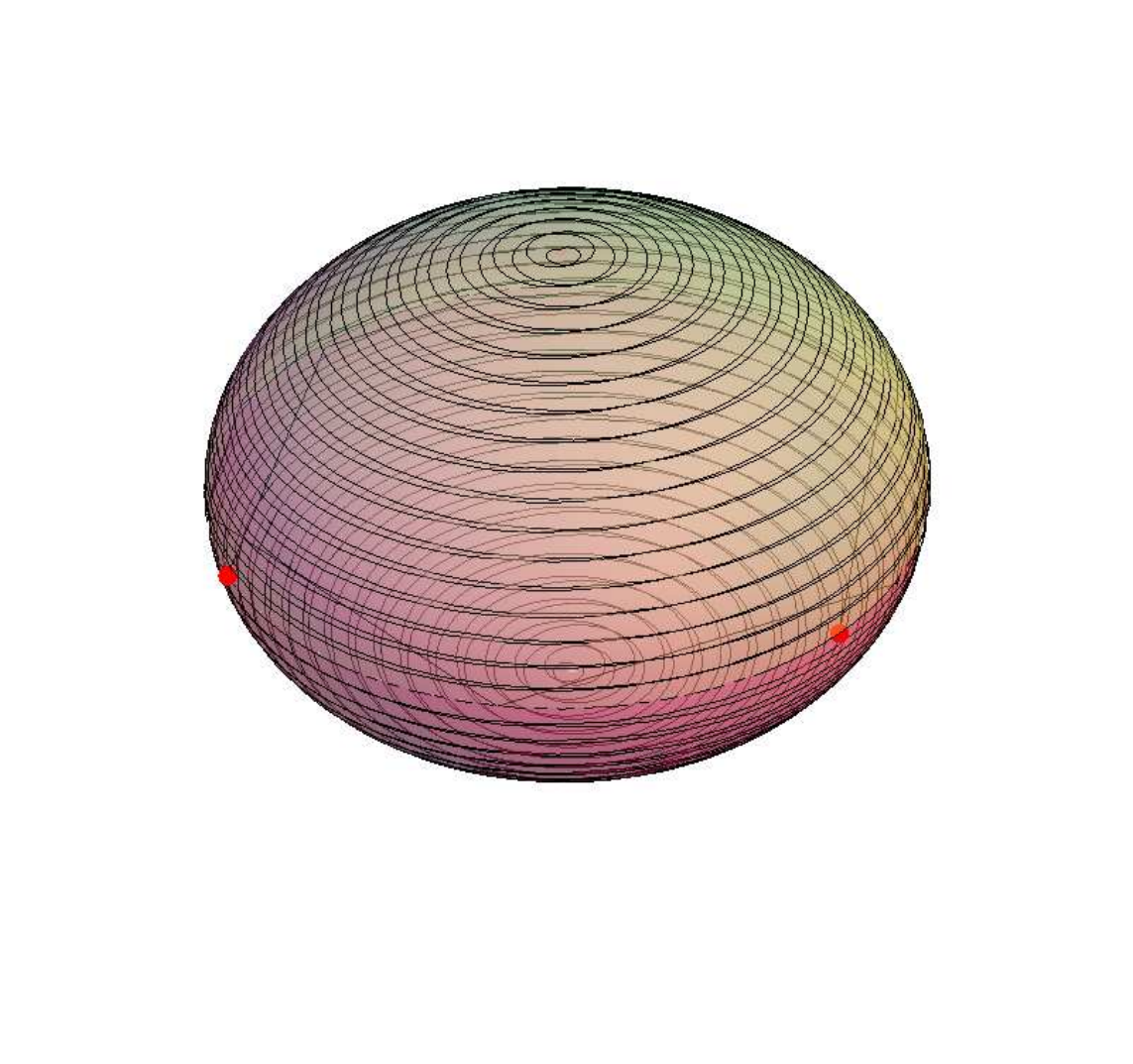}&\includegraphics[width=0.45\textwidth]{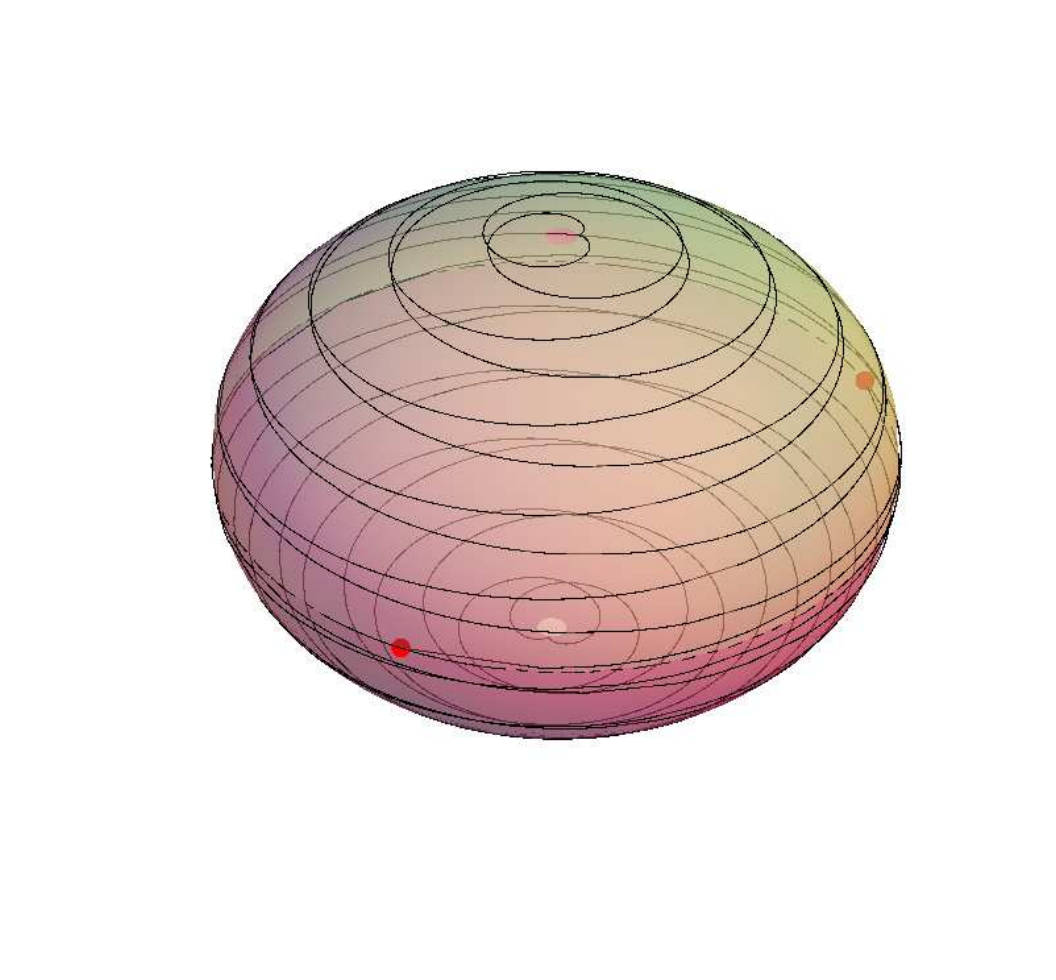}\\
		$a=1.0001$, one latitudinal oscillation&$a=1.001$, one latitudinal oscillation\\
		\hline
	\end{tabular}
\vspace{1cm}
\center (\textit{Figure continued})

\end{figure*}

\begin{figure*}[h]
\begin{tabular}[t]{|c|c|}
		\hline
		\includegraphics[width=0.45\textwidth]{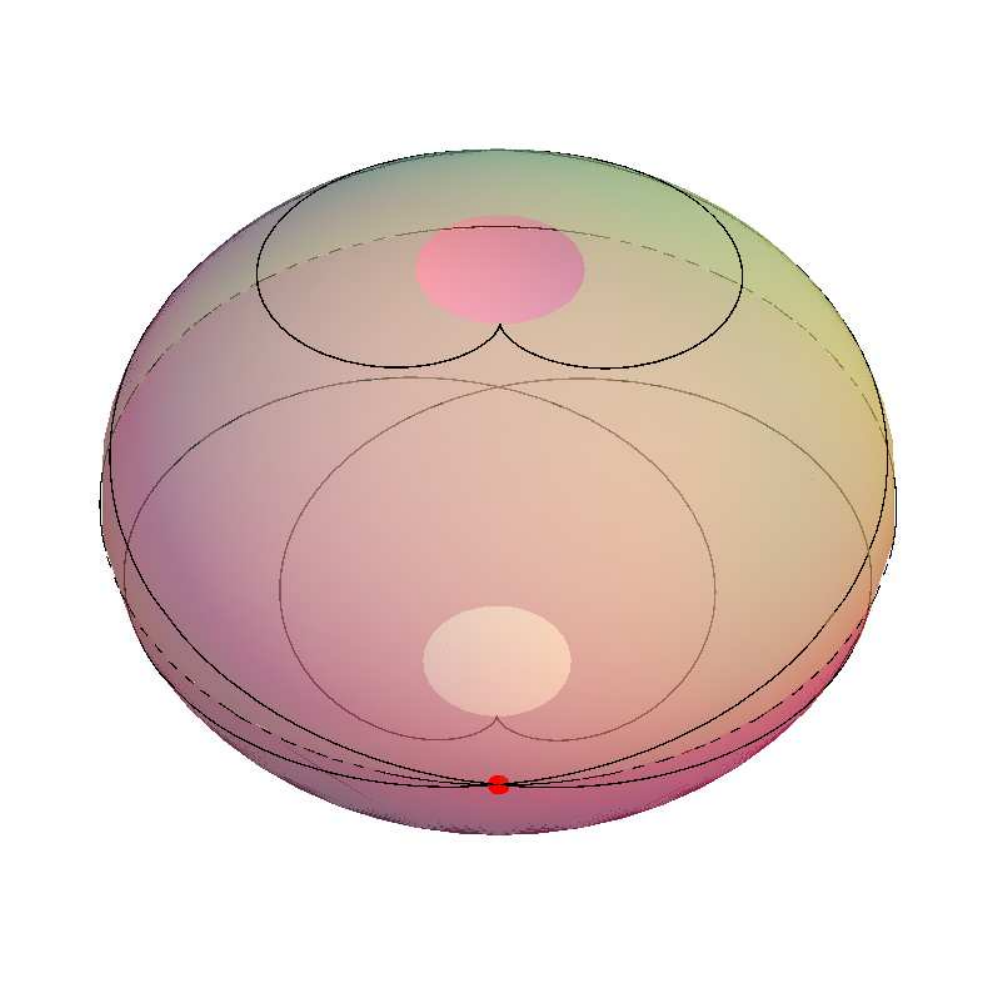}&\includegraphics[width=0.45\textwidth]{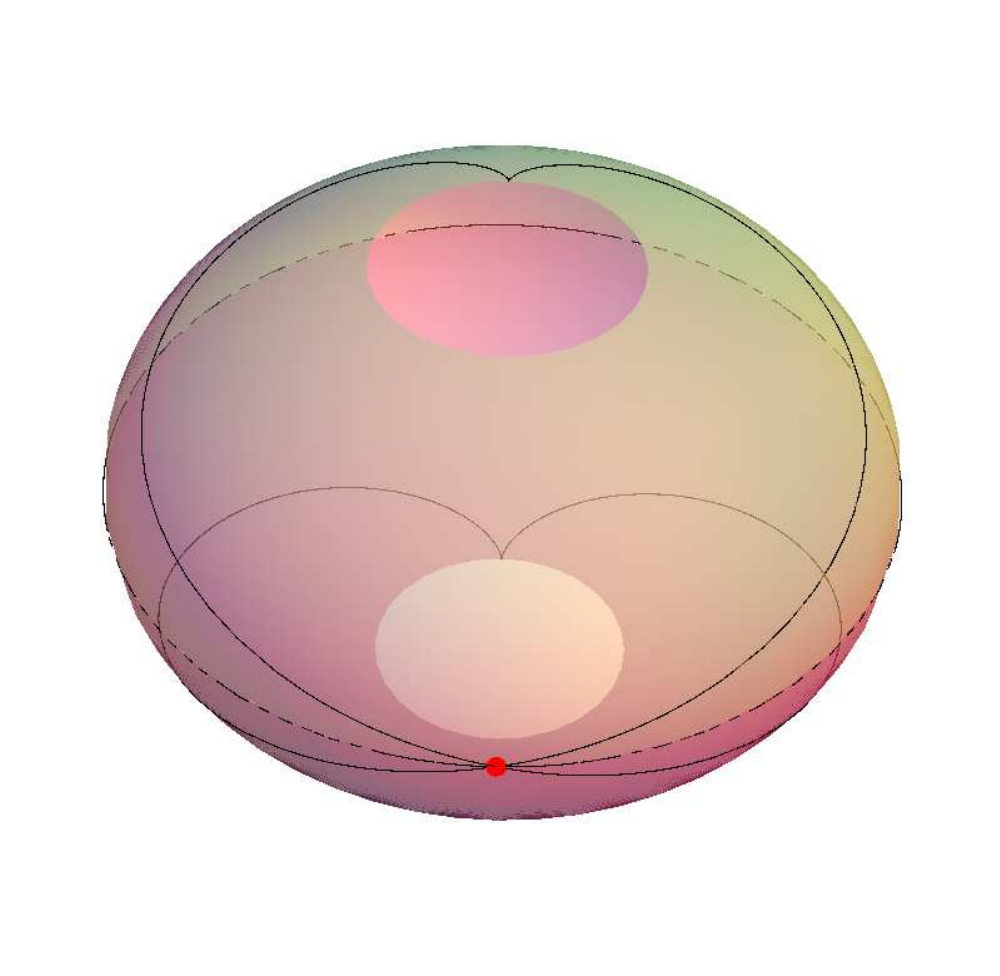}\\
		$a=\sqrt{25/24}=1.02$, one latitudinal oscillation&$a=\sqrt{9/8}=1.06$, one latitudinal oscillation\\
		\hline
		\includegraphics[width=0.45\textwidth]{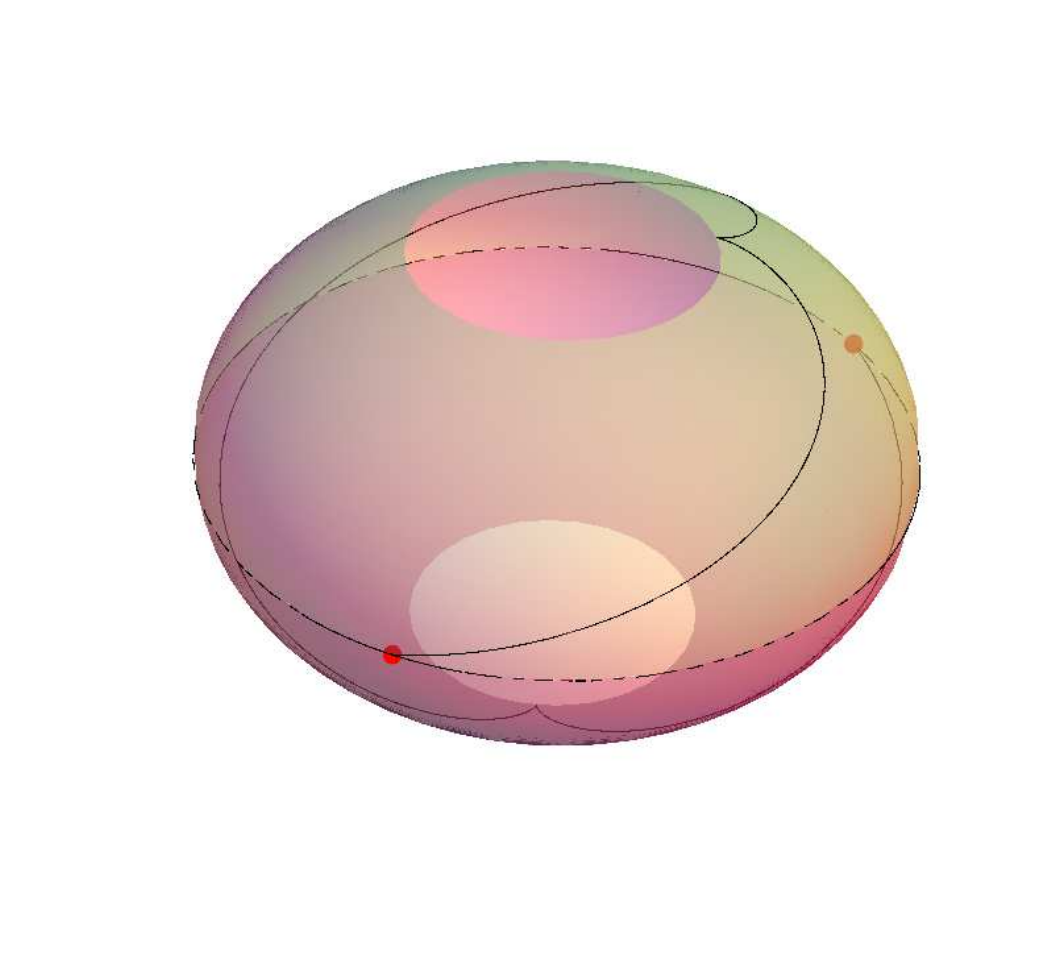}&\includegraphics[width=0.45\textwidth]{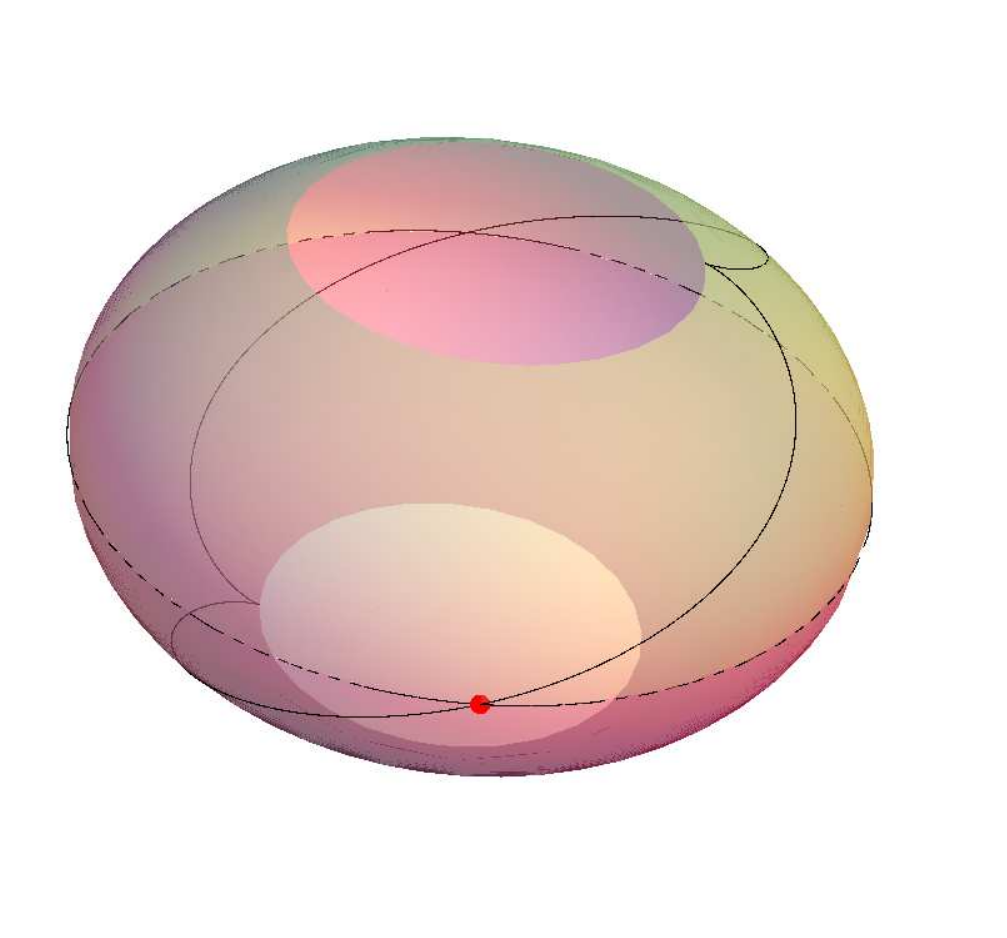}\\
		$a=1.1$, one latitudinal oscillation&$a=\sqrt{4/3}=1.15$, one latitudinal oscillation\\
		\hline	
		\includegraphics[width=0.45\textwidth]{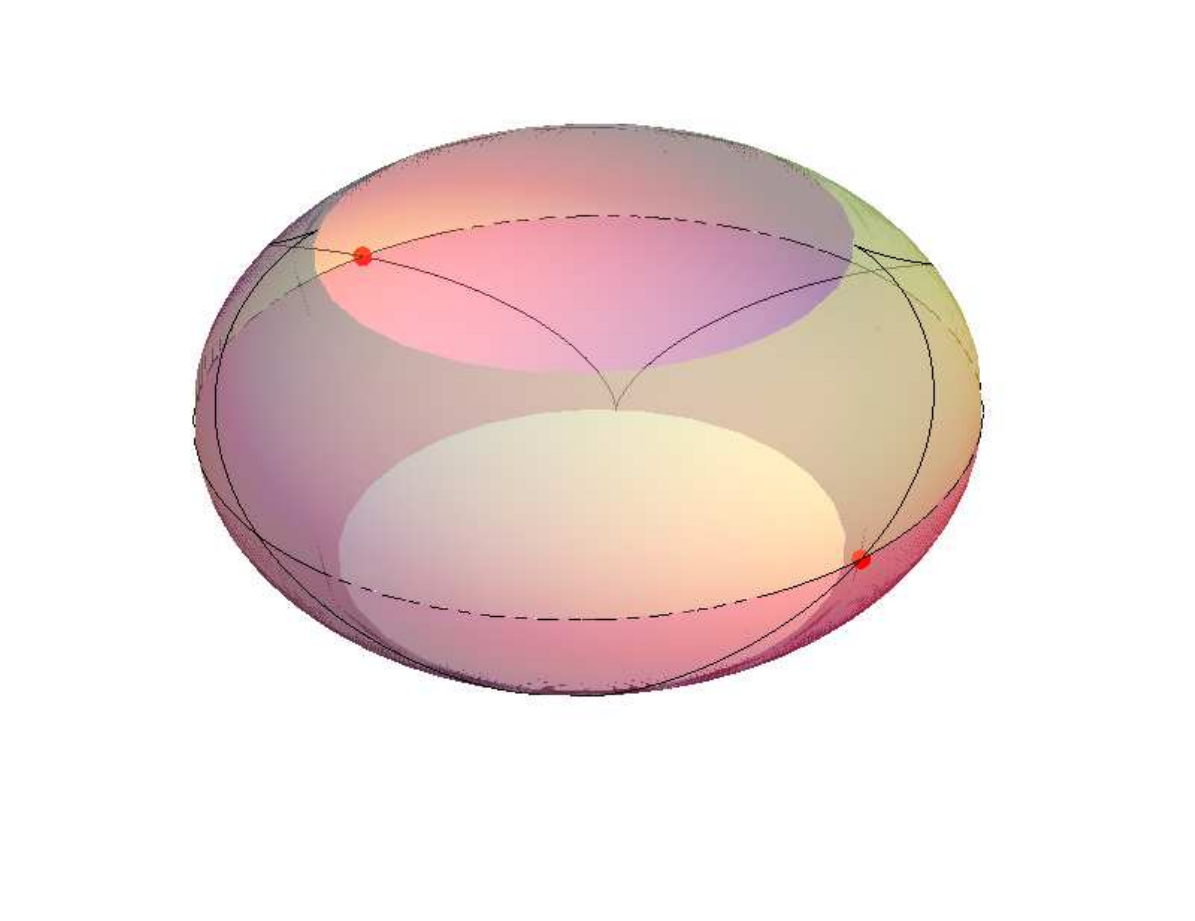}&\scalebox{.8}{\includegraphics[trim=0 -2cm 0cm -3cm]{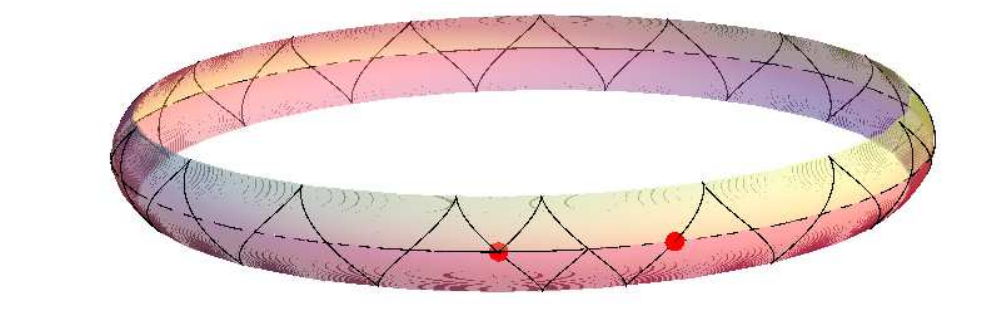}}\\
		$a=\sqrt{9/5}=1.34$, two latitudinal oscillations&$a=3$, seventeen latitudinal oscillations
		\\
		\hline
	\end{tabular}
	\caption{The SPOs with zero energy, located solely at radius $r=1$, are presented for KNS spacetimes with increasing value of the rotational parameter $a$. The turning points of the latitudinal and the azimuthal motion coalesce in these cases, hence the turning loops have shrinked into peaks. In the attached Table \ref{tab_zenorbs} we demonstrate the descending total change in azimuth per one latitudinal oscillation, and the descending range in the latitudinal motion, with increasing spin parameter $a$. In some cases closed spherical orbits emerge due to properly chosen parameters of the SPOs.}\label{phot_paths_zen}
\end{figure*}
\clearpage

\begin{table*}[h]
	\caption{Characteristics of the spherical photon orbits with zero energy in Fig.\ref{phot_paths_zen}.}\label{tab_zenorbs}
	\begin{tabularx}{\textwidth}{XXXX}
		\hline
		$a$&$\Phi^2/Q$&$\Delta \phi/2\pi$&$\theta_{min}$\\
		\hline
		\hline
		$1.0001$&$0.0002$&$69.7$&$0.8^\circ$\\
		$1.001$&$0.002$&$21.4$&$2.6^\circ$\\
		$1.02$&$0.042$&$4$&$11.5^\circ$\\
		$1.06$&$0.125$&$2$&$19.5^\circ$\\
		$1.1$&$0.21$&$1.4$&$24.6^\circ$\\
		$1.15$&$0.33$&$1$&$30.0^\circ$\\
		$1.34$&$0.8$&$0.5$&$41.8^\circ$\\
		$3$&$8$&$0.06$&$70.5^\circ$\\
		\hline
		
	\end{tabularx}	
\end{table*}
\twocolumngrid
\end{widetext}

\subsection{Summary of properties of spherical photon orbits}

We can summarize properties of the SPOs in the following way. The covariant energy $\cale>0$ ($\cale<0$) have the SPOs at $r>1$ ($r<1$) for all KBHs and KNSs; there is SPO with $\cale=0$ at $r=1$. The other properties depend on the dimensionless spin $a$.  

In the KBH spacetimes, the stable SPOs are located under the inner event horizon at $0<r<r_{ms-}$, the unstable ones at $r_{ms-}<r<r_{ph0}$, all being co-rotating (prograde), none of these orbits can be polar. Above the outer event horizon only unstable SPOs spread in the interval $r_{ph+}\leq r\leq r_{ph-}$. They are purely co-rotating for $r_{ph+}<r<r_{pol-}$; with turning point in the azimuthal direction but globally counter-rotating (retrograde) for $r_{pol-}<r<3$; and purely counter-rotating at $3<r<r_{ph-}$. One polar orbit exist above the outer horizon of the KBH spacetimes. 

In the KNS spacetimes the SPOs exist at $0<r<r_{ph-}$, being stable/unstable for $r\lessgtr r_{ms+}$. For the KNS spacetimes with $1<a<a_{\Delta\phi=0(min)}$ the SPOs at $r<1$ are purely co-rotating; for $1<r<r_{pol+}$ they have turning point in the azimuthal direction but are co-rotating globally; between the inner stable and the outer unstable polar orbit, i. e., at $r_{pol+}<r<r_{pol-}$, they are purely co-rotating; at $r_{pol-}<r<3$ they have turning point in the azimuthal direction and are counter-rotating globally; for $3<r<r_{ph-}$ they are purely counter-rotating. Two polar orbits can exist in such KNS spacetimes. 

There exists extremely small interval of the spin parameter $a_{\Delta\phi=0(min)}<a<a_{pol(max)}$, for which the KNS spacetimes possess two polar orbits and one orbit of zero nodal shift $r_{\Delta\phi=0}(a)$ -- for $1<r<r_{\Delta\phi=0}(a)$ there are orbits with turning point in the azimuthal motion, being co-rotating in global; such orbits occur also for $r_{\Delta\phi=0}<r<r_{pol+}$, being globally counter-rotating; at the radii $r_{pol+}<r$, the same behaviour occurs as in the previous case. 

\begin{figure}[htbp]
	
	\centering
	\begin{tabular}{c}
	
	\includegraphics[width=0.45\textwidth]{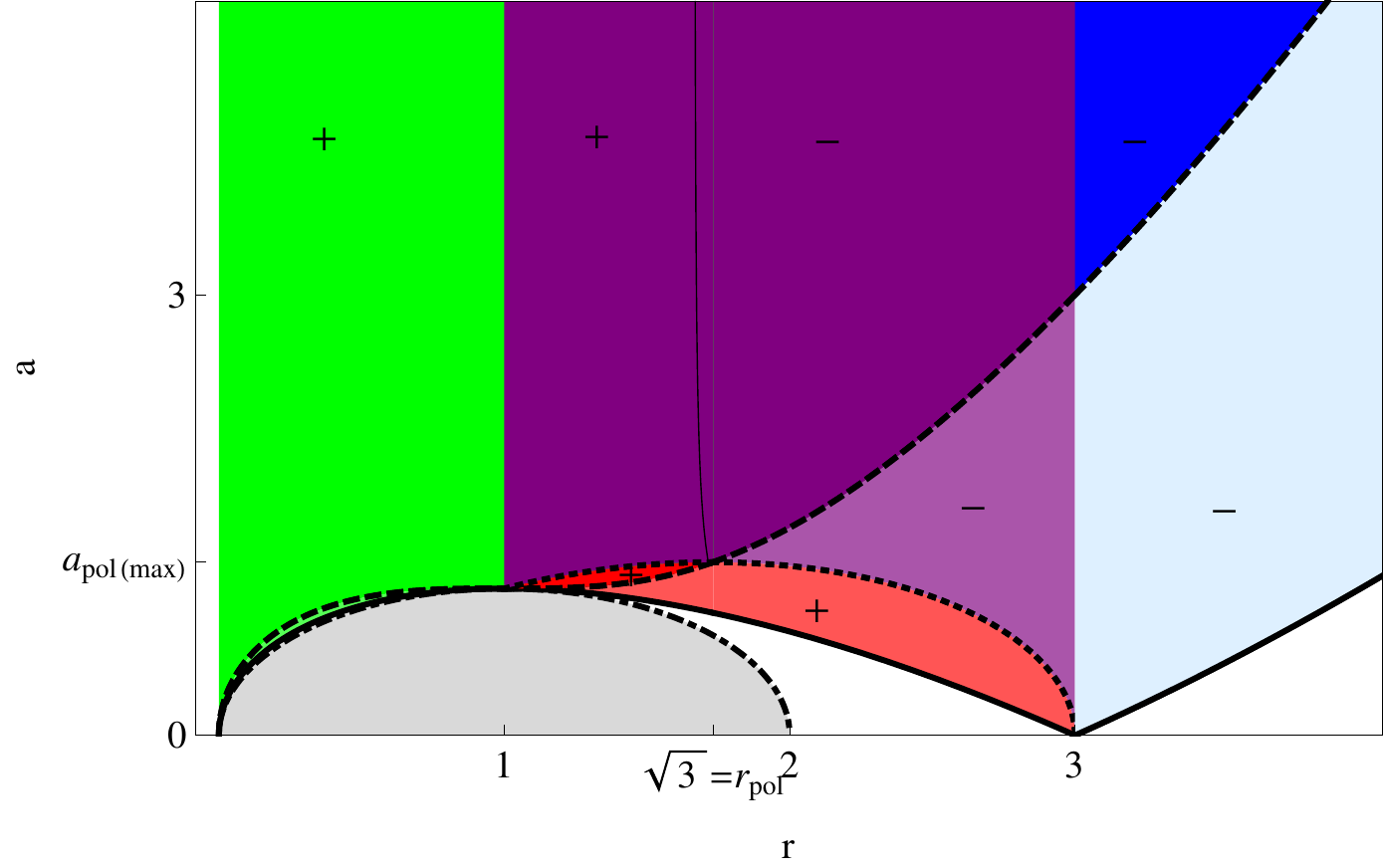}\\
	\\
	\includegraphics[width=0.45\textwidth]{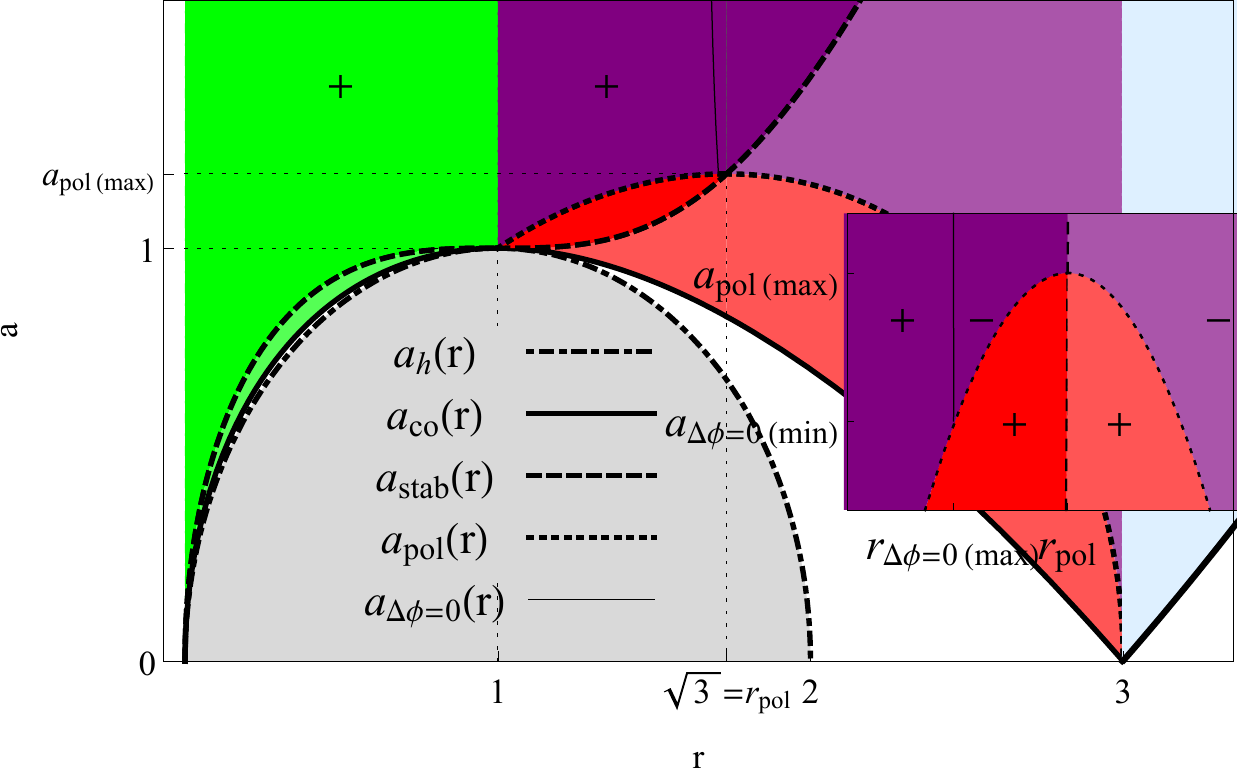}\\
	
	\end{tabular}
	\caption{Characteristic functions $a(r)$ with supplementary informations about stability and latitudinal and azimuthal motion for wider interval of the parameter $a$ (above) and in greater detail (bellow). Significance of the characteristic functions is given in Fig. \ref{Fig_ar}. Gray shading demarcates the dynamic region, colouring highlights region of the SPOs. Orbits with $\cale<0,$ all being purely prograde, i. e., without turning points in $\phi$-direction, are highlighted in green; purely prograde orbits with $\cale>0$ in red; orbits with turning points in $\phi$-direction and with $\cale>0$ in purple and purely retrograde with $\cale>0$ in blue. Full/faint hues correspond to stable/unstable orbits. Thin black curve is the zero nodal shift function $a_{zs}(r)$ dividing regions of globally prograde/retrograde orbits marked by $+/-$ sign.}\label{Figure 1ref}

\end{figure}

For the KNS spacetimes with $a>a_{pol(max)}$, the SPOs at $r<1$ are purely corotating; at $1<r<3$ there are orbits with turning points of the azimuthal motion, which can be both co-rotating or counter-rotating globally, in dependence on the spin $a$; spherical orbits at $r>3$ are purely counter-rotating. For the KNS spacetimes with $a>3$ there is $3<r_{ms+}$, hence stable purely counter-rotating spherical orbits exist at $3<r<r_{ms+}$. 
All the above results are clearly illustrated in Fig. \ref{Figure 1ref}.

\vspace{1cm}
\clearpage

\section{Classification of the Kerr spacetimes due to properties of the spherical photon orbits} 

Using the knowledge of the behaviour of the above described functions, we are now able to summarize the properties of the SPOs in dependence on the value of the Kerr spacetime spin parameter $a$, giving thus the corresponding classification of these spacetimes. The following description of the individual classes is supplemented by an explicit illustrations of the spatial structures of the SPOs, which can regarded as a spatial representation of the Fig. \ref{Figure 1ref}. In the figures we use the so called Kerr-Schild coordinates $x,y,z$ that are connected to the Boyer-Lindquist coordinates $r,\theta$ by the relations 
\be
x^2 + y^2 = (r^2 + a^2)\sin^2\theta ,\quad z^2 = r^2\cos^2\theta.
\ee
We present a meridional sections of the SPOs with the $y$- coordinate being suppressed, hence, the surfaces of constant Boyer-Lindquist radius are depicted as an oblate ellipses. 

\begin{description}
	\item \textbf{Class I}: Kerr black holes $0<a<1$ (see Fig. \ref{Class_I}) endowed with the SPOs of two families. First family is limited by radii $0<r<r_{ph0}<r_{-},$ where
	\be
	r_{ph0}\equiv 4\sin^2 [\frac{1}{6}\arccos(1-2a^2)] \label{rph0}
	\ee
	is the radius of co-rotating equatorial circular photon orbit located under the inner black hole horizon $r_{-}$ (green dot). These are the orbits with negative energy $\cale<0$; at $0<r<r_{ms-}$ they are stable (rich green area), for $r_{ms-}<r<r_{-}$ they are unstable with respect to radial perturbations (light green area). The radius $r_{ms-}$, given by (\ref{rms-}), denotes marginally stable spherical orbit with negative energy. All the first family orbits are prograde and span small extent in latitude in vicinity of the equatorial plane -- its maximum at $r=r_{ms-}$ is approaching the value $\theta_{min(z\cale)}(a=1)=\arccos\sqrt{2\sqrt{3}-3}=47.1^\circ$ as $r\to 1$ when $a\to 1$.  Here and in the following, we restrict our discussion on the 'northern' hemisphere ($0\leq \theta \leq \pi/2$), the situation in the 'southern' hemisphere is symmetric with respect to the equatorial plane. Second family of the SPOs spreads between the inner co-rotating (red dot) and outer counter-rotating (blue dot) equatorial circular orbits with radii given by (\ref{bh_circ_orbs}). There is no turning point of the azimuthal motion for orbits with radii $r_{ph+}<r<r_{pol-}$, where $r_{pol-}$ is the radius of the polar spherical orbit (purple ellipse) given by (\ref{rpol12}), and all such photons are prograde. At radii $r_{pol-}<r<3$ (black dotted incomplete ellipse), there exist orbits with one turning point of the azimuthal motion in each hemisphere, such that the photons become retrograde as they approach the symmetry axis. At radii $3<r<r_{ph-}$, all spherical orbits are occupied by retrograde photons with no turning point of the azimuthal motion.
	\begin{figure}[h]
		\includegraphics[width=0.45\textwidth]{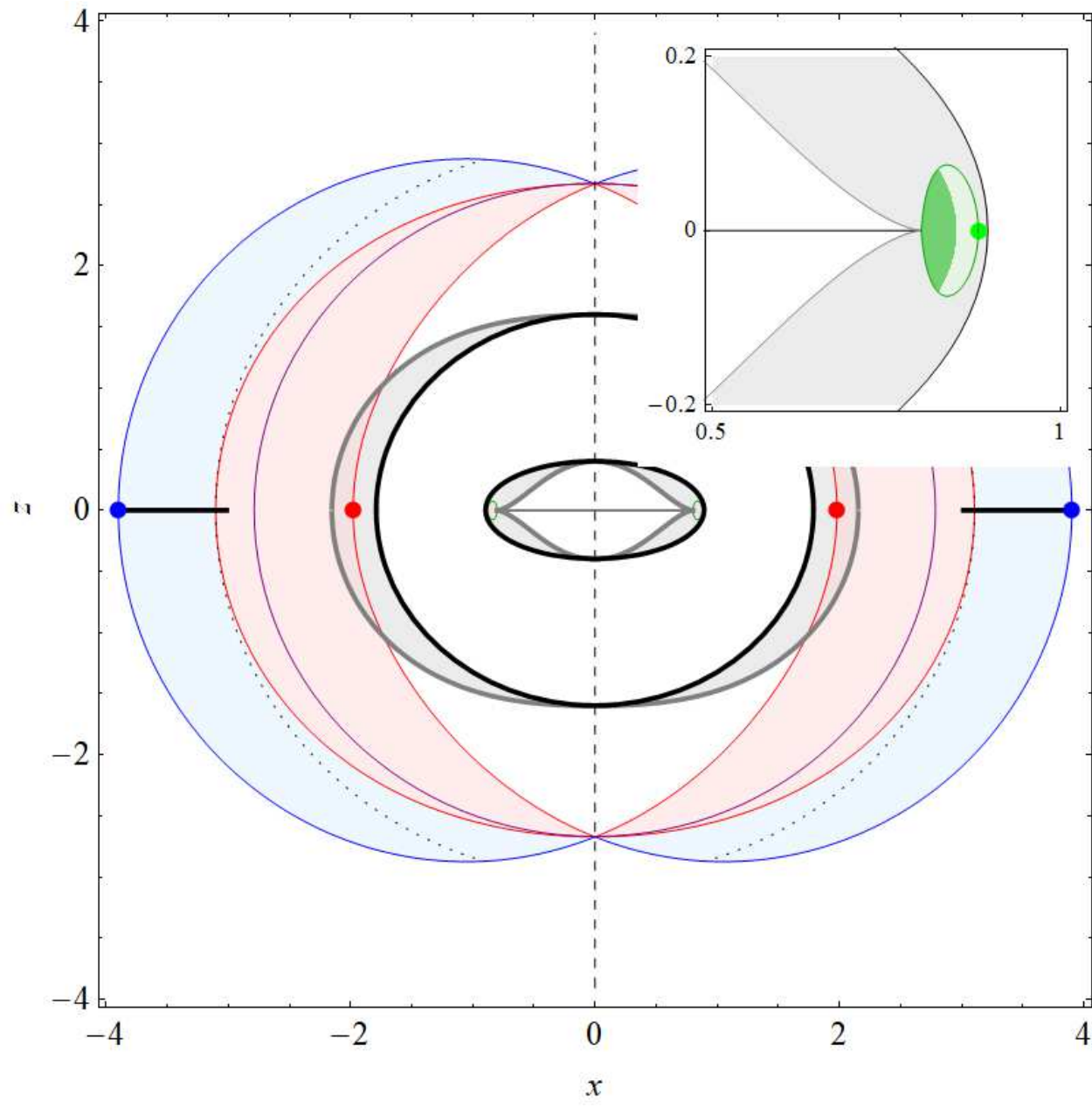}
		\caption{Structure of the SPOs in the KBH spacetime with $a=0.8$ corresponding to the Class I. Colouring is as in Fig. \ref{Figure 1ref} - in green it is depicted the region of the spherical orbits with negative energy $\cale$, in red the region of photon prograde motion in the azimuthal direction, and in blue the region of the retrograde azimuthal motion. Note that not all the orbits are purely prograde, nor purely retrograde, but can have a turning point of the azimuthal motion, hence the colouring of a particular curve representing some fixed Boyer-Lindquist radius $r$ can be change. Full/light hues correspond to stable/unstable orbits. Grey shading demarcates the area where $g_{tt}\geq 0$ (ergosphere), grey curve is its boundary (ergosurface). Purple curve is the polar orbit. Loci of turning points of the azimuthal motion are designated by red curves, loci of turning points of the latitudinal motion are distinguished by blue curves. In all figures some outstanding radii are highlighted, namely $r=1$ (full black ellipse), $r=3$ (black dotted incomplete ellipse). The ring singularity is depicted by the horizontal abscissa, the spin axis by the vertical dashed line. The bold dots represent the photon equatorial circular orbits, namely, the co-rotating orbits with negative energy $\cale<0$ (green), the co-rotating orbits with $\cale>0$ at $r=r_{ph+}$ (red), and the counter-rotating orbits at $r=r_{ph-}$ (blue). In addition, we included images of a possible Keplerian accretion discs, represented by the first family of the equatorial circular orbits of the test particles \cite{Stu:1980:BULAI:} (bold black horizontal abscissas). Their inner edge is the marginally stable circular orbit.}\label{Class_I}
	\end{figure}
		
	\item \textbf{Class II}: Extreme KBH with $a=1$ (Fig. \ref{Class_II}). A family of stable prograde spherical orbits with negative energy occurs at radii $0<r<1$ near equatorial plane \footnote{From the perspective of the locally non-rotating observer such photons appear to be retrograde, see \cite{Char-Stu:2017:EPJC:}.}, where also non-spherical bound photon orbits with two turning points of the radial motion exist. Such orbits are not present in any kind of the most general case of the Kerr-Newmann-(anti) de~Sitter black hole spacetimes. At radii $1 < r < r_{ph-}=4$, there are orbits with positive energy and properties similar to the second family orbits in the previous KBH case. The special case of $r=1$ apparently corresponds to the SPO with zero energy, which is marginally stable and prograde. This zero energy orbit has minimum latitude $\theta_{min(z\cale)}(a=1)=47.1^\circ$; note that for $a>1$, there is $\theta_{min(z\cale)}(a)=\arccos{(1/a)}$. The SPOs at radii $0<r\leq1$ have the same properties as those that occur in all KNS spacetimes, and we shall not repeat them in the following cases. Similarly, for all KNS spacetimes the orbits at $r>1$ have positive energy.
	
	\begin{figure}[h]
		\includegraphics[width=0.45\textwidth]{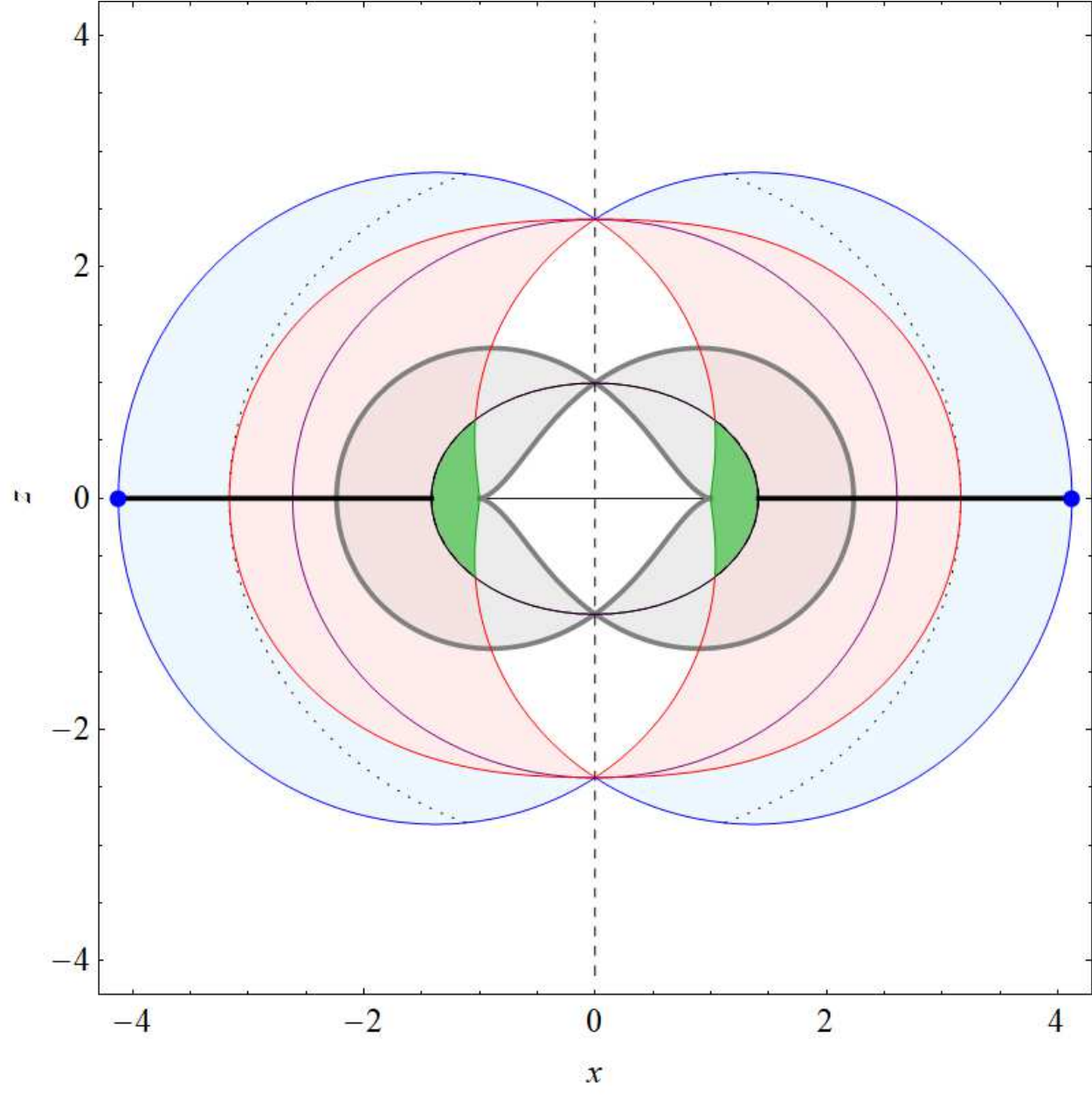}
		\caption{Structure of the SPOs in the extreme Kerr BH $a=1$.}\label{Class_II}
	\end{figure}
	
	\item \textbf{Class III}: KNS spacetimes with $1<a<a_{\Delta \phi=0(min)}=1.17986$ (see Fig. \ref{Class_III}). Two polar SPOs appear at radii $r_{pol+}, r_{pol-}$ (inner and outer purple ellipse, respectively) given by Eq. (\ref{rpol12}). Photons at $1<r<r_{pol+}$ are stable, they have $\cale>0$ and one turning point of the azimuthal motion in each hemisphere. The latitudinal coordinate where the SPO has the azimuthal turning point is given by Eq. (\ref{thaztp}), the minimum allowed latitude reads $\theta_{min}=\arccos \sqrt{m_{\theta}(r;a)}$. At the region $\theta_{min}\leq \theta<\theta_{\phi}$, the photon motion is in negative $\phi$-direction (deep blue area), while for $\theta_{\phi}\leq \theta\leq \pi/2$ it is in positive $\phi$-direction (deep red area). The break point dividing the globally prograde orbits from the globally retrograde ones is at $r=r_{pol+}$. The motion constants $\ell_{sph}\to -\infty$, $q_{sph} \to \infty$, as $r\to 1$ from the right, and $\ell_{sph}=0$, $q_{sph}=27$ at $r=r_{pol+}$. The orbits at the radii $r_{pol+}<r<r_{pol-}$ are prograde, with no change in the azimuthal direction, for $r_{pol+}<r<r_{ms+}$ being stable (deep red area), for $r_{ms+}\leq r\leq r_{pol-}$ being unstable (light red area). The function $m_{\theta}(r;a)$ has a local minimum at $r=r_{ms+}$, hence, the marginally stable SPO is of the least extent in the latitude (black dashed curve), contrary to the case of the marginally stable SPOs with negative energy at $r_{ms-}$ at the KBH spacetimes, where they have the widest extent (see detail in Fig. \ref{Class_I}). The motion constant $q_{sph}$ of photons on this orbit corresponds to the local minimum $0<q_{sph(min)}<27$ of the function defined in (\ref{qsph}), and the local maximum $0<\ell_{sph(max)}$ of the function defined by (\ref{lsph}). The turning point of the azimuthal motion appears for the SPOs at the radii $r_{pol-}<r<3$, and such SPOs appear to be retrograde as whole. The motion constants for $r=r_{pol-}$ are $q_{sph(min)}<q_{sph}<27$, and $\ell_{sph}=0$. At $r=3$, there is $q_{sph}=27$, corresponding to the local maximum of (\ref{qsph}), i.e., to the photons crossing the equatorial plane with zero velocity component in the $\phi$-direction (\cite{Teo:2003:GenRelGrav:}), and $\ell_{sph}<0$. The SPOs in the region $3<r<r_{ph-}$ are purely retrograde with $\ell_{sph}<0$ and $q_{sph}\to 0$ as $r \to r_{ph-}$.
	
	\begin{figure}[h]
		\includegraphics[width=0.45\textwidth]{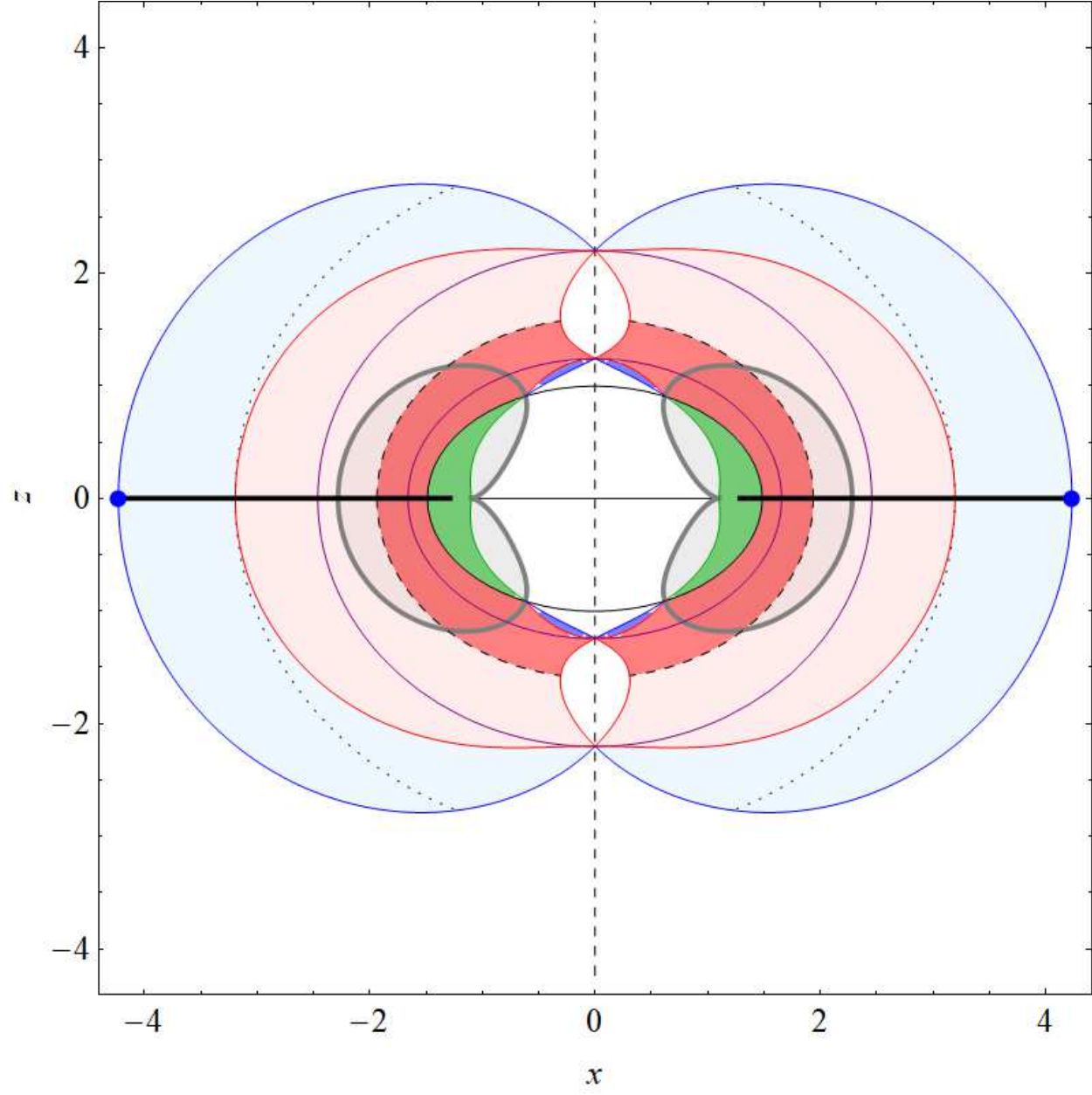}
		\caption{Structure of the SPOs in the KNS spacetime with $a=1.1$ corresponding to the Class III.}\label{Class_III}
	\end{figure}
	         
	\item \textbf{Class IV}: KNS spacetimes with $a_{\Delta\phi=0(min)}\leq a<a_{pol(max)}=1.17996$. In the limit case $a=a_{\Delta\phi=0(min)}$, there exist radius $r=r_{\Delta\phi=0(max)}=1.7147$ of spherical orbit with zero nodal shift $\Delta \phi=0$, infinitesimally distant from the $4\pi$-discontinuity point $r_{pol+}$ (see Fig. \ref{shift_nodes}d). The radius $r_{pol+}$ therefore corresponds to special case of polar oscillatory orbit. For $a_{\Delta\phi=0(min)}<a<a_{pol(max)}$, the radius which separates the globally prograde SPOs from the globally retrograde ones is at $r=r_{\Delta\phi=0}\lessapprox r_{\Delta\phi=0(max)}<r_{pol+}$ (see the detail of Fig. \ref{Figure 1ref}). For $r_{\Delta\phi=0}<r<r_{pol+}$ there appear globally retrograde orbits. The other properties remain the same as in previous case and the SPOs structure is represented by the Fig. \ref{Class_III}.
	
	\item \textbf{Class V}: KNS spacetimes with $1.17996=a_{pol(max)}\leq a<3.$ For $a=a_{pol(max)}$, the two polar SPOs coalesce at $r=r_{pol}=\sqrt{3}$ (see Fig. \ref{Class_V} above). 
	
	\begin{figure}[h]
		\includegraphics[width=0.45\textwidth]{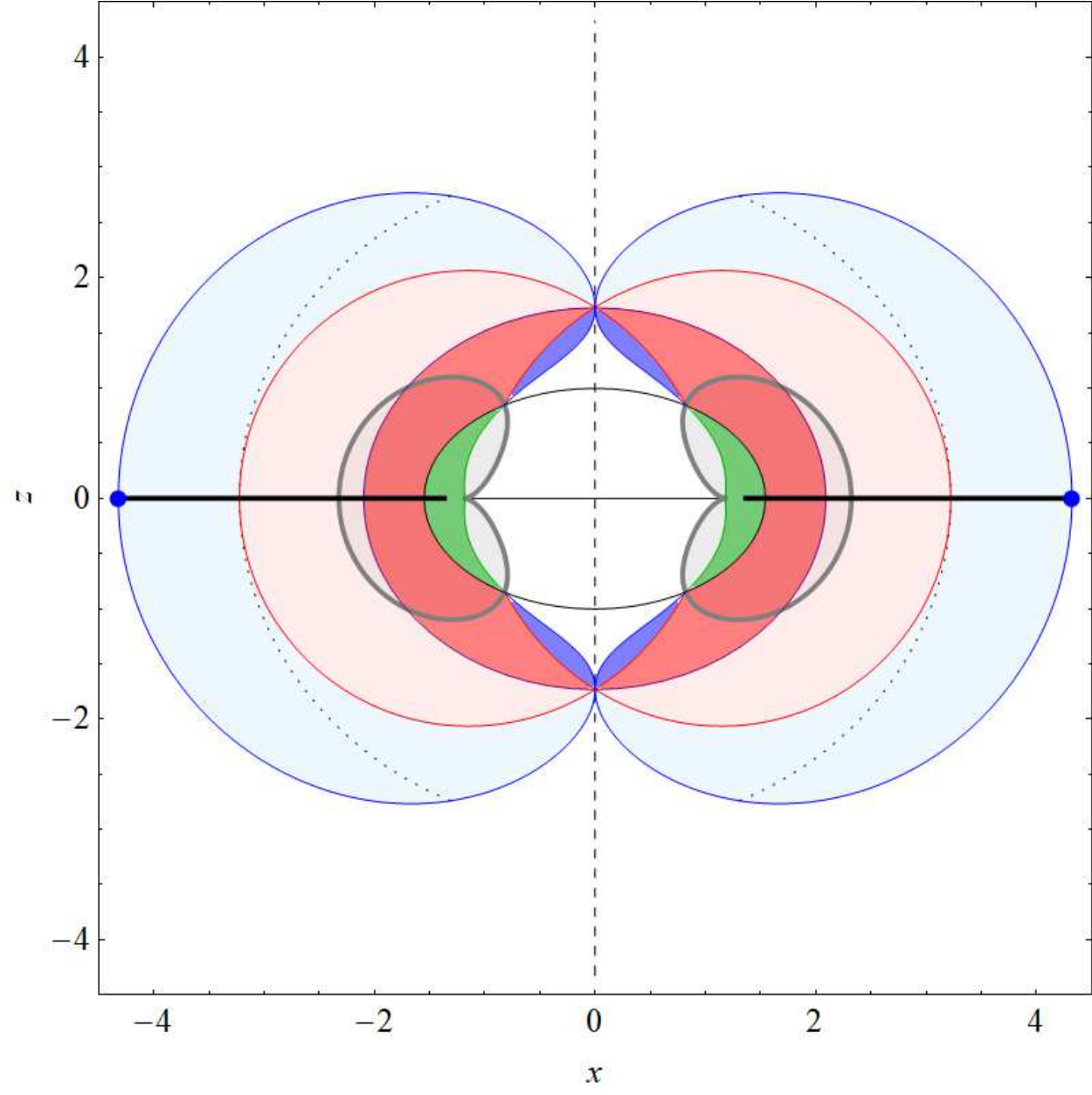}
		\includegraphics[width=0.45\textwidth]{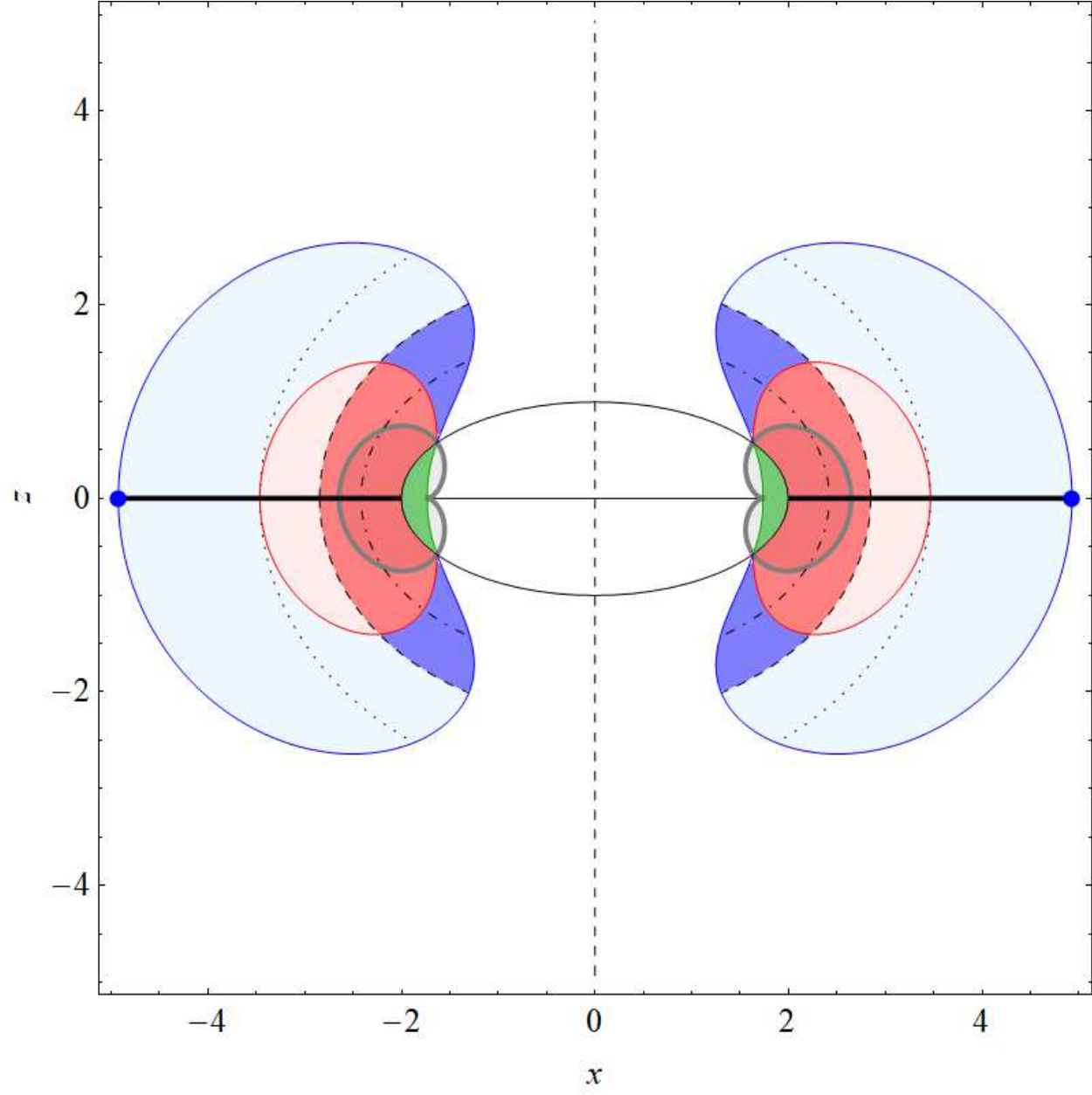}
		\caption{Structure of the SPOs in the KNS spacetime with $a=a_{pol(max)}$ (above) and  with $a=1.7$ (bellow) corresponding to the Class V.}\label{Class_V}
	\end{figure}
	For KNS spacetimes with $a>a_{pol(max)}$, there are no spherical polar orbits. As a consequence, no purely prograde spherical orbits, neither stable nor unstable, are possible. The radius which separates the globally prograde spherical orbits from the globally retrograde ones is at $r<r_{\Delta\phi=0(max)}$ (black dot-dashed curve in Fig. \ref{Class_V} bellow) and it slowly decreases as $a\to \infty$ (see Fig. \ref{Fig_rzs_a}). In addition to this, the discussion is qualitatively same as in the previous case.
		
	\item \textbf{Class VI}: Kerr naked singularity spacetimes with $a\geq 3.$ For $a=3$ the local extrema of the function $q_{sph}(r;a)$ coalesce at the inflex point at $r=3=r_{ms+}$ with $q_{inf}=27$, where it becomes local minimum $q_{sph,min}=27$ for $a>3$. The local maximum is then at $r_{ms+}>3$ (see Fig.\ref{Fig_qsph_lsph}), which is also locus of the local maximum of latitudinal turning function $m_{\theta}(r;a)$ (Fig.\ref{Fig_lat_az}f) at the radius of the marginally stable spherical orbit. The orbits at $3<r<r_{ms+}$ are stable and purely retrograde; in the range $r_{ms+}<r<r_{ph-}$, there are unstable retrograde orbits (see Fig.\ref{Class_VI}).
	
	\begin{figure}[h]
		
		\includegraphics[width=7.5cm]{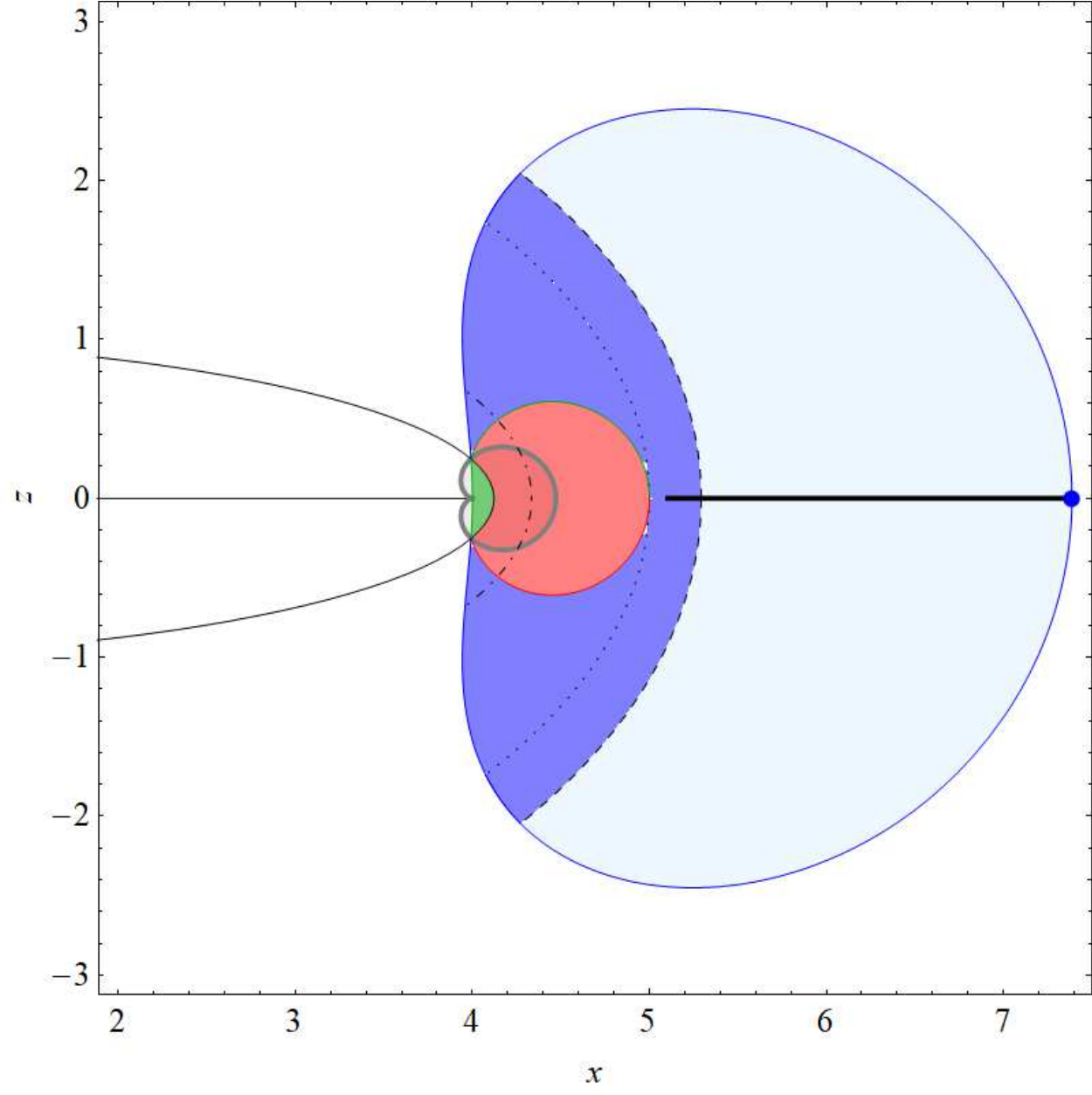}
		\caption{Enlargement of part of structure of the SPOs in the KNS spacetime with  $a=4$ corresponding to the Class VI. The spin axis is outside of the drawing. Note the reversed order of orbits $r=3$ (black dotted) and $r=r_{ms+}$ (black dashed).}\label{Class_VI}
	\end{figure} 
	 
\end{description}

We present a systematic construction of the spherical photon trajectories for the outstanding radii and other appropriately chosen representative radii for the KNS spacetime of the Class III with dimensionless spin parameter $a=1.1$ in Fig. \ref{phot_paths}. Basic characteristics of these orbits, i. e., their radii $r$, impact parameters $\ell, q$, total nodal shift $\Delta_{\phi}$, minimum attained latitude $\theta_{min}$, latitude of change in the $\phi$-direction $\theta_{\phi}$, sign of energy $\cale$ and type, are presented in Tab. \ref{tab_charact}. 

\begin{figure*}[h]
	\flushleft  \underline{\textbf{Class III: $a=1.1$}}\\
\centering
	\begin{tabular}[t]{|c|c|}
	
		\hline
		\includegraphics[width=7cm]{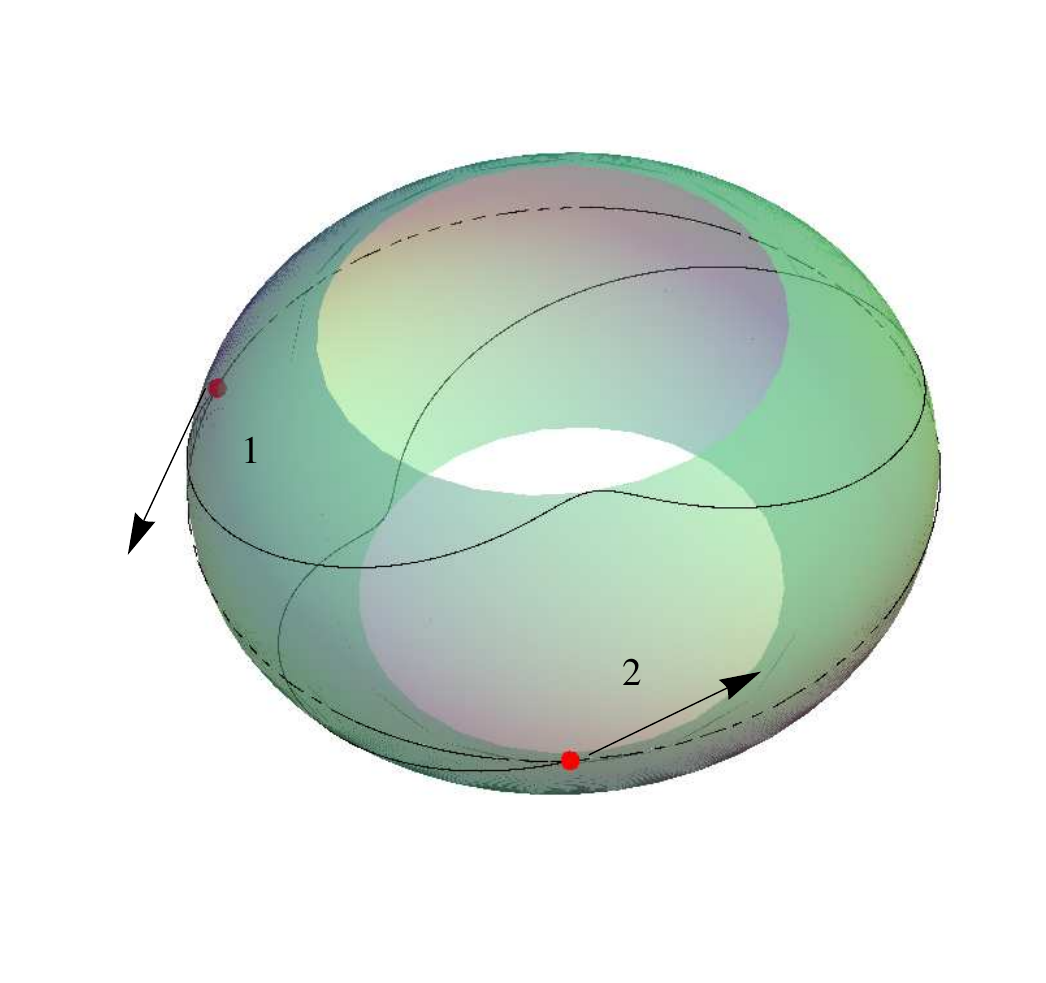}&\includegraphics[width=7cm]{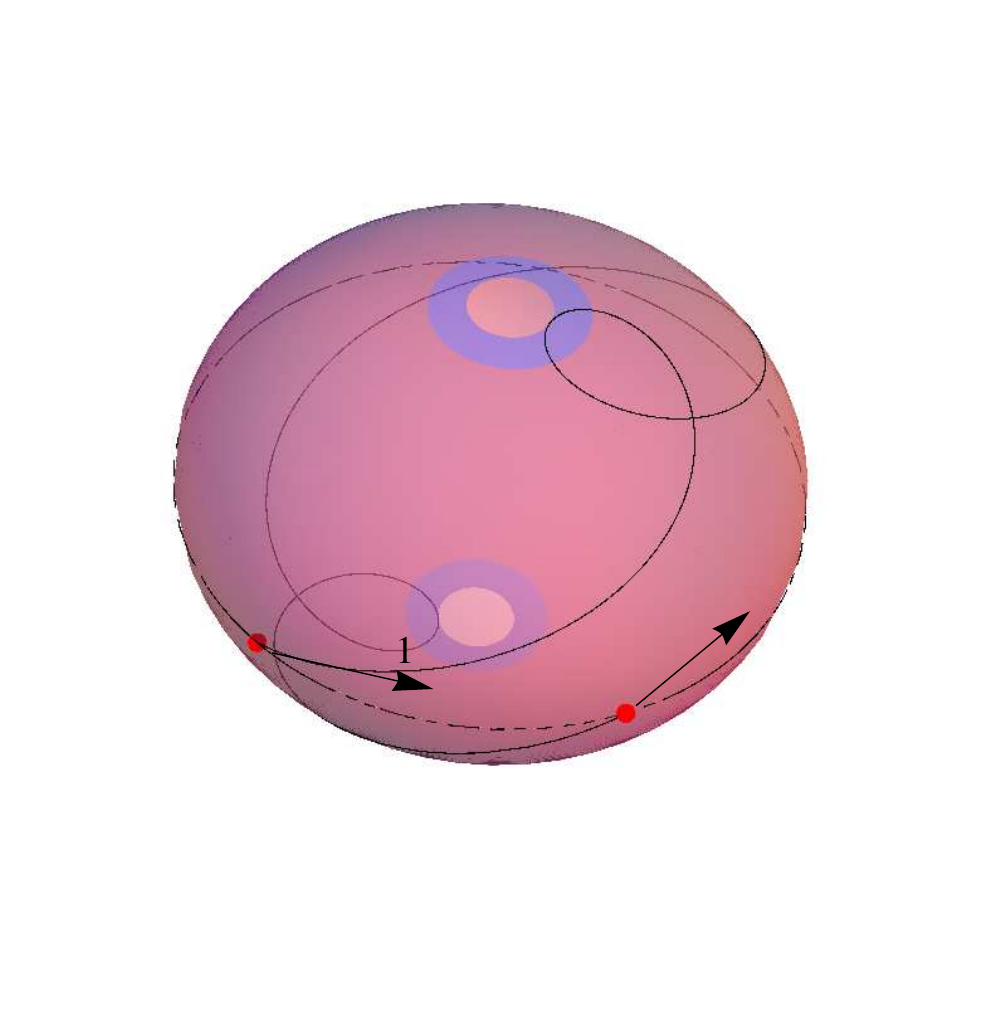}\\
		$r=0.9$, one latitudinal oscillation&$r=1.15$, one latitudinal oscillation\\
		\hline
		\includegraphics[width=7cm]{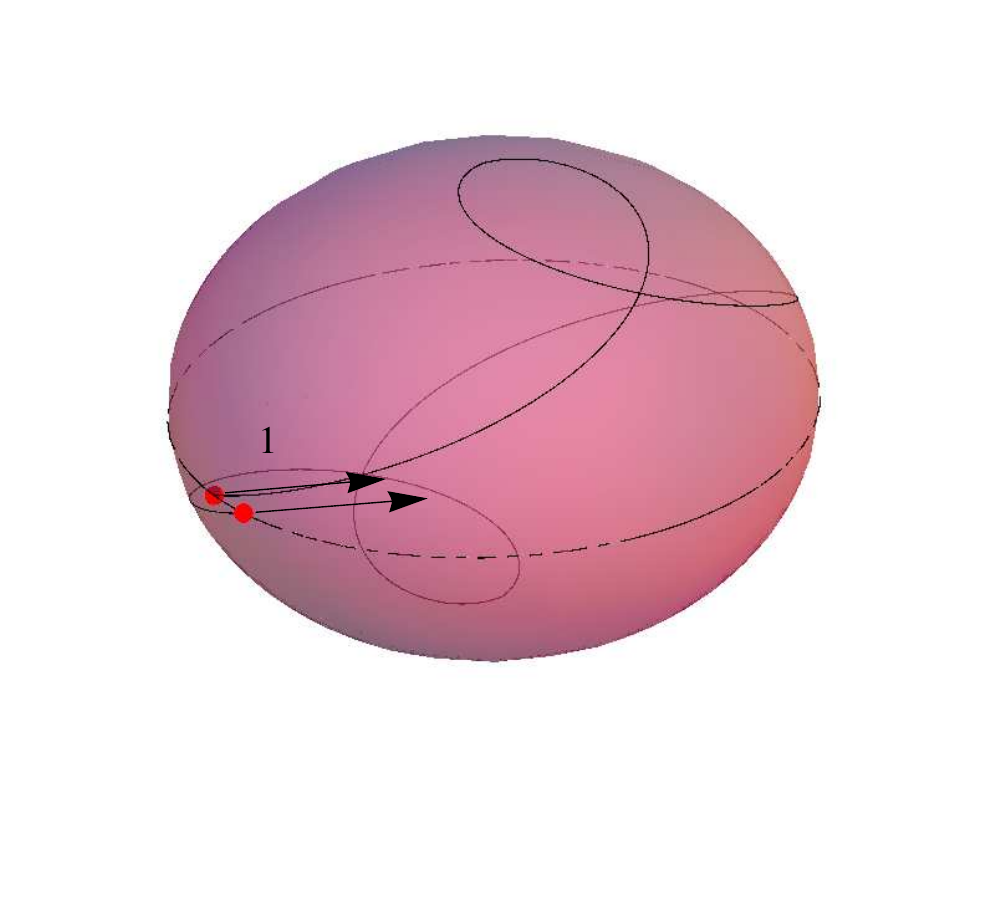}&\includegraphics[width=7cm]{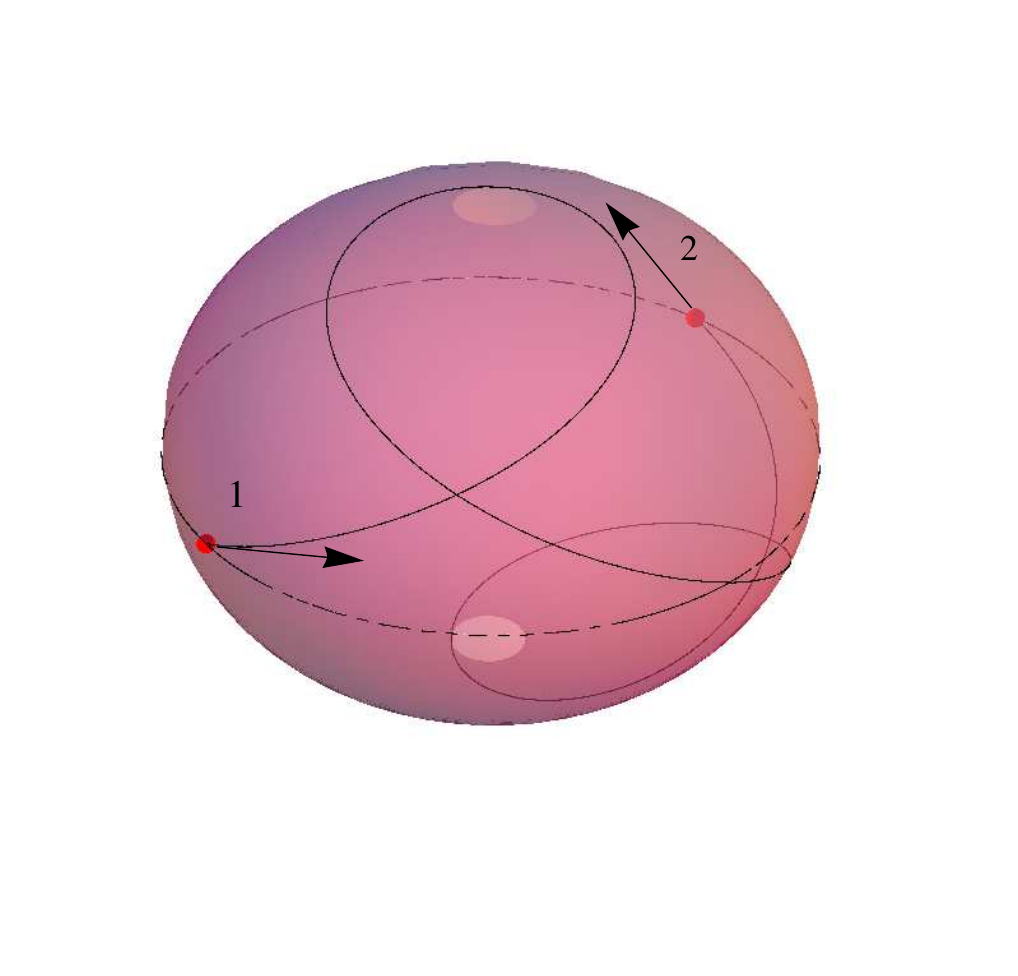}\\
		$r=r_{pol+}=1.24261$, one latitudinal oscillation&$r=1.4$, one latitudinal oscillation\\
		\hline
		\includegraphics[width=7cm]{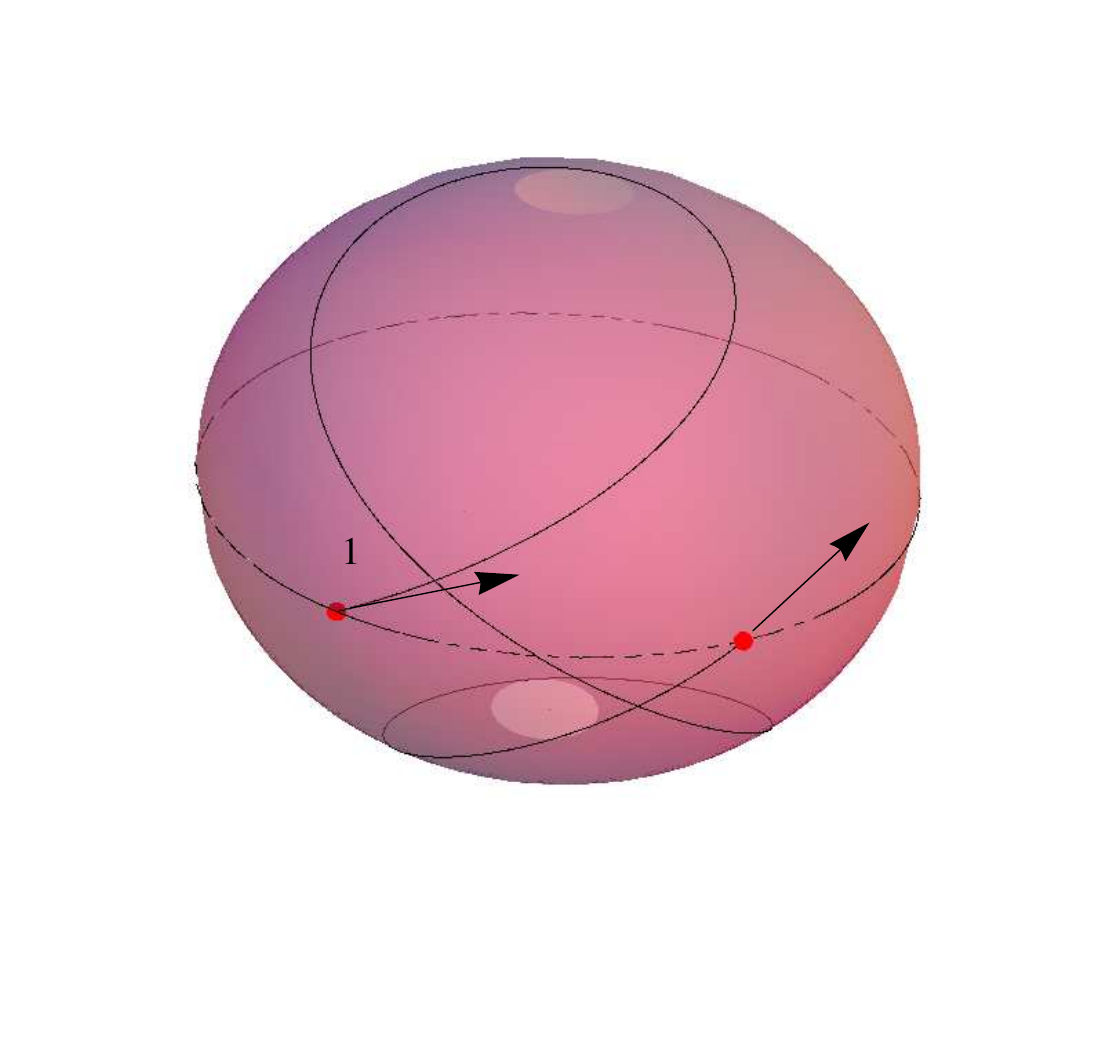}&\includegraphics[width=7cm]{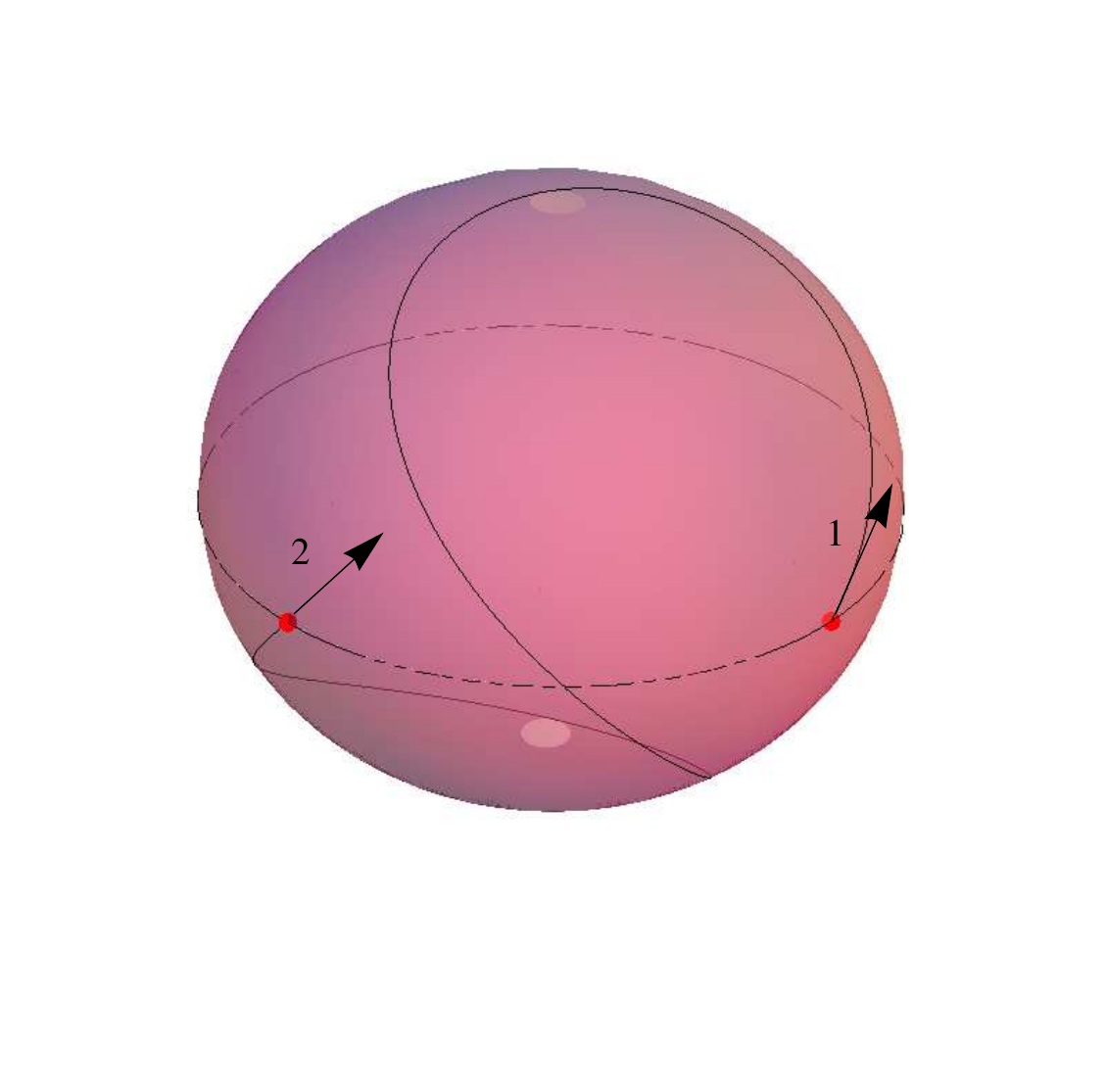}\\
		$r=r_{ms}=1.54439$, one latitudinal oscillation&$r=2$, one latitudinal oscillation\\
		\hline
		\end{tabular}
		\center (\textit{Figure continued})
\end{figure*}

\begin{figure*}[h]
	\begin{tabular}[t]{|c|c|}
		\hline	
		\includegraphics[width=7cm]{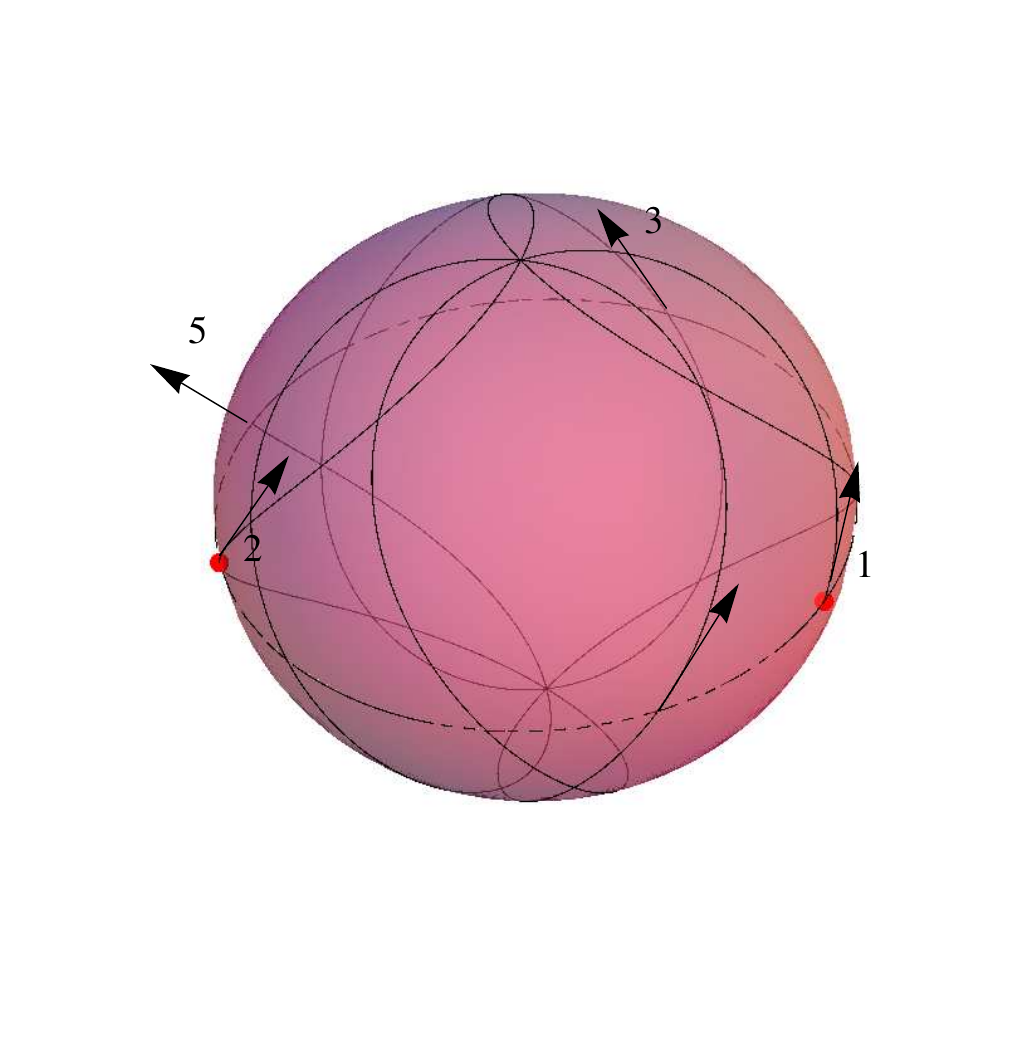}&\includegraphics[width=7cm]{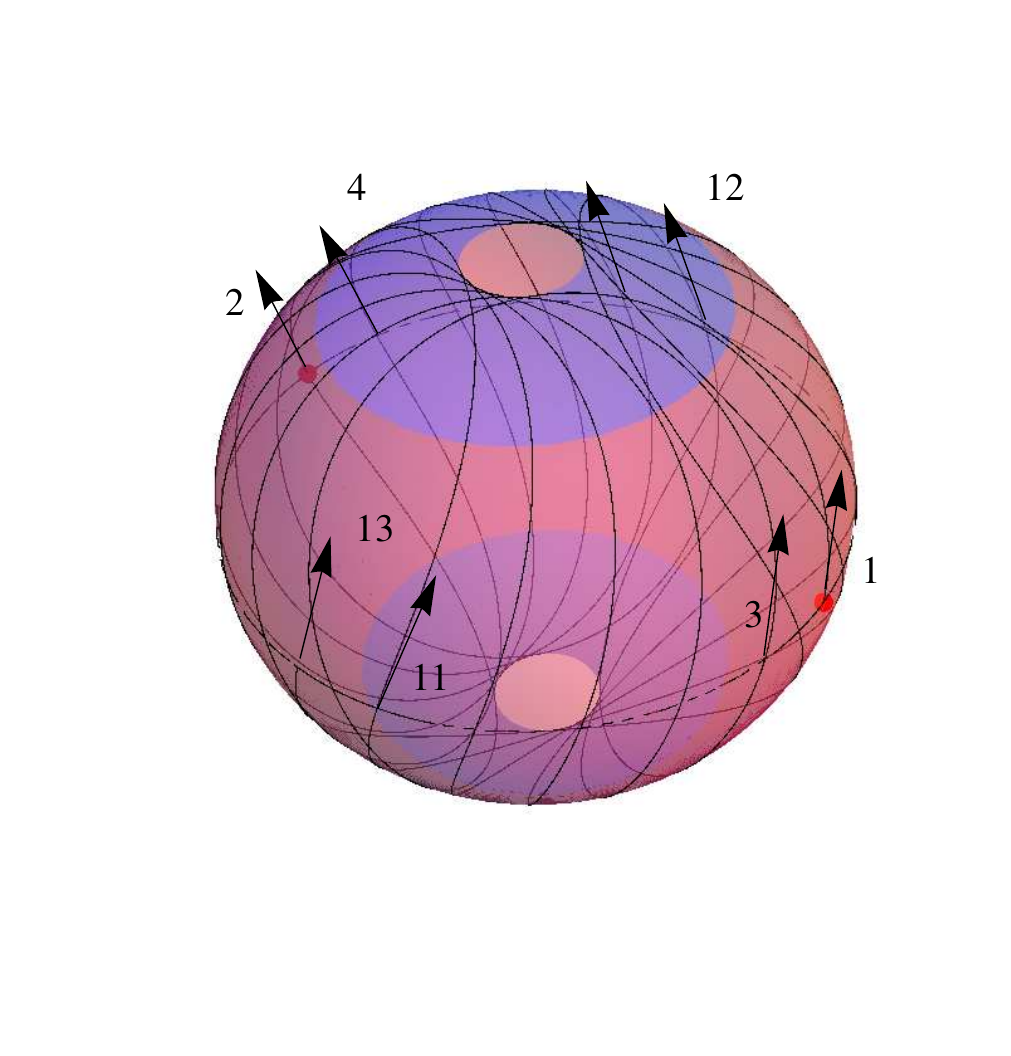}\\
		$r=r_{pol-}=2.2$, four latitudinal oscillations&$r=2.6$, twelve latitudinal oscillations\\
		\hline
		\includegraphics[width=7cm]{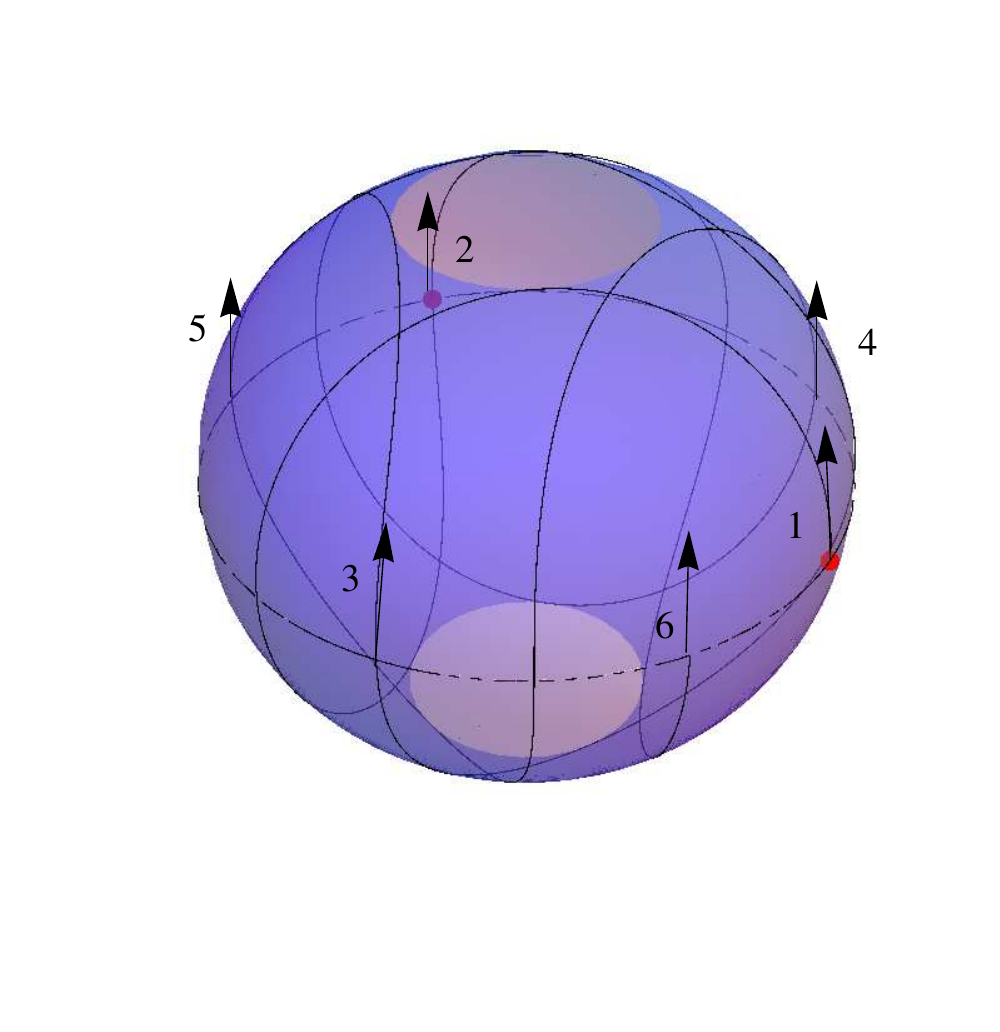}&\includegraphics[width=7cm]{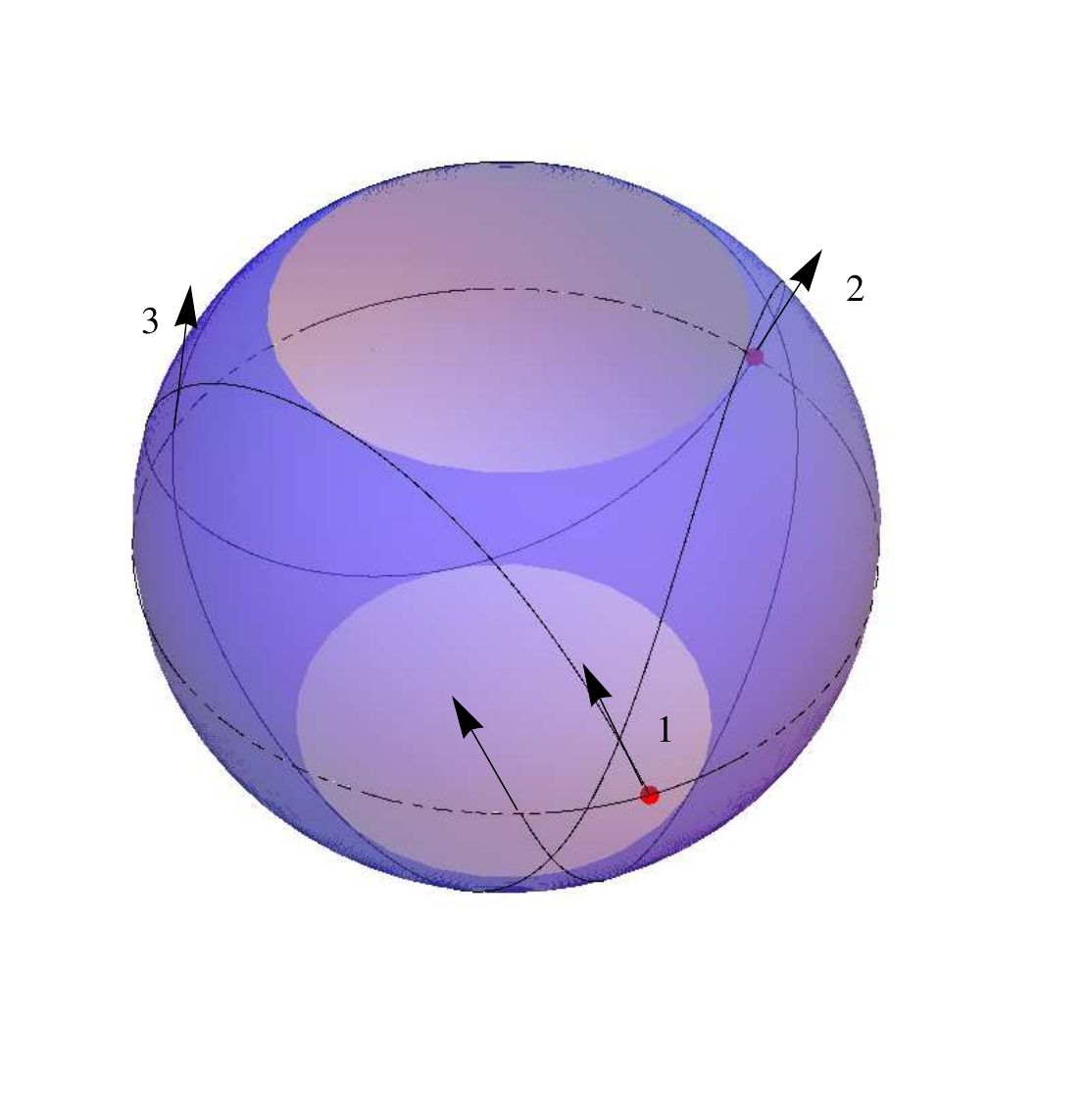}\\
		 $r=3$, five latitudinal oscillations&$r=3.4$, three latitudinal oscillation3\\ 
		 \hline
	\end{tabular}
	\caption{Various types of the SPOs in the KNS spacetime with $a=1.1.$  Stability/instability of the SPOs and their other characteristics are presented in the attached Table \ref{tab_charact}.  }\label{phot_paths}
\end{figure*}

\begin{table*}[h]
	\caption{Characteristics of the spherical photon orbits in Fig.\ref{phot_paths} .}\label{tab_charact}
	\begin{tabularx}{\textwidth}{XXXXXXXX}
	\hline
	$r$&$\ell$&$q$&$\Delta_{\phi}$&$\theta_{min}$&$\theta_{\phi}$&$\sign \cale$&type\\
	\hline
	\hline
	$0.9$&$5.4$&$52.5$&$469^\circ$&$36.7^\circ$&-&-1&stable\\
	$1.15$&$-0.9$&$50.5$&$426^\circ$&$7.4^\circ$&$14.1^\circ$&+1&stable\\
	$r_{pol+}=1.24$&$0$&$27$&$728^\circ$&$0^\circ$&$0^\circ$&+1&stable,in.polar\\
	$1.4$&$0.5$&$17.8$&$912^\circ$&$6.9^\circ$&-&+1&stable\\
	$r_{ms}=1.54$&$0.66=\ell_{max}$&$16=q_{min}$&$781^\circ$&$9.1^\circ$&-&+1&marg.stable\\
	$2$&$0.33$&$18.8$&$630^\circ$&$4.3^\circ$&-&+1&unstable\\
	$r_{pol-}=2.2$&$0$&$21$&$227^\circ$&$0^\circ$&$0^\circ$&$+1$&unst.,out.polar\\
	$2.6$&$-0.9$&$25$&$-189^\circ$&$10.4^\circ$&$38.0^\circ$&$+1$&unstable\\
	$3$&$-2.2=-2a$&$27=q_{max}$&$-223^\circ$&$22.6^\circ$&$90^\circ$&$+1$&unstable\\
	$3.4$&$-3.8$&$24$&$-246^\circ$&$37.0^\circ$&-&$+1$&unstable\\
	\hline

	\end{tabularx}	
\end{table*}

\clearpage

\section{Spherical photon orbits related to the Keplerian disks and possible observational consequences}

It is well known that in the KBH spacetimes the SPOs define the light escape cones in the position of an emission, i.e., they represent a boundary between the photons captured by the black hole and the photons escaping to infinity. In case of the KNS spacetimes, there is in addition a possibility of existence of trapped photons, which remain imprisoned in the vicinity of the ring singularity \cite{2005ragt.meet..143S, Stu-Sche:2010:CLAQG:}. Such a "trapping" region spreads in the neighbourhood of the stable SPOs, where small radial perturbations cause that the photons oscillate in radial direction between some pericentre and apocentre, but remain trapped in the gravitational field -- efficiency of the trapping process was for the Kerr superspinars (naked singularity spacetimes) studied in detail \cite{Stu-Sche:2010:CLAQG:} where also possible self-irradiation of accreting matter was briefly discussed. \footnote{Recall that even general self-irradiation (occulation) of an accretion disk orbiting a black hole could have significant influence on the optical phenomena related to the accretion disks, as demonstrated for the first time in \cite{Bao-Stu:1992:ApJ:}.} 
Note that the self-irradiation of the disk is possible also due to the unstable SPOs, but they could be send away from the sphere (to infinity) due to any small perturbative influence -- for this reason we focus our attention on the influence of the stable spherical photons on the Keplerian disk in a special and observationally interesting case of the oscillatory photon orbits that return to a fixed azimuthal position, and the self-irradiation thus occurs repeatedly at the fixed position relative to distant observers.

\subsection{Connection of the spherical photon orbits and the stable circular geodesics}

The existence of the trapped photon orbits motivates us to examine the spread of the SPOs in relation to a possible distribution of radiating matter related to accretion disks. Of special interest are the stable equatorial circular orbits of test particles, which are assumed to be governing the structure of the thin accretion (Keplerian) disks. Particularly important are the marginally stable circular orbits, representing the inner boundary of the Keplerian disks -- if a mass element of a Keplerian disk reaches the marginally stable orbit after loss of the energy and angular momentum due to viscous friction, any additional loss of its energy causes its direct free fall onto the naked singularity (black hole). The thorough discussion of the equatorial circular orbits in the field of the Kerr naked singularities was done in \cite{Stu:1980:BULAI:}. Since the SPOs can exist only at radii $r<r_{ph-}$, we can restrict our attention to the equatorial circular orbits of the so called first family, which are co-rotating at large distance from the KNS, but they could become counter-rotating at close vicinity of the ring singularity; the unstable orbits of the first family could extend down to the ring singularity \cite{Stu:1980:BULAI:}. On the contrary, the circular orbits of the second family, which are all counter-rotating, are located at $r>r_{ph-}$ and hence are irrelevant for the study of the interaction of the SPOs with the Keplerian disks. The marginally stable circular orbits of the first family are determined by \cite{1972ApJ...178..347B, Stu:1980:BULAI:}
\be
r_{msc}=3+Z_{2}-\sqrt{(3-Z_{1})(3+Z_{1}+2Z_{2})},\label{rmsc}
\ee     
where
$$Z_{1}=1+\sqrt[3]{1-a^2}(\sqrt[3]{1+a}+\sqrt[3]{1-a}),$$ $$Z_{2}=\sqrt{3a^2+Z^2_{1}}.$$ We introduce the subscript 'msc' in order to distinguish the marginally stable SPOs. 
In the case of thick accretion disks governed by axi-symmetric toroidal structures of perfect fluid, the radiation matter could reach the radius of the marginally bound orbit at \cite{1972ApJ...178..347B, Stu:1980:BULAI:} 
\be
r_{mb}=2 + a +2\sqrt{1 + a}.
\ee
In the present paper, we focus our attention on the Keplerian disks, because the case of toroidal accretion configurations requires more detailed study of the relation of the SPOs and the orbiting matter, as in some Kerr spacetimes the existence of the SPOs could be excluded by the presence of the accretion torus. 

The stable circular orbits of the first family are located at $r>r_{msc}$. It is relevant from the astrophysical point of view that these stable circular orbits interfere with the area of the stable SPOs at $r<r_{ms}$ for KNS spacetimes with rotation parameter $1\leq a \leq a_{i}=4.468$, since just in these spacetimes there is $r_{ms}>r_{msc}$. The limits $a=1, a_{i}$ correspond to equality $r_{ms}=r_{msc}=1$ for $a=1$, and $r_{ms}=r_{msc}=3.667$ for $a=a_{i}$. Moreover, due to the results of \cite{Stu:1980:BULAI:}, it follows that even the circular orbits with negative energy with respect to infinity ($\cale<0$) lie at the region of stable spherical orbits. Their radii satisfy inequality $r_{zE1}<r<r_{zE2}$,  where 
\bea
r_{zE1}&=&\frac{8}{3}\cos^2(\frac{\pi}{3}+\frac{1}{3}\arccos \sqrt{\frac{27}{32}}a),\\
r_{zE2}&=&\frac{8}{3}\cos^2(\frac{\pi}{3}-\frac{1}{3}\arccos \sqrt{\frac{27}{32}}a)\label{rze}
\eea
are radii of the zero energy orbits.

We can expect observationally significant optical and astrophysical effects, particularly in the region of negative energy orbits connected with an interplay between the possible extraction of the rotational energy of the KNS (Kerr superspinar) and a subsequent trapping of the radiated energy, both for the radiated electromagnetic and gravitational waves. Both forms could substantially violate the structure and stability of the accretion disk, generating thus an observationally relevant feedback, which we presume to be the subject of our further study. Of course, similar effects can be expected in the whole overlap region of the stable circular orbits of matter and the stable SPOs, where repeated re-absorption of the radiated heat, and self-illumination and self-reflection phenomena take place, as indicated in the preliminary study of the Kerr superspinars in \cite{Stu-Sche:2010:CLAQG:}. The illustration of the overlap region of the stable circular orbits with the stable SPOs, and the other relevant orbits in dependence on the Kerr spacetime rotation parameter $a$ is given in Fig. \ref{radfunc}. 

Now we concentrate attention on the special case of the repeated irradiation of a Keplerian disk at an azimuthal position fixed to the distant observers that could be observationally relevant, enabling even an efficient determination of spacetime parameters of the KNS (Kerr superspinar) spacetimes allowing existence of this effect.

\begin{widetext}
	\onecolumngrid
	
	\begin{figure*}[h]
		\centering
		\begin{tabular}{c}
			\includegraphics[width=\textwidth]{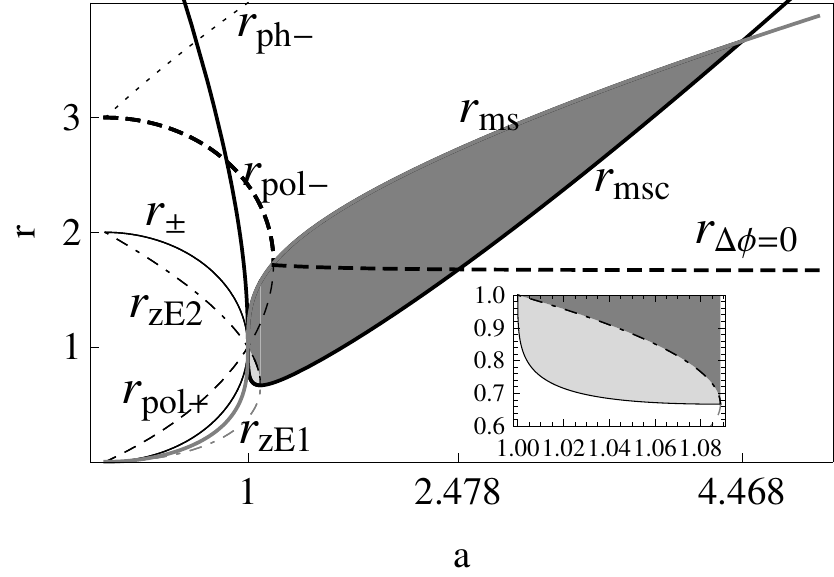}
		\end{tabular}	 
		\caption{Relative position of the equatorial circular orbits of the test particles and the spherical photon orbits. The overlap region of the stable circular orbits of test particles and stable photon spherical orbits is highlighted in grey, the light grey shading correspond to radii of the circular orbits with negative energy. The function $r_{\Delta \phi =0}$ (bold dashed curve) corresponds to radii of the closed spherical photon orbits with zero total change in azimuth $\Delta \phi =0.$ It interferes with the area of possible occurrence of the thin Keplerian accretion disc, hence, some self-illumination effects manifesting itself as periodic light echo are plausible here. The remaining functions are defined in the text.     }\label{radfunc}
	\end{figure*} 
	\twocolumngrid
\end{widetext} 

\subsection{Time periods of the latitudinal oscillations and time sequences related to the oscillatory orbits}

In order to have a clear observational signature of the effects of the SPOs, we have to calculate dependence of the time period of the nodal motion, i.e., the time interval measured by a distant static observer between the two subsequent crossing of the equatorial plane by the photons following the spherical orbits. We first give a general formula and then we discuss the case of the closed spherical orbits finishing their azimuthal motion at the starting point in the equatorial plane. Such orbits could be observationally relevant, as the time interval of the nodal motion could give exact information on the Kerr naked singularity (black hole) spin, if its mass is determined by an independent method. 

The time interval that elapses during one latitudinal oscillation can be computed by combining equations (\ref{Cart_M}) and (\ref{Cart_t}). The searched period of the latitudinal motion at a fixed radius $r$ can be expressed by the integral
\be
\Delta t = 4\int \limits_0^{m_{+}} \frac{\Sigma (r^2+a^2)-2ar[\ell_{sph}-a(1-m)]}{2\Delta \sqrt{M(m,a,\ell_{sph},q_{sph})}}, \label{tau_int}
\ee
where $\ell_{sph}$ and $q_{sph}$ are taken at the fixed radius $r$. Of course, the nodal period, i.e. the period between two subsequent crossing of the equatorial plane, is half of the time interval given by Eq.(\ref{tau_int}). The nodal period can be expressed, using the standard procedures (see Gradshtein, S., Ryzhik,M., Tables of integrals, series and products), in terms of the elliptic integrals by the formula
\bea
\Delta t_{nod} &=& 2\frac{r^2(r^2+a^2)-2ar(\ell_{sph}-a)}{a \Delta \sqrt{m_{+}-m_{-}}}K(\frac{m_{+}}{m_{+}-m_{-}})\nonumber \\
&+&4a[\frac{m_{-}}{\sqrt{m_{+}-m_{-}}}K(\frac{m_{+}}{m_{+}-m_{-}})\nonumber \\
&+&\sqrt{m_{+}-m_{-}}E(\frac{m_{+}}{m_{+}-m_{-}})].\label{tau}
\eea
Here $m_{\pm}$ are the positive/negative roots of $M(m)$, $K(s)$ denotes the complete elliptic integral of the first kind given by (\ref{K(m)}) and 
\be
E(s)=\int \limits_0^{\pi/2}\sqrt{1-s\sin^2\theta}\din \theta
\ee 
is the complete elliptic integral of the second kind.

\begin{figure}[h]
	\centering
	\includegraphics[width=8cm]{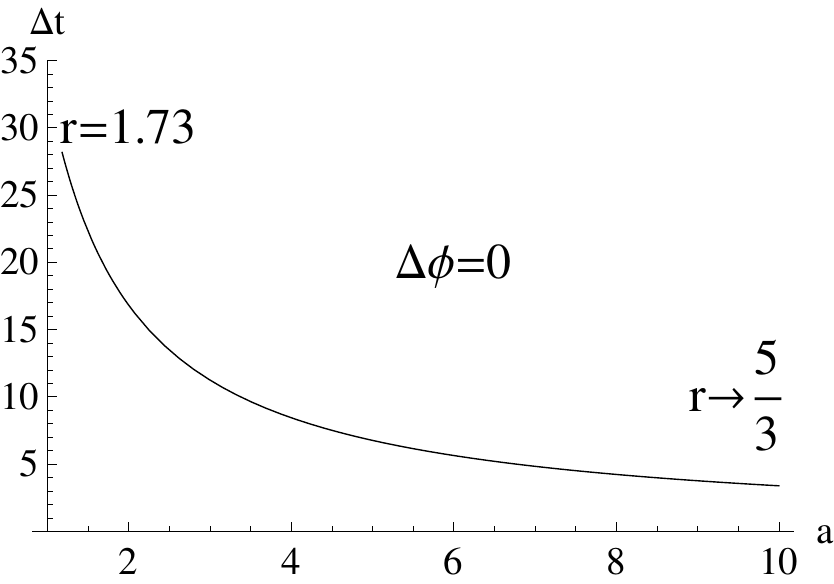}
	\caption{Dependence of the period of the photon circulation over a closed spherical orbit with zero nodal shift on the spin parameter $a.$ The graph is defined for $a>1.18.$ Such orbits are of octal shape (see case $k=0$ in Fig. \ref{clos_orb}).} \label{fig_del_t_0}
\end{figure}

Using the general formula for the time period of the nodal motion, we are able to calculate the time period of the special nodal motion of closed oscillatory orbits with vanishing change of the azimuthal coordinate for one node. We give the time periods $\Delta t_{nod=0}$ in dependence on the dimensionless spin parameter $a$ of the spacetime in Fig. \ref{fig_del_t_0}.

Notice that there is relatively strong decline of the time period $\Delta t_{nod=0}$ with increase of the spin parameter $a$, although the dependence of the radius of these orbits, $r_{\Delta\phi=0}$, on the spin $a$ demonstrates only very slow descent, as shown in Figs. \ref{Fig_rzs_a}, \ref{radfunc}. Numerical computation reveals that
\be
\lim_{a\rightarrow \infty}r_{\Delta\phi=0}(a)=5/3. \label{lim_rzs}
\ee 
We have found that the condition $r_{\Delta\phi=0}>r_{msc}$ is satisfied for the KNS spacetimes with dimensionless spin in the interval $a\in (a_{pol(max)}=1.7996, 2.47812)$, therefore, just for this interval of the KNS spacetimes we can consider the effect of the irradiation at a fixed azimuth. Observation of such an effect enables to estimate the spin parameter $a$, if the mass parameter of the spacetime can be fixed by an independent method. \footnote{We could slightly extend the range of the considered spin of the KNS spacetime, down to the value of $a_{\Delta \phi=0(min)}=1.79857$ when the SPOs with $\Delta \phi = 0$ start to appear; however, in the case of KNS spacetimes with $a\in(a_{\Delta \phi=0(min)},a_{pol(max)})$ there is an extended region of spherical orbits allowing for $\Delta \phi \sim 2\pi$ that invalidates applicability of the orbits with $\Delta \phi = 0$.} In this case we have considered only the spherical orbits with positive covariant energy $\cale>0$, as the oscillatory orbits with azimuthal turning point are necessarily located at $r>1$. 

Now we have to distinguish the case of the SPOs demonstrating $\Delta \phi = 0$ with the turning point of the azimuthal motion, and the analogous cases of the periodic SPOs where $\Delta \phi = k2\pi$, with $k$ being an integer, giving also the return to the fixed azimuthal coordinate in the equatorial plane where the Keplerian disk is located. 

\begin{figure}[h]
	\centering
	\includegraphics[width=8cm]{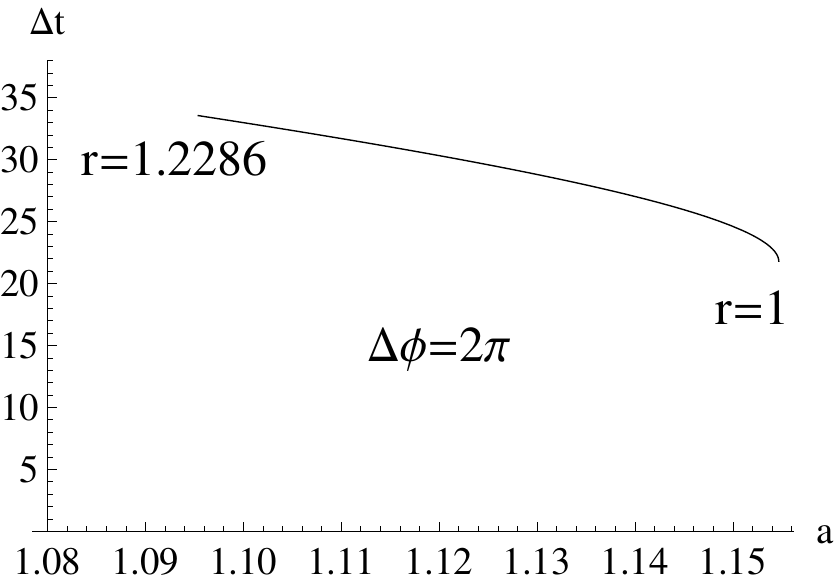}
	\caption{Dependence of time periods of the closed SPOs with $\Delta \phi=2\pi$ on the spin parameter $a$ defined for $a\in (1.095,1.155).$ } \label{fig_del_t_2pi_n}
\end{figure}

The case of the SPOs with $\Delta t_{nod=2\pi}$ is given in Fig. \ref{fig_del_t_2pi_n}. We found by numerical methods that the SPOs with $\cale>0$ and $\Delta \phi=2\pi$ are limited by the radii $r=1.22854$ for $a=1.09533$ and, by solving Eq. \ref{Del_Phi1}, by the radii $r=1$ for $a=\sqrt{4/3}=1.1547.$ A representative of such orbits is depicted in Fig. \ref{clos_orb} (see case $k=1$).

\begin{figure}[h]
	\centering
	\includegraphics[width=8cm]{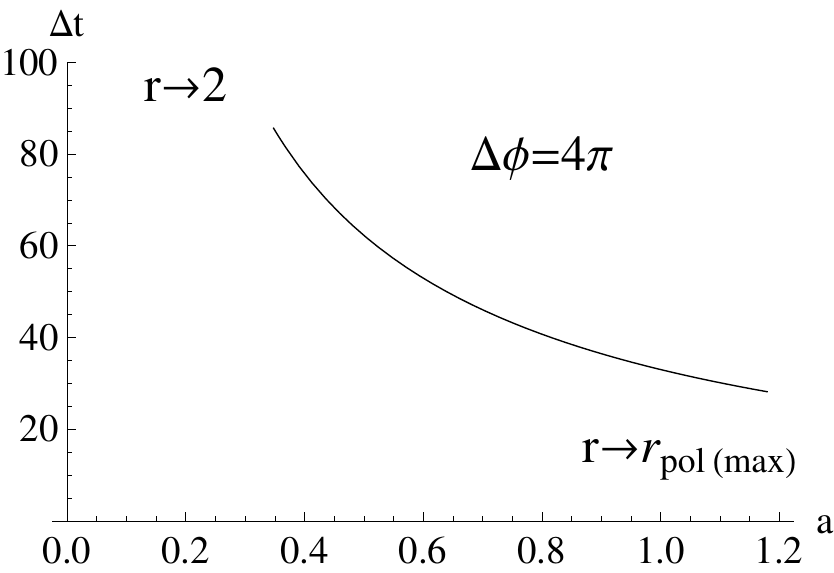}
	\caption{Case $\Delta \phi=4\pi.$ The shape of such orbits represents second picture in Fig. \ref{clos_orb}.} \label{fig_del_t_4pi}
\end{figure}

The case of the orbits with $\cale>0$ demonstrating $\Delta t_{nod=4\pi}$ is illustrated in Fig. \ref{fig_del_t_4pi}. These orbits exist for the spin parameter $0<a\leq a_{\Delta\phi=0(min)}=1.17986$ with $r\to 2$ as $a\to 0$, and $r=r_{\Delta\phi=0(max)}=1.7147$ as $a=a_{\Delta\phi=0(min)}$. This limit case is depicted in Fig. \ref{phot_trace_rzs_azs}a, for the values from the inside of the interval the Fig. \ref{clos_orb}, case $k=2$, is representative. In the figures we depict together with the graphs the radii corresponding to the limits of their definition range.

Now we can consider also the SPOs at $r<1$ (with $\cale<0$), in the KNS spacetimes with spin $1<a<a_{pol(max)}$, since we are not limited by the necessity of having the azimuthal turning point. Of course, for the KNS spacetimes allowing for existence of closed spherical orbits of all three types (or two of them), we can use the time sequences (the ratio of the time delays) corresponding to the relevant types of the closed orbits to obtain relevant restrictions on the allowed values of the dimensionless spin $a$ independently of their mass parameter. For comparison, we also compute the time intervals for the closed spherical orbits in Fig.\ref{clos_orb} in associated table \ref{tab_clos_orbs}. 

To estimate the astrophysical relevance of the time delay effect, let us present the time interval for the circular photon orbit in case of the Schwarzschild black hole: in the limit $a\to 0$ the formula (\ref{tau}) for $r=3$ gives the result $\Delta t_{Schwarz}=32.6484$, in accordance with the exact solution $\Delta t_{Schwarz}=6\sqrt{3}\pi$ obtained by integration from the Schwarzschild line element after inserting $\din s=\din r=\din \theta=0$ and $\theta=\pi/2$. The nodal time delay is governed by half of the quantity: $\Delta t_{nod(Schwarz)}=16.3242$. 

In order to have dimensional estimates governing the time delays, we have to express the resulting time delays in the dimensional form that takes in the standard units the form
\be
       \Delta t_{nod(dim)} =  \frac{GM}{c^3}  \Delta t_{nod} . 
\ee 
Therefore, in the systems with stellar mass compact object (KNS) we can estimate the time delay of the level of $\Delta t_{nod(dim)} \sim 0.01s$, but in the case of the supermassive object in the Galaxy centre we can estimate $\Delta t_{nod(dim)} \sim 10^3s \sim 1/4\,hour$, while in the case of the galaxy M87 the central object can demonstrate time delay by three orders higher $\sim 10^6s \sim 12\,days$. Clearly, from the point of view of the observational abilities of the recent observational techniques, the most convenient candidate for testing the time delay effect seems to be the Galaxy centre SgrA* object. 

\section{Concluding remarks}
We have shown that in the field of the KNS spacetimes there exist a variety of the SPOs, which are not present in the KBH spacetimes. Existence of spherical photon orbits stable relative to the radial perturbations has been demonstrated. The photons at spherical orbits located at $r>1$ are of standard kind, having covariant energy $\cale>0$, as outside the black hole horizon, but at $r<1$ they must have $\cale<0$; for the special position at $r=1$, the photons at spherical orbits have $\cale=0$. \footnote{The photons at the spherical orbits located under the inner black hole horizon have also $\cale<0$.} 

The character of the spherical motion in the field of Kerr naked singularities is more complex in comparison with those of spherical orbits above the outer horizon of Kerr black holes -- along with the orbits purely co-rotating or counter-rotating relative to distant observers, also orbits changing orientation of the azimuthal motion occur in the field of Kerr naked singularities. Existence of polar spherical orbits reaching the symmetry axis of the Kerr spacetime is limited to the naked singularity spacetimes with spin $a<a_{pol(max)}=1.17996$. On the other hand, the KNS spacetimes with $a>a_{pol(max)}=1.17996$ allow for existence of oscillatory orbits with azimuthal turning point that return in the equatorial plane to the fixed original azimuthal coordinate as related to distant observers. 

From the astrophysical point of view it seems that the most interesting and relevant is the existence of the closed spherical photon orbits with zero total change in azimuth, which intersect themselves in the equatorial plane, where the stable circular orbits of massive particles can take place. This effect is potentially of high astrophysical relevance as it enables a relatively precise estimation of the dimensionless spin of the KNS spacetimes allowing their existence, if their mass parameter is known due to other phenomena. This is the case of the KNS spacetimes with the spin parameter $a\in (a_{pol(max)}=1.7996, 2.47812)$ (see Fig. \ref{radfunc}). These phenomena could be extended for the case of the closed, periodic orbits demonstrating the azimuthal angle changes $\Delta\phi=2k\pi$ with integer k, when the orbits are again closed at fixed azimuth as related to distant observers. 

In the case of the effects related to a fixed azimuth as related to distant observers it is in principle possible to admit some event (e.g. a collision) in the Keplerian accretion disc causing a release of energy in the form of electromagnetic radiation that would be partly and repeatedly returned back to the same radius as the original event, and at the same azimuth as observed by a distant observer, initializing repetition of the effects under slightly modified internal conditions. It can be expected that such periodic light echo could be characteristic for particular KNS spacetime. Detailed study of the related phenomena will be the task of our future work. 

\section*{Acknowledgements}
Authors acknowledge the Czech Science Foundation Grant No. 16-03564Y.


\bibliographystyle{abbrv}

\end{document}